\documentclass[prd,
,superscriptaddress,
nofootinbib,%
tightenlines ]{revtex4}
\usepackage{epsfig}
\usepackage{mathrsfs}

\usepackage{amsfonts}
\usepackage{color}
\usepackage{amssymb}
\newcommand{\gA}{\stackrel{\circ}{g}_{\!A}}
\newcommand{\mkrig}{{\stackrel{\circ}{m}}}

\usepackage{citesort}
\usepackage{siunitx}
\usepackage{amsfonts}
\usepackage{amsmath}
\allowdisplaybreaks

\usepackage{graphicx}
\usepackage{amssymb}

\newcommand{\ben}{\begin{displaymath}}
\newcommand{\een}{\end{displaymath}}
\newcommand{\be}{\begin{equation}}
\newcommand{\ee}{\end{equation}}
\newcommand{\bea}{\begin{eqnarray}}
\newcommand{\eea}{\end{eqnarray}}
\newcommand{\eq}{\begin{eqnarray}}
\newcommand{\en}{\end{eqnarray}}
\newcommand{\dxn}{{\cal D}x_n}

\newcommand{\dxthree}{{\cal D}x_3}
\newcommand{\dxfour}{{\cal D}x_4}
\newcommand{\ei}{e^{iL{\bf n}{\bf r}}}
\newcommand{\sumn}{\sum_{{\bf n}\neq 0}}
\newcommand{\pisq}{\pi^2}
\newcommand{\Ka}{K_0(|{\bf n}|L\sqrt{g})}
\newcommand{\Kb}{K_1(|{\bf n}|L\sqrt{g})}
\newcommand{\Kc}{K_2(|{\bf n}|L\sqrt{g})}
\newcommand{\intL}{\int_V\frac{d^nk}{(2\pi)^ni}\,}
\newcommand{\dintL}{\overline{\int_V}\frac{d^nk}{(2\pi)^ni}\,}

\begin{document}
\title{Finite volume corrections to forward Compton scattering off the nucleon}

\author{J.~Lozano}
\affiliation{Helmholtz-Institut f\"ur Strahlen- und Kernphysik (Theorie) and Bethe Center for Theoretical Physics,
  Universit\"at Bonn,
53115 Bonn, Germany\\\vspace*{.1cm}}
\author{A.~Agadjanov\thanks{Email: agadjanov@uni-mainz.de}}
\affiliation{PRISMA$^+$ Cluster of Excellence and Institut f\"ur Kernphysik,
Johannes Gutenberg-Universit\"at Mainz, D-55099 Mainz, Germany\\\vspace{.1cm}}
\author{J.~Gegelia}
\affiliation{Ruhr-University Bochum, Faculty of Physics and Astronomy,
Institute for Theoretical Physics II, D-44870 Bochum, Germany\\\vspace*{.1cm}}
\affiliation{Tbilisi State  University,  0186 Tbilisi, Georgia\\\vspace*{.1cm}}
\author{U.-G.~Mei{\ss}ner}
\affiliation{Helmholtz-Institut f\"ur Strahlen- und Kernphysik (Theorie) and Bethe Center for Theoretical Physics,
  Universit\"at Bonn,
  53115 Bonn, Germany\\\vspace*{.1cm}}
 \affiliation{Institute for Advanced Simulation, Institut f\"ur Kernphysik
   and J\"ulich Center for Hadron Physics, Forschungszentrum J\"ulich, D-52425 J\"ulich,
Germany\\\vspace*{.2cm}}
\affiliation{Tbilisi State  University,  0186 Tbilisi, Georgia\\\vspace*{.1cm}}
\author{A.~Rusetsky}
\affiliation{Helmholtz-Institut f\"ur Strahlen- und Kernphysik (Theorie) and Bethe Center for Theoretical Physics,
  Universit\"at Bonn,
53115 Bonn, Germany\\\vspace*{.1cm}}
\affiliation{Tbilisi State  University,  0186 Tbilisi, Georgia\\\vspace*{.1cm}}

\begin{abstract}

\bigskip\bigskip
  
We calculate the spin-averaged amplitude for doubly virtual forward Compton
scattering off nucleons in the framework of manifestly Lorentz-invariant
baryon chiral perturbation theory at complete one-loop order $O(p^4)$. The calculations are
carried out both in the infinite and in a finite volume. The obtained results allow for a
detailed estimation the finite-volume corrections to the amplitude which can be extracted
on the lattice using the background field technique.

\end{abstract}

\maketitle

\section{Introduction}

Recent years have observed a rapidly increasing interest in  calculations
of  nucleon structure functions on the lattice. Different algorithms,
which enable one to extract these quantities from lattice measurements, have been proposed.
For example, in Ref.~\cite{Ji:2013dva} a method for a direct
calculation of the quark and gluon distribution functions on  Euclidean lattices by Lorentz-boosting of
the nucleons was suggested. Recently, a lattice calculation of the Euclidean
four-point function, describing
the virtual Compton amplitude, and its relation to the leptoproduction cross section
has been considered~\cite{Fukaya:2020wpp}.
In the present paper we shall concentrate on an alternative proposal which is
based on the use of the background field technique (or, the Feynman-Hellmann method)
for measuring the forward doubly virtual Compton scattering amplitude off
nucleons, see Refs.~\cite{Horsley:2012pz,Chambers:2014qaa,Chambers:2015bka,Chambers:2017dov,Can:2020sxc,Agadjanov:2016cjc,Agadjanov:2018yxh}.
This amplitude is directly related
to the moments of the structure functions. For a review of the present status
of lattice studies of the structure functions, see, e.g., Ref.~\cite{Lin:2017snn}.

\medskip

In Refs.~\cite{Horsley:2012pz,Chambers:2014qaa,Chambers:2015bka,Chambers:2017dov,Can:2020sxc,Agadjanov:2016cjc,Agadjanov:2018yxh}
a comprehensive theoretical assessment of the feasibility
of the extraction of the Compton amplitude has been carried out. Here, one has to note that
a similar technique has been already successfully used for the extraction of the magnetic moments
and polarizabilities of certain hadrons~\cite{Savage:2016kon,Chang:2015qxa,Detmold:2006vu}. The study of 
Compton scattering, however, implies another level of sophistication. Namely, whereas the static
characteristics
of the nucleon can be measured in constant background magnetic and electric fields, the dependence
of the forward Compton amplitude on the photon virtuality, $q^2=-Q^2$, cannot be studied similarly.
Therefore, one has to use periodic background fields (in space), which enable one to consider
non-zero values of the photon three-momentum, while the time-component of
the photon momentum $q$ stays zero. Several subtle issues had to be addressed
in this context, for example, a consistent realization of the periodic background field on a finite
lattice~\cite{Davoudi:2015cba}, or the zero-frequency limit~\cite{Agadjanov:2018yxh}.
It must also be mentioned that, according to Ref.~\cite{Hu:2007ts},
the interpretation of the lattice measurements, which are done in a finite
volume, might be ambiguous both for constant and periodic fields.
More precisely, the quantity that is obtained as a
result of such a measurement (for instance, the polarizability)
could be different from what one has previously identified as a finite-volume
counterpart of the polarizability. This point of view has been
countered in Ref.~\cite{Agadjanov:2018yxh}, where it has been argued that the finite-volume
lattice results allow for a unique interpretation in terms of  well-defined
physical quantities (at least when the photon virtuality is not zero).
Since this issue is important in the context of the
problem considered here, we shall briefly address it later. 

\medskip

It should be stressed that the measurement of the forward Compton amplitude
on the lattice is a useful endeavor by itself, even beyond its relation
to the nucleon structure functions. Indeed, let us point out
 that the forward Compton amplitude represents an important building block in many
long-standing fundamental problems that have recently come under a renewed
scrutiny. In particular, the knowledge of this amplitude is needed for the 
evaluation of the Lamb shift in  muonic hydrogen~\cite{Carlson:2011zd},
as well as the proton-neutron mass difference. The study of the latter problem
has a decades-long history~\cite{Cottingham:1963zz,Gasser:1974wd}, but still
continues to attract quite some interest that is reflected in a string of
recent publications~\cite{WalkerLoud:2012bg,Gasser:2015dwa,Erben:2014hza,Thomas:2014dxa,Gasser:2020mzy,Gasser:2020hzn}.
To a large part, this upsurge of interest can be related to the fact that
the present lattice studies are in a position to
address the calculation of the purely electromagnetic proton-neutron mass shift
in QCD plus QED and hence the results of phenomenological determinations can be directly confronted with
lattice data. Further, in Refs.~\cite{Gasser:2020mzy,Gasser:2020hzn}, under the assumption that the
high-energy behavior of the Compton amplitude is fully determined by Reggeon exchange (the so-called
Reggeon dominance hypothesis), a sum rule has been derived that involves this amplitude in a particular
kinematics
(a variant of this sum rule has been known in the literature already for fifty years~\cite{Elitzur:1970yy}).
Notably, the latter enables one to express the Compton amplitude through the experimentally measured
electroproduction cross sections. Calculations on the basis of the above sum rule have been performed
recently~\cite{Gasser:2020hzn}, where the uncertainties emerging from the use of all
presently available experimental input have been thoroughly analyzed.
A direct evaluation of the Compton amplitude on the lattice would allow one to compare the outcomes
of these two different theoretical calculations. Should it happen that the results are very different,
this could be attributed to the failure of the Reggeon dominance hypothesis, i.e.,  to the existence of
so-called fixed poles in the Compton amplitude. At present, we are not aware of any mechanism within QCD
that would lead to such poles. Hence, their discovery would challenge our understanding of the asymptotic
behavior of QCD and stimulate a quest
for new mechanisms, which are responsible for this behavior.

\medskip

One of the most important questions, which so far has not been addressed in the context
of the extraction of the Compton  amplitude from lattice data,
is the issue of the finite-volume corrections to the physical  quantities of
interest. It is very important to estimate, prior to performing any
calculations on the lattice, how large lattices should be used to suppress
the unwanted finite-volume artifacts. Note that, even though on
general grounds these artifacts are exponentially small, due to possible large
prefactors they might be still substantial for the presently used lattice sizes.
The systematic study of this problem that is carried out in what follows
within the framework of Baryon Chiral Perturbation Theory (BChPT) at 
order $O(p^4)$ is intended to fill this gap.

\medskip

The layout of the paper is the following. In  Sect.~\ref{sec:def}
we collect all definitions and input, which will be needed later.
This concerns both purely infinite-volume calculations as well as the
finite-volume setting used on the lattice for the extraction of the Compton amplitude.
Further, the calculation of the Compton amplitude is carried out 
in the infinite as well as in a finite volume. Namely,  Sect.~\ref{sec:infinite}
contains the full expression of the infinite-volume Compton amplitude
at $O(p^4)$ 
in BChPT. Also, a comparison to the results
available in the literature is carried out. The expression
of the finite-volume amplitude at the same order is given in Sect.~\ref{sec:finite}.
In Sect.~\ref{sec:numerics}, the results of the numerical estimations of
the finite-volume artifacts are discussed.
Sect.~\ref{sec:concl} contains our conclusions.

\section{Basic definitions and notations}
\label{sec:def}

\subsection{Doubly-virtual Compton scattering in forward direction
  in the infinite volume}
\label{DVCS}
 
In this paper we follow the notations of Ref.~\cite{Gasser:2015dwa}.
In order to render the paper self-contained, below we collect all formulae
that will be used in the infinite-volume calculations.
The Compton scattering amplitude is defined as:
\begin{equation}
  \hat T^{\mu\nu}(p',s',q'| p,s,q) = \frac{i}{2} \int d^4 x\, e^{i q\cdot x}
  \langle p',s' | T j^\mu(x) j^\nu(0) | p,s\rangle \,,
\label{ComptAmpl}
\end{equation}
where $(p',s')$ and $(p,s)$ are the four-momenta and spin projections of incoming and outgoing
nucleon, respectively, and $q$ and $q'$ are the momenta of the (virtual) photons in the initial and
final state, respectively.  Further, $j^\mu$ denotes the electromagnetic current.
The state vectors of the nucleon are normalized as:
\begin{equation}
 \langle p', s' | p, s\rangle = 2p^0 (2\pi)^3 \delta^{(3)} ({\bf p}\,'-{\bf p}\,)\delta_{s's}\,.
\label{ComptAmpl1}
\end{equation}
We define the unpolarized forward scattering amplitude as an average over the
nucleon spins:
\begin{equation}
T^{\mu\nu}(p,q) = \frac{1}{2} \sum_s \hat T^{\mu\nu}(p,s,q| p,s,q) \,.
\label{UnpolComptAmpl1}
\end{equation}
Using Lorentz-invariance, current conservation and parity, this amplitude can be expressed through
two invariant amplitudes: 
\begin{eqnarray}\label{eq:K1K2}
T^{\mu\nu}(p,q) & = & T_1(\nu,q^2) K_1^{\mu\nu} + T_2(\nu,q^2) K_2^{\mu\nu} \,,\nonumber\\[2mm]
K_1^{\mu\nu} & = & q^\mu q^\nu - g^{\mu\nu} q^2\,,\nonumber\\[2mm]
K_2^{\mu\nu} & = & \frac{1}{m^2}  \left\{ \left( p^\mu q^\nu + p^\nu q^\mu \right) p\cdot q
-g^{\mu\nu} (p\cdot q)^2 -p^\mu p^\nu q^2\right\}  \,,
\label{UnpolComptAmplPar}
\end{eqnarray}
where $\nu = p\cdot q/m$ and $m$ is the nucleon mass.

\medskip

At this place we mention that the choice of the invariant amplitudes is not
unique. In the literature, another choice is often made,
using the set $\hat T_1$, $\hat T_2$ with
$\hat T_1=q^2T_1+\nu^2T_2$ and $\hat T_2=-\nu^2T_2$. This alternative choice,
however, produces kinematic singularities, which complicate
the discussion of the asymptotic behavior of these amplitudes. As a result, the issue with
the fixed poles may become obscure. For further details on this subject
we refer the reader to Refs.~\cite{Gasser:2015dwa,Gasser:2020hzn} and also
to~\cite{Leutwyler:2015jga,Hoferichter:2019jhr}. Further, in
Refs.~\cite{Gasser:2020mzy,Gasser:2020hzn}, another set of invariant amplitudes
$\bar T=T_1+\frac{1}{2}\,T_2$, $T_2$ has been introduced instead of $T_1$, $T_2$.
The advantage of using this set consists in the fact that the leading asymptotic behavior of
$\bar T$ at large values of $Q^2$ is governed by  spin-0 operators, whereas
the contribution from the spin-2 operators in the operator product expansion cancels in this
particular linear combination. A thorough discussion of this question is given in Ref.~\cite{Gasser:2020hzn}.
Here, we only mention that the set $\bar T,T_2$ is obviously free from the kinematic singularities,
as well as the set $T_1,T_2$.

\medskip

Further, the invariant amplitudes can be split into the elastic and inelastic
(or, equivalently, into the Born and non-Born) parts.
Again note that such a splitting is not uniquely defined and the definition
of Ref.~\cite{Gasser:2015dwa} differs from the ones used in
Refs.~\cite{Drechsel:2002ar,Alarcon:2013cba}\footnote{See, e.g.,
  Refs.~\cite{Scherer:1996ux,Gasser:2020hzn} for a 
  general discussion of the issue of non-uniqueness of the Born part
  of the Compton scattering amplitude.}. Here we would like to mention only
that the definition, which will be used in the following, unambiguously follows
from the requirement that the elastic amplitude vanishes in the limit
$\nu\to\infty$,  for fixed $q^2$, and thus obeys an unsubtracted dispersion relation
in the variable $\nu$. Under this requirement, the elastic part is given by:
\begin{eqnarray}
T_1^{\sf el}(\nu,q^2) & = & \frac{4 m^2 q^2 \left\{  G_E^2(q^2) - G_M^2(q^2) \right\}  }{(4m^2 \nu^2-q^4)(4m^2-q^2)}\,,
\nonumber\\
T_2^{\sf el}(\nu,q^2) & = & - \frac{4 m^2  \left\{ 4 m^2 G_E^2(q^2) - q^2 G_M^2(q^2) \right\}  }{(4m^2 \nu^2-q^4)(4m^2-q^2)}\,,
\label{T1T2FF}
\end{eqnarray}
where $G_E,G_M$ denote the electric and magnetic (Sachs) form factors of the nucleon.

\medskip

The inelastic invariant amplitudes are defined as $T_i^{\sf inel}=T_i-T_i^{\sf el}$,
with $i=1,2$. The amplitudes $T_i^{\sf inel}$ obey dispersion relations in the variable $\nu$:
\eq
T_1^{\sf inel}(\nu,q^2)&=&T_1^{\sf inel}(\nu_0,q^2)+2(\nu^2-\nu_0^2)\int_{\nu_{\sf th}}^\infty
\frac{\nu'd\nu'\,V_1(\nu',q^2)}{({\nu'}^2-\nu_0^2)({\nu'}^2-\nu^2-i\varepsilon)}\, ,
\nonumber\\[2mm]
T_2^{\sf inel}(\nu,q^2)&=&2\int_{\nu_{\sf th}}^\infty\frac{\nu'd\nu'\,V_2(\nu',q^2)}{{\nu'}^2-\nu^2-i\varepsilon}\, .
\en
Here, one has already taken into account the fact that, according to  Regge
theory, the dispersion relations for $T_1^{\sf inel},\,T_2^{\sf inel}$ require
one subtraction and no subtractions, respectively. The lower integration
limit is equal to $\nu_{\sf th}=(W_{\sf th}^2-m^2-q^2)/(2m)$, with
$W_{\sf th}=m+M_\pi$, where $M_\pi$ is the pion
mass. The quantities $V_1, \, V_2$ denote the absorptive parts of $T_1^{\sf inel},\,T_2^{\sf inel}$. They can
be expressed through the experimentally observed total (transverse, longitudinal)  electroproduction cross
sections $\sigma_T(\nu,q^2),\,\sigma_L(\nu,q^2)$.

\medskip

The choice of the subtraction point $\nu_0$ is arbitrary. In the literature,
the choice $\nu_0=0$ is often used. The quantity $S_1^{\sf inel}(q^2)=T_1^{\sf inel}(0,q^2)$ is usually
referred to as the subtraction function. Analogously,
one can define the full subtraction function that includes the elastic part as well: $S_1(q^2)=S_1^{\sf el}(q^2)
+S_1^{\sf inel}(q^2)=T_1(0,q^2)$.
At $q^2=0$ the inelastic part of the subtraction function is given by
\eq
S_1^{\sf inel}(0)=-\frac{\kappa^2}{4m^2}-\frac{m}{\alpha}\,\beta_M\, ,
\en
where $\kappa,\,\beta_M$ denote the anomalous magnetic moment and the
magnetic polarizability of the nucleon, respectively,
and  $\alpha \simeq 1/137$ is the electromagnetic fine-structure constant.

\medskip

Recently, a different subtraction function was introduced in Refs.~\cite{Gasser:2020mzy,Gasser:2020hzn} .
The subtraction point has been chosen at $\nu_0=iQ/2$,  where $Q=\sqrt{-q^2}$. The new
subtraction function is expressed through the amplitude $\bar T$:
\eq
\bar S(q^2)=\bar T^{\sf inel}(\nu_0,q^2)\, .
\en
At $Q^2=0$ one has:
\eq
\bar S(0)=-\frac{\kappa^2}{4m^2}+\frac{m}{2\alpha}\,(\alpha_E-\beta_M)\, .
\en
The two subtraction functions are closely related to each other. Namely,
the difference $S_1^{\sf inel}(q^2)-\bar S(q^2)$ is given 
through a convergent integral over the experimentally measured
electroproduction cross sections. Hence, it suffices to calculate one
of these subtraction functions. Since the choice $\nu_0=0$, in contrast to
$\nu_0=iQ/2$, can be implemented on the lattice in a straightforward
manner~\cite{Agadjanov:2016cjc,Agadjanov:2018yxh}, we stick to this choice.

\subsection{Extraction of the subtraction function on the lattice}

Below, we shall collect all formulae which are needed for the extraction
of the subtraction function on the lattice with the use of the background
field method. More details are contained in the original
papers~\cite{Agadjanov:2016cjc,Agadjanov:2018yxh}. Here, we consider
the nucleon placed in a periodic magnetic field on the lattice with a spatial size $L$
(the temporal size of the lattice is assumed to be much larger and is effectively set to infinity).
The configuration of the magnetic field is chosen as:
\eq
   {\bf B}=\bigl(0,0,-eB\cos(\omega {\bf n}{\bf x})\bigr)\, ,\quad\quad
   {\bf n}=(0,1,0)\, ,
\en
where $e$ denotes the proton charge.
Because of the periodic boundary conditions, the available values of $\omega$
are quantized:\footnote{In fact, this is one of the possible realizations of
the external field on the lattice. An alternative implies the quantization of the magnitude of the field,
rather than its frequency~\cite{Davoudi:2015cba}. However, as was demonstrated in
Ref.~\cite{Agadjanov:2018yxh}, the present realization provides an optimal framework for the
extraction of the subtraction function at non-zero values of the momentum transfer.}
\eq
   \omega=\frac{2\pi n}{L}\, .
\en
The energy levels of a nucleon in the magnetic field depend on the projection of the nucleon spin
along the $z$-axis.
In Ref.~\cite{Agadjanov:2018yxh} it has been shown that the spin-averaged level shift in the magnetic field
with a given configuration is given by:
\eq\label{eq:deltaE}
\delta E=-\frac{1}{4m}\,\biggl(\frac{eB}{\omega}\biggr)^2T_L^{11}(p,q)+O(B^3)\, ,\quad\quad p^\mu=(m,{\bf 0})\, ,
\quad q^\mu=(0,0,\omega,0)\, .
\en
Note that here $T_L^{11}(p,q)$ denotes the $11$-component of the full Compton scattering
amplitude in a finite volume (in other words, $T_L^{11}(p,q)$ includes both inelastic and elastic parts).
Further, $q^2=-\omega^2$. Hence, placing a  nucleon in the periodic magnetic field enables one to extract the
amplitude at non-zero (albeit discrete) values of $q^2<0$.
The other variable is $\nu=p\cdot q/m=0$ in the given kinematics. Thus, in order to
obtain a non-zero values of $\nu$, one has to put the nucleon in a moving frame.

\medskip
   
Note also that due to the lack of Lorentz invariance on a finite
hypercubic lattice, the decomposition of this amplitude
into invariant amplitudes in a form given in Eq.~(\ref{eq:K1K2}) does in general not
hold. However, all quantities in Eq.~(\ref{eq:deltaE}) are well-defined in a finite volume. For example,
in perturbation theory, $T^{11}(p,q)$ is given by a sum of all diagrams at a given order, where
all momentum integrations are replaced by finite-volume momentum sums.
In the infinite-volume limit one has:
\eq
\lim_{L\to\infty}T_L^{11}(p,q)=T^{11}(p,q)=-\omega^2S_1(q^2)\, .
\en
The finite-volume corrections in the above formula are suppressed by a factor $\exp(-M_\pi L)$, multiplied
by powers of $L$. As already mentioned in the introduction, despite the exponential
factor, the corrections can  still be sizable for the present-day lattices. Last but not least, let us
stress once more that, for all values of $L$,
Eq.~(\ref{eq:deltaE}) enables one to extract a perfectly well-defined
quantity $T_L^{11}(p,q)$, which in the infinite volume limit yields the
quantity $S_1(q^2)$ that we are after. This demonstrates explicitly
that in this setup one could avoid any ambiguous
interpretation of the results as mentioned in Ref.~\cite{Hu:2007ts}.

\medskip

We conclude this section by briefly specifying the scope and aims of the present paper.
It is clear that the extraction of the Compton amplitude on the lattice can be carried
out only in a restricted kinematical domain. For instance, if the variable
$\nu$ lies above the pion production threshold $\nu_{\sf th}=(2m)^{-1}(W_{\sf th}^2-m^2-q^2)$, 
then the extracted matrix element does not possess an infinite-volume limit. This can be seen
immediately since, in the infinite-volume limit, the corresponding amplitude
is complex, whereas the amplitude extracted from an Euclidean lattice is always real. Hence, in order
to arrive at the infinite-volume amplitude, one has either to take into account the proper
Lellouch-L\"uscher factor
in analogy with Ref.~\cite{Lellouch:2000pv}, see also Refs.~\cite{Hansen:2012tf,Briceno:2012yi,Briceno:2014uqa,Briceno:2015csa,Bernard:2012bi,Agadjanov:2014kha,Agadjanov:2016fbd} (a first step in this direction
has been made in Ref.~\cite{Briceno:2019opb}),
or to use an approach that resembles the optical potential method of Ref.~\cite{Agadjanov:2016mao}, see
also Refs.~\cite{Hansen:2017mnd,Bulava:2019kbi}.
All this is very complicated and  not even needed to achieve the goals we have stated in the
beginning. Indeed, given the subtraction function, which is obtained from the Compton amplitude at $\nu=0$,
one may restore the whole Compton amplitude by using dispersion relations.
The whole uncertainty  related to the fixed poles then resides in the
subtraction function, and the rest is uniquely determined by analyticity, unitarity and the
experimental input.

\subsection{Effective Lagrangian}
\label{effective_Lagrangian}

In this paper the forward Compton scattering amplitude will be calculated in BChPT
at order $p^4$, both in the infinite and in a finite volume.
Below we specify the effective Lagrangian with the pions and nucleons, which
is needed for such a calculation.  The leading-order effective Lagrangian
of pions interacting with external sources has the form \cite{Gasser:1984yg}: 
\begin{equation}
{\cal L}_\pi^{(2)} =  \frac {F^2}{4}\, \langle D_\mu U  (D^\mu U)^\dagger \rangle + \frac{F^2}{4}\,
\langle\chi U^\dagger +U \chi^\dagger\rangle \,,
\label{PionAction}
\end{equation}
where $\chi= 2 B(s+i p)$, $D_\mu U=\partial_\mu U -i r_\mu U +i U l_\mu $
and the $2\times 2$ matrix $U$ represents the pion field. The parameter
$B$ is related to the quark condensate, $F$ is the pion decay constant in the two-flavor chiral limit,
and $s$, $p$, $l_\mu =v_\mu-a_\mu $ and $r_\mu =v_\mu + a_\mu $ are external sources. The notation
$\langle\cdots\rangle$ denotes the trace in flavor space.

\medskip

The full order four effective Lagrangian of nucleons interacting with
pions and external fields is given in Ref.~\cite{Fettes:2000gb}. Below we
specify only those terms, which contribute in our calculations:\footnote{Note
that our definitions of the low-energy constants agree to those of Ref.~\cite{Fettes:2000gb} except for
$c_6$ and $c_7$, which are related as $c_6=c_6^{FMMS}/4\mkrig$ and
$c_7=(c_6^{FMMS}+c_7^{FMMS})/2\mkrig$ (here $c_6^{FMMS}$ and $c_7^{FMMS}$ denote the corresponding couplings of
Ref.~\cite{Fettes:2000gb}), see Ref.~\cite{Djukanovic:2008skh}.}
\begin{eqnarray}
  {\cal L}_{\rm \pi N}&=&{\cal L}_1+{\cal L}_2+{\cal L}_3+{\cal L}_4+\cdots\, ,
  \nonumber\\[2mm]
{\cal L}_1& = & \bar\Psi \, i \gamma^\mu D_\mu \Psi-\mkrig \bar\Psi\Psi
+\frac{\gA}{2}\, \bar\Psi \gamma^\mu \gamma_5 u_\mu \Psi\, ,
\nonumber\\[2mm]
{\cal L}_2&=& c_1 \langle \chi_+\rangle  \bar\Psi  \Psi
- \frac{c_2}{8 \mkrig^2}  \langle u^\alpha u^\beta \rangle
\left( \bar\Psi \left\{ D_\alpha, D_\beta\right\}  \Psi+{\rm h.c.}
 \right)
+ \frac{c_3}{2} \, \langle u^\mu u_\mu\rangle  \bar\Psi  \Psi  
\nonumber\\[2mm]
                      &+&
                          i\frac{c_4}{4}\,\bar\Psi[u_\mu,u_\nu]\sigma^{\mu\nu}\Psi
+ \frac{c_6}{2} \bar\Psi \,\sigma^{\mu\nu} \tilde F^+_{\mu\nu} \Psi
+ \frac{c_7}{8} \bar \Psi \,\sigma^{\mu\nu} 
\langle F^+_{\mu\nu} \rangle \Psi  + \cdots\, ,
\nonumber\\[2mm]
{\cal L}_3&=&  \biggl(\frac{i d_6}{2\mkrig} \bar\Psi [D^\mu, \tilde F^+_{\mu\nu}] D^\nu \Psi
+ {\rm h.c.}\biggr)
+ \biggl(\frac{i d_7}{2\mkrig} \bar\Psi [D^\mu, \langle  F^+_{\mu\nu} \rangle ] D^\nu \Psi
+ {\rm h.c.}\biggr)  +\cdots\, ,
           \nonumber\\[2mm]
{\cal L}_4& =& -\frac{e_{54}}{2} \bar\Psi [D^\lambda, [D_\lambda,  \langle  F^+_{\mu\nu} \rangle ]] \sigma^{\mu\nu}  \Psi - \frac{e_{74}}{2} \bar\Psi [D^\lambda, [D_\lambda,  \tilde  F^+_{\mu\nu} ]] \sigma^{\mu\nu}  \Psi +
e_{89} \bar\Psi  \langle  F^+_{\mu\nu} \rangle\langle  F^{+\mu\nu} \rangle   \Psi  \nonumber\\[2mm]
&-& \biggl(\frac{e_{90}}{4\mkrig^2}  \bar\Psi  \langle  F^+_{\lambda\mu} \rangle\langle  F^{+\lambda\alpha}  \rangle g_{\alpha\nu} D^{\mu\nu} \Psi  + {\rm h.c.}\biggr)  
+e_{91} \bar\Psi  \tilde  F^+_{\mu\nu} \langle  F^{+\mu\nu} \rangle   \Psi
\nonumber\\[2mm]
&-& \biggl(\frac{e_{92}}{4\mkrig^2}  \bar\Psi  \tilde  F^+_{\lambda\mu} \langle  F^{+\lambda \alpha} \rangle  g_{\alpha\nu} D^{\mu\nu} \Psi  + {\rm h.c.}\biggr)
+ e_{93} \bar\Psi  \langle \tilde  F^+_{\mu\nu} \tilde  F^{+\mu\nu} \rangle   \Psi
\nonumber\\[2mm]
&-& \biggl(\frac{e_{94}}{4\mkrig^2}  \bar\Psi  \langle \tilde  F^+_{\lambda\mu}  F^{+\lambda \alpha} \rangle  g_{\alpha\nu} D^{\mu\nu} \Psi  + {\rm h.c.}\biggr) 
-\frac{e_{105}}{2} \bar\Psi  \langle  F^+_{\mu\nu} \rangle \langle  \chi_+ \rangle  \sigma^{\mu\nu} \Psi
\nonumber\\[2mm]
&-&\frac{e_{106}}{2} \bar\Psi  \tilde  F^+_{\mu\nu}  \langle  \chi_+ \rangle  \sigma^{\mu\nu} \Psi
- \biggl(\frac{e_{117}}{8 \mkrig^2}  \bar\Psi  \langle  F^-_{\lambda\mu}  F^{-\lambda\alpha} + F^+_{\lambda\mu}  F^{+\lambda\alpha}   \rangle g_{\alpha\nu} D^{\mu\nu} \Psi  + {\rm h.c.}\biggr)
\nonumber\\[2mm]
&+&  \frac{e_{118}}{2}  \bar\Psi  \langle  F^-_{\mu\nu}  F^{-\mu\nu} + F^+_{\mu\nu}  F^{+\mu\nu} \rangle \Psi  +\cdots\,  ,
\label{MAction}
\end{eqnarray}
where
\begin{eqnarray}
\sigma^{\mu\nu} &=& \frac{i}{2} \left( \gamma^\mu \gamma^\nu -\gamma^\nu \gamma^\mu \right)\,,\nonumber\\[2mm]
\tilde X &=&  X-\frac{1}{2} \langle X\rangle\, ,\nonumber\\[2mm]
D_\mu \Psi &=& \partial_\mu\Psi + \left(\Gamma_\mu  -i v_\mu^{(s)}\right)\Psi, \nonumber\\[2mm]
\Gamma_\mu & = & \frac{1}{2} \left[ u^\dagger \partial_\mu u  +u \partial_\mu u^\dagger -i (u^\dagger r_\mu u+u l_\mu u^\dagger )\right]\, ,\nonumber\\[2mm]
D^{\mu\nu} & = & D^\mu D^\nu +D^\nu D^\mu \,,\nonumber\\[2mm]
u_\mu & = & i \left[ u^\dagger \partial_\mu u  - u \partial_\mu u^\dagger -i (u^\dagger r_\mu u - u l_\mu u^\dagger )\right]\,,\nonumber\\[2mm]
  F_{\mu\nu}^\pm &=& u F_{L\mu\nu} u^\dagger\pm u^\dagger F_{R\mu\nu} u\,,
                     \nonumber\\[2mm]
F_{R\mu\nu} &=& \partial_{\mu} r_\nu - \partial_\nu r_\mu-i[r_\mu,r_\nu] \,, \nonumber\\[2mm]
F_{L\mu\nu} &=& \partial_{\mu} l_\nu - \partial_\nu l_\mu-i[l_\mu,l_\nu] \,,  \nonumber\\[2mm]
\chi_+ & = & u^\dagger \chi u^\dagger+u \chi^\dagger u\,.
\label{adddefs}
\end{eqnarray}
In the above expressions, $\Psi$ denotes the nucleon field, $u=\sqrt{U}$ contains pion
fields, $v^{(s)}_\mu$ is the part of the vector current that is proportional to the
unit matrix in the flavor space, $\mkrig$ and $\gA$ denote the nucleon mass and the axial-vector coupling
constant in the chiral limit,  
and $c_i,d_i,e_i$ are the low-energy constants at $O(p^2)$, $O(p^3)$ and $O(p^4)$,
respectively. It is assumed that the above Lagrangian is used to generate
Feynman diagrams, which will be evaluated by using the EOMS renormalization scheme.
Acting in this manner,  there is no need to explicitly display the counterterms  of the
effective Lagrangian that remove the power-counting breaking contributions from the loop diagrams.
Hence, the numerical values of the finite parts of the LECs correspond to the EOMS scheme.

\medskip

To obtain the expressions that correspond to diagrams with external photons,
we need to substitute $s={\cal M}$, $p=a_\mu=0$,
$v_\mu = -e \tau_3 {\cal A}_\mu/2$ and $v^{(s)}_\mu = -e {\cal A}_\mu/2$,
where ${\cal A}_\mu$ is the electromagnetic field. We work in the isospin
limit $m_u=m_d=\hat m$, and $M^2=2 B \hat m$ is the pion mass at  leading
order.

\section{Infinite volume}
\label{sec:infinite}

\subsection{The workflow}

We calculate doubly-virtual Compton scattering on the proton and
on the neutron separately up-to-and-including $O(p^4)$.
There are tree and one-loop diagrams contributing to this process
at the given accuracy.  Standard power-counting rules of low-energy chiral
effective field theory apply to these diagrams \cite{Weinberg:1991um,Ecker:1994gg}. More specifically,
we are assigning chiral order $-2$ to pion propagators, nucleon propagators
count as of order $-1$, the interaction vertices originating from the effective Lagrangian 
of the order $N$ count also as of order $N$, and the integrations over
loop momenta are assigned order $4$.  While the power counting rules are directly applicable
to the tree diagrams, the loop diagrams of our 
manifestly Lorentz-invariant formalism contain pieces that violate the
counting rules. However, power-counting violating terms are polynomials
in the quark masses and external momenta and thus can be systematically absorbed
in the redefinition of the parameters  of the effective Lagrangian. In our
calculations, we use dimensional regularization supplemented with the EOMS
scheme~\cite{Gegelia:1999gf,Fuchs:2003qc}. In this scheme, the polynomials
in quark masses and external momenta which break the power counting up to a given chiral order,
are dropped from each diagram. This naturally guarantees that the
renormalized one-loop diagrams satisfy power counting\footnote{
  The terms which break power counting can be also systematically removed by using the
  heat kernel method, see Refs.~\cite{Du:2016ntw,Du:2016xbh}. }. 
Note that, in difference to the results of the heavy baryon formalism~\cite{Jenkins:1990jv,Bernard:1992qa},
our expressions for loop diagrams contain an infinite number of higher-order
terms. 

\medskip 

In the calculations of the diagrams we used the programs
FeynCalc \cite{Mertig:1990an,Shtabovenko:2016sxi} and
X-package \cite{Patel:2016fam}. We have verified that the sum of all
tree and one-loop diagrams satisfy  current conservation.  
This guarantees that the whole unpolarized  amplitude is parameterized
in terms of two invariant functions $T_1$ and $T_2$. However, this does not
apply to the individual diagrams. In order to extract the contributions
of separate diagrams to $T_1$ and $T_2$, we single out the contributions
to the coefficients of the structures  $q^\mu q^\nu$ and $-q^2p^\mu p^\nu/m^2$,
since these appear exclusively in $K_1^{\mu\nu}$ and $K_2^{\mu\nu}$,
respectively, see Eq.~(\ref{eq:K1K2}). The individual contributions, which are
listed below, should be interpreted in this sense.
Finally, using the LSZ scheme, we add these contributions and multiply the result with the
residue of the nucleon propagator at the pole corresponding to one-nucleon state.
Acting in this manner, one gets the
full expressions of $T_1(\nu,q^2)$ and $T_2(\nu,q^2)$ we are interested in.

\medskip

Next, we need an algorithm for the separation of the elastic contributions
from $T_1$ and $T_2$. For instance, let us first extract the $s$-channel pole. To this
end, we multiply the full expressions for $T_1(\nu,q^2)$, $T_2(\nu,q^2)$
by $2m\nu+q^2$ and then substitute $\nu=-q^2/(2m)$. Apparently, as a result of
this procedure, one obtains the residue of the $s$-channel pole. In order to determine the residue
of the $u$-channel pole, we multiply the amplitudes by $2m\nu-q^2$. Finally,
adding both pole terms together, we arrive at the elastic amplitudes
$T_1^{\sf el}(\nu,q^2)$ and $T_2^{\sf el}(\nu,q^2)$.
These vanish in the limit when $\nu\to\infty$ and $q^2$ stays fixed. This
exactly coincides with our definition of the elastic amplitudes.

\subsection{The amplitude in the infinite volume}

\begin{figure}[t!]
	\centering

        \includegraphics[width=0.9\linewidth]{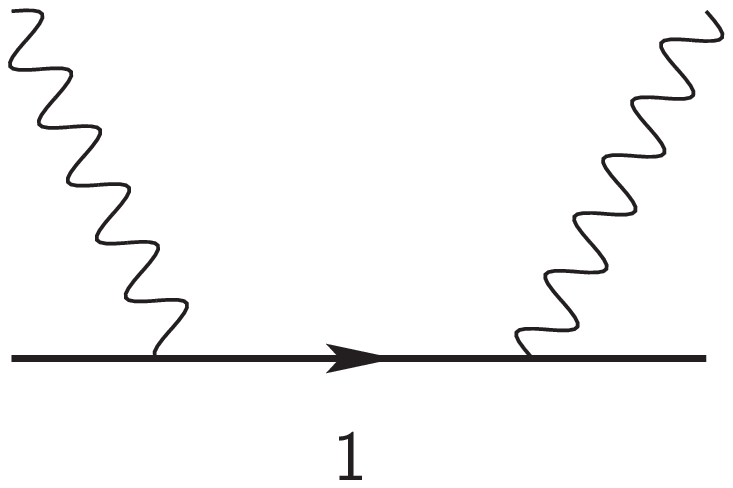}
	\caption{Tree diagram contributing at $O(p)$. Solid and wiggly lines
          denote  nucleons and photons, respectively. The crossed diagram is not shown.
        }
	\label{fig:O1}
\end{figure}

\begin{figure}[t!]
	\centering

        \includegraphics[width=0.9\linewidth]{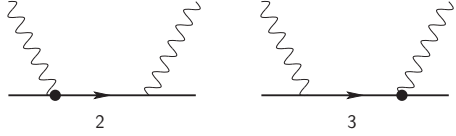}
	\caption{Tree diagrams contributing at $O(p^2)$. Solid and wiggly lines
          denote nucleons and photons, respectively. The filled  circles
          are  vertices from the second-order Lagrangian ${\cal L}_2$. Crossed diagrams are not shown.}
	\label{fig:O2}
\end{figure}

\begin{figure}[t!]
	\centering
        \includegraphics[width=0.9\linewidth]{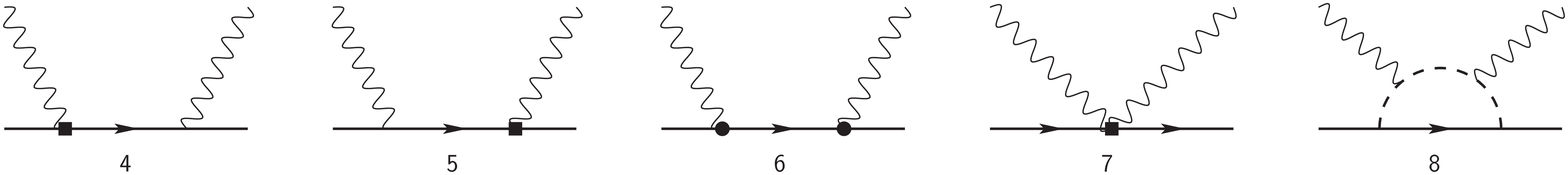}

        \includegraphics[width=0.9\linewidth]{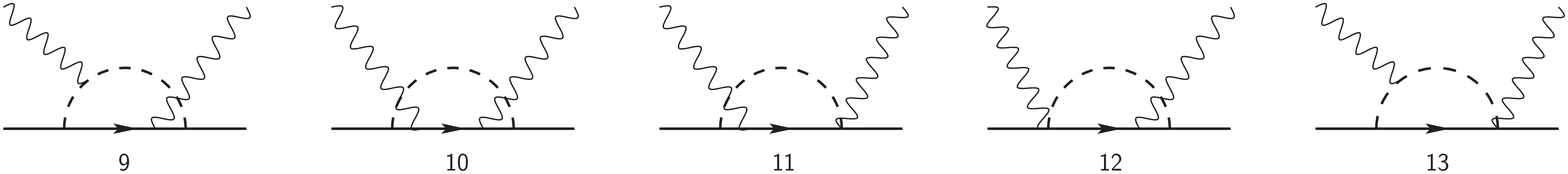}

        \includegraphics[width=0.9\linewidth]{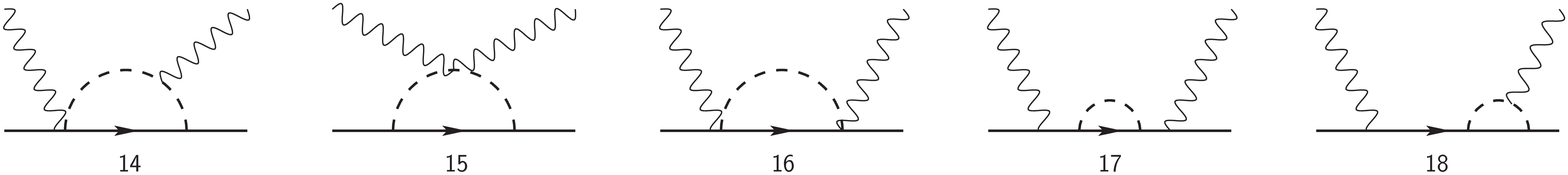}

        \includegraphics[width=0.9\linewidth]{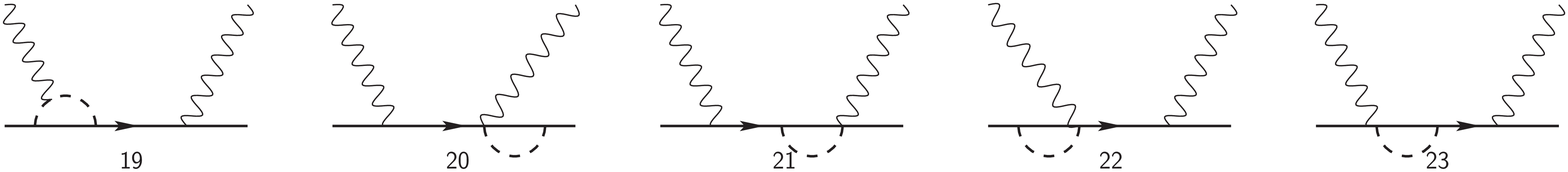}

        \includegraphics[width=0.9\linewidth]{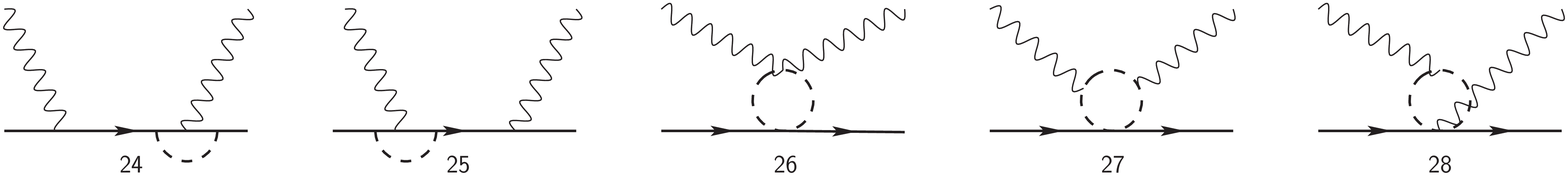}

        \includegraphics[width=0.9\linewidth]{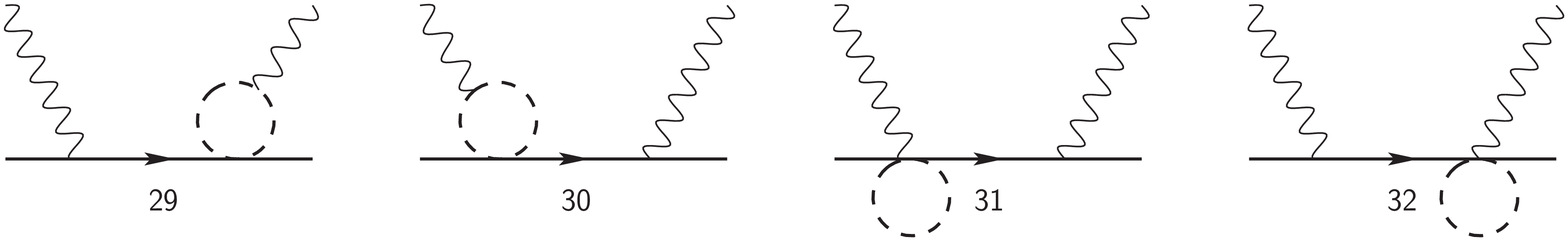}

      \caption{Tree and loop diagrams contributing at $O(p^3)$. Solid, dashed
          and wiggly lines denote nucleons, pions and photons, respectively. Filled circles and squares
          represent vertices from  ${\cal L}_2$ and ${\cal L}_3$, respectively.
          Crossed diagrams are not shown.}
        
	\label{fig:O3}
\end{figure}

\begin{figure}[t!]
	\centering
        \includegraphics[width=0.9\linewidth]{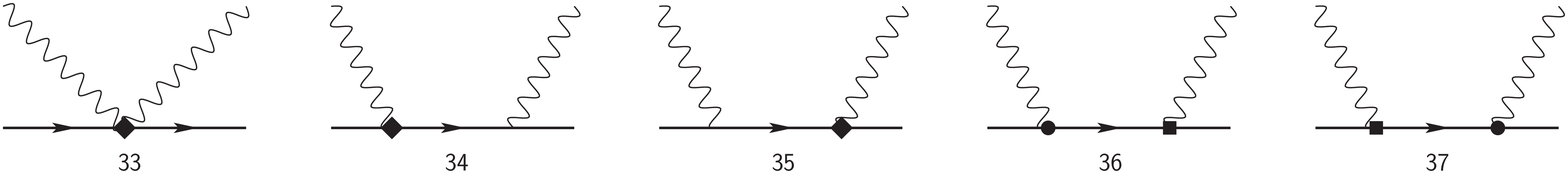}

        \includegraphics[width=0.9\linewidth]{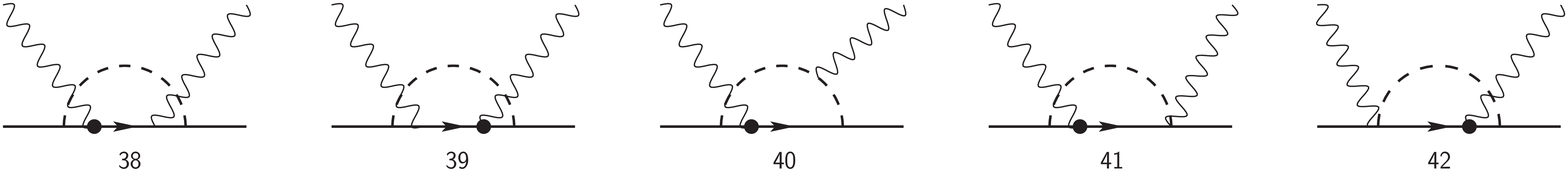}

        \includegraphics[width=0.9\linewidth]{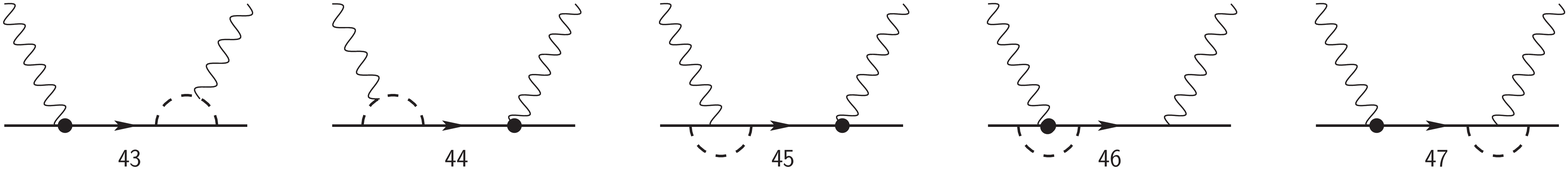}

        \includegraphics[width=0.9\linewidth]{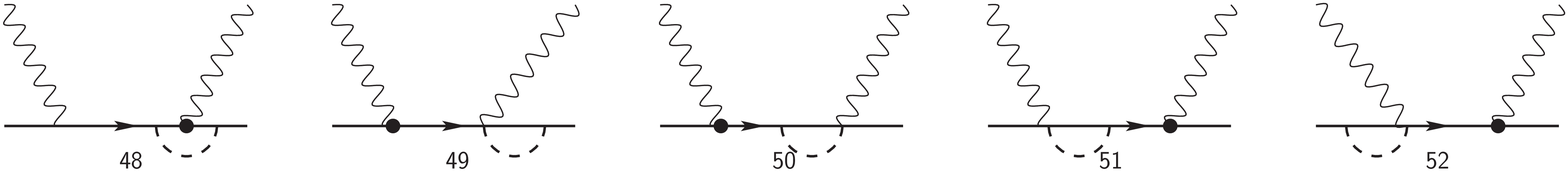}

        \includegraphics[width=0.9\linewidth]{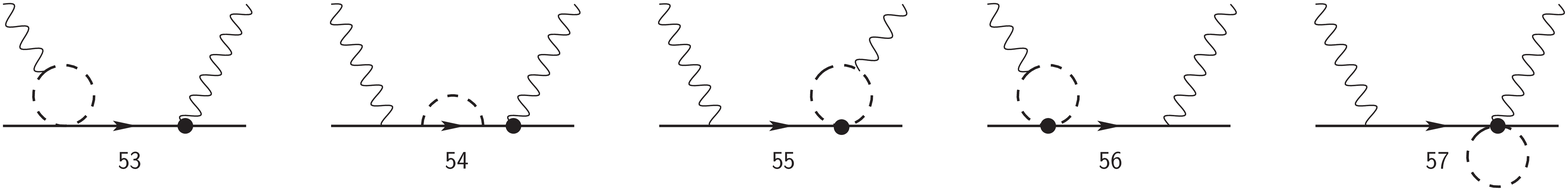}

        \includegraphics[width=0.9\linewidth]{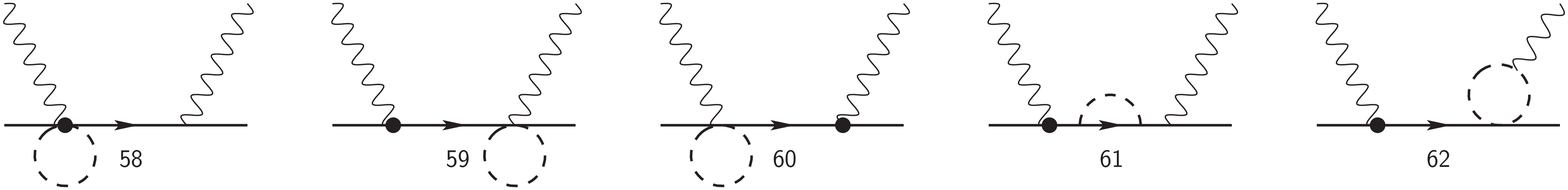}

        \includegraphics[width=0.9\linewidth]{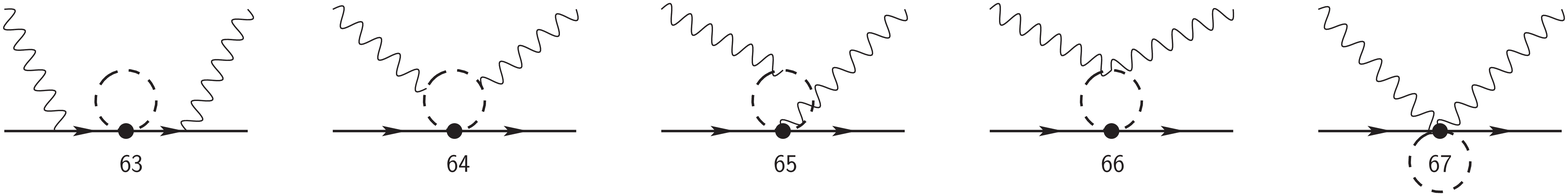}

	\caption{Tree and loop diagrams contributing at $O(p^4)$. Solid, dashed
          and wiggly lines denote nucleons, pions and photons, respectively. Filled circles, squares and diamonds
          represent vertices from  ${\cal L}_2$, ${\cal L}_3$ and ${\cal L}_4$, respectively.
          Crossed diagrams are not shown.
}
	\label{fig:O4}
\end{figure}

The invariant amplitudes $T_1$, $T_2$ up-to-and-including $O(p^4)$ are given
as sums over the contributions of the  diagrams shown in Figs.~\ref{fig:O1}-\ref{fig:O4}:
\eq
T_i(\nu,q^2)=Z_N\biggl(T_i^1(\nu,q^2)+\sum_{a=2}^3T_i^a(\nu,q^2)
+\sum_{a=4}^{32}T_i^a(\nu,q^2)+\sum_{a=33}^{67}T_i^a(\nu,q^2)\biggr)\, ,\quad\quad i=1,2\, . 
\en
The different terms in the above equation are the contributions at $O(p)$, $O(p^2)$,
$O(p^3)$ and $O(p^4)$, respectively.
The enumeration of the contributions corresponds to the one of the diagrams
shown in Figs.~\ref{fig:O1}-\ref{fig:O4}. 
Here, we remind the reader that, under the individual
contributions to the invariant
amplitudes $T_1$, $T_2$, we understand the scalar factors that multiply
the structures $q^\mu q^\nu$ and $-q^2 p^\mu p^\nu/m^2$, respectively.
Further, $Z_N$ is the residue of the nucleon propagator at the pole:
\eq
Z_N&=&1-\frac{3g_A^2}{4F^2}\,\biggl(2m^2\bigl(2M^2B_0'(m^2,M,m)+B_1(m^2,M,m)\bigr)
+M^2B_0(m^2,M,m)+A_0(m)\biggr)
\nonumber\\[2mm]
&& ~~ +\frac{6c_2}{mF^2}\,A_{00}(M)+O(p^5)\, ,
\en
which counts as $Z_N=1+O(p^2)$. For this reason, one has to take this factor
into account only together with the tree-level diagrams at $O(p)$ and $O(p^2)$.
Note also that, to this accuracy, one may replace $\gA,\mkrig,F$ by their physical
values $g_A,m,F_\pi$. The loop functions which enter the above expression are
tabulated in Appendix~\ref{app:integrals}. The derivative in the function
$B_0$ (denoted by the prime) is taken with respect to the first argument.

\medskip

In the Appendices~\ref{app:infinite-tree} and \ref{app:infinite-loop} we
list the individual contributions to $T_1(\nu,q^2)$ and $T_2(\nu,q^2)$ for
$\nu=0$. The full expressions for a generic $\nu$ are much more complicated
and we do not display them here explicitly. These can be extracted
from the Mathematica notebook, which is posted on the archive as an
ancillary file. Note also that
the expressions, given in these appendices, should be understood as $2\times 2$
matrices in the isospin space, folded by the isospin wave functions of
a proton or a neutron (not shown separately). For example, the factor
$1+\tau^3$ is equal to 2 and 0 for the proton and the neutron, respectively.

\subsection{Numerical input}

In order to make numerical predictions, one has to fix
the values of the low-energy constants (LECs) that enter the amplitudes. It should be
noted that, albeit these LECs do not depend on the quark masses by definition, such a
dependence sneaks in when these are determined through the fit of the amplitudes
calculated in a given chiral order to the experimental data. In certain cases, such
remnant quark mass dependence can be numerically significant and, in addition,
lead to the different results when different schemes, say, the Infrared Regularization (IR)
or the EOMS scheme, are used. This fact should be kept in mind once  input from
different fits is used in the amplitudes.

\medskip

The LECs that appear in the amplitudes fall into different groups. We shall use
$g_A=1.2672$ and $F_\pi=92.3\,\mbox{MeV}$ throughout the paper, and $M_\pi,m$
will be identified with the charged pion and the proton masses, respectively.
The order $p^2$ LECs $c_i$ are studied in most detail and rather precise values for these
are available in the literature. Moreover, different fits
(see, e.g., Refs.~\cite{Becher:2001hv,Meissner:2005ba,Krebs:2012yv})
yield  results which are compatible with each other
at an accuracy that is sufficient for our purposes.
The recent and very precise determination of $c_{1,2,3,4}$ from  $\pi N$ input has been
performed in Ref.~\cite{Hoferichter:2015hva} using the matching of the chiral
representation to the solution of the Roy-Steiner equations:
\eq
c_1&=&(-1.11\pm 0.03) \,\mbox{GeV}^{-1}\, ,\quad
c_2=(3.13\pm0.03) \,\mbox{GeV}^{-1}\, ,\nonumber\\[2mm]
c_3&=&(-5.61\pm 0.06) \,\mbox{GeV}^{-1}\, ,\quad
c_4=(4.26\pm  0.04)\,\mbox{GeV}^{-1}\, .
\en
In the following calculations, we shall use these values.
Owing to the fact that the quoted uncertainties are
so small, one may neglect their impact on the total uncertainty and safely
stick to the central values.

\medskip
Turning to the LECs $c_{6,7}$, we note that these appear in the amplitudes in combination
with the $O(p^4)$ LECs:
\eq
\tilde c_6=c_6-4M^2e_{106}\, ,\quad\quad
\tilde c_7=c_7-4M^2e_{105}\, ,
\en
where $M^2=M_\pi^2$ at this order. These are exactly the contributions which appear in
the anomalous magnetic moments of the proton and the neutron,
$\kappa_p=1.793\, ,~\kappa_n=-1.913$, and can be fitted
to the latter. A consistent extraction of these couplings has been performed in
Ref.~\cite{Fuchs:2003ir}, which gives the results for the IR and EOMS schemes
separately (no errors are attached to their results). Here, we quote the result
for the EOMS scheme only, as this is used in our calculation:
\eq
\tilde c_6=1.26\,\mbox{GeV}^{-1}\, ,\quad
\tilde c_7=-0.13\,\mbox{GeV}^{-1}\, .
\en
Next, the $O(p^3)$ LECs $d_{6,7}$ and $O(p^4)$ LECs $e_{54,74}$
enter the expression of the electric and magnetic radii of the proton and the neutron,
and can be fitted by using experimental data on the nucleon electromagnetic form factors.
This was done in Ref.~\cite{Fuchs:2003ir}, where the values of these LECs
are given, again separately for different schemes and with no uncertainties attached:
\eq
d_6=-0.69\,\mbox{GeV}^{-2}\, ,\quad d_7=-0.50\,\mbox{GeV}^{-2}\, ,
\en
and
\eq
e_{54}=0.19\,\mbox{GeV}^{-3}\, ,\quad e_{74}=1.59\,\mbox{GeV}^{-3}\, .
\en
Note that the values for $d_6,d_7$ are consistent with the earlier determination
in Ref.~\cite{Kubis:2000zd}.

Let us now turn to the last group of the $O(p^4)$ LECs, which are related to the nucleon polarizabilities.
At order $p^3$, the polarizabilities are predictions free of LECs~\cite{Bernard:1991rq}.
The $O(p^4)$ LECs that contribute to the amplitudes at $q^2=0$
can be related to the polarizabilities, and we use (experimental or lattice)
input for the latter. This allows to determine four
linearly independent combinations of the $O(p^4)$ LECs which,
according to Ref.~\cite{Djukanovic:2008skh}, can be defined as:
\eq\label{eq:exy}
e_x^\pm=2e_{90}+e_{94}+e_{117}\pm e_{92}\, ,\quad
e_y^\pm=2e_{89}+e_{93}+e_{118}\pm e_{91}\, .
\en
In Ref.~\cite{Djukanovic:2008skh} the results of three different
fits for $e^+_x,e^+_y$ are presented. In the following, however, we shall not use these results.

\medskip

In order to carry out the comparison with the results from the literature, we 
perform a numerical evaluation of the subtraction functions
$S_1^{\sf inel}(q^2)$
and $\bar S(q^2)$, which can be written as:
\eq\label{eq:subtr-subtr}
S_1^{\sf inel}(q^2)&=&-\frac{\kappa^2}{4m^2}-\frac{m}{\alpha}\,\beta_M
+(S_1^{\sf inel}(q^2)-S_1^{\sf inel}(0))\, ,
\nonumber\\[2mm]
\bar S(q^2)&=&-\frac{\kappa^2}{4m^2}+\frac{m}{2\alpha}\,(\alpha_E-\beta_M)
+(\bar S(q^2)-\bar S(0))\, .
\en
In other words, in order to minimize the uncertainty, we aim at a
description of the $q^2$-dependence of the subtraction functions only,
their values at $q^2=0$ are considered as input. Stated differently, up-to-and-including
order $p^4$, the $q^2$-dependence of the subtraction functions $S_1(q^2)$ and
$\bar S(q^2)$ (but not their normalization at $q^2=0$) is determined by the LECs, which
are rather well known from the fit to the data on the low-energy $\pi N$ scattering and
nucleon electromagnetic form factors. Thus, the $q^2$-dependence can be determined
very accurately from BChPT.

\medskip

The experimental values for 
the electric and magnetic polarizabilities are summarized in the recent paper by
Melendez {\it et al.,}~\cite{Melendez:2020ikd}
(see e.g. Refs.~\cite{McGovern:2012ew,Myers:2014ace,OlmosdeLeon:2001zn,Levchuk:1999zy} for some earlier work):
\eq
\mbox{proton:}&&\quad  \alpha_E^p+\beta_M^p=14.0 \pm 0.2\, , \quad\alpha_E^p-\beta_M^p=7.5 \pm 0.9\, ,
\nonumber\\[2mm]
\mbox{neutron:}&&\quad  \alpha_E^n+\beta_M^n=15.2\pm 0.4\, , \quad \alpha_E^n-\beta_M^n=7.9 \pm 3.0
\, .
\en
For the difference proton-neutron one gets:
\eq\label{eq:modelA}
 \alpha_E^{p-n}+\beta_M^{p-n}=-1.20\pm 0.45\, ,\quad  \alpha_E^{p-n}-\beta_M^{p-n}=-0.4\pm 3.1\, ,\quad \beta_M^{p-n}=-0.4\pm 1.6\, .
\en
All quantities are given in units of $10^{-4}\,\mbox{fm}^3$. 

\medskip

In Ref.~\cite{Gasser:2015dwa}, using Reggeon dominance, the isovector electric and  magnetic polarizabilities
have been predicted with an accuracy that supersedes the experimental precision.
For instance, the value for the electric polarizability, extracted from the recent
Review of Particle Physics~\cite{Zyla:2020zbs},
is given by $\alpha_E^{p-n}=-0.6(1.2)$. On the other hand, using Reggeon dominance
and the experimental value for
$\alpha_E^{p-n}+\beta_M^{p-n}$ from Ref.~\cite{Melendez:2020ikd}, which was determined by
using the Baldin sum rule,
one gets:
\eq\label{eq:modelB}
\alpha_E^{p-n}=-1.7\pm 0.4\, ,\quad
\beta_M^{p-n}=0.5\pm 0.6\, ,\quad
\alpha_E^{p-n}-\beta_M^{p-n}=-2.2\pm 0.9\, .
\en
Finally, recently a very accurate lattice calculation of the magnetic polarizability has become
available~\cite{Bignell:2020xkf}:
\eq
\beta_M^p=2.79\pm 0.22^{+13}_{-18}\, ,\quad
\beta_M^n=2.06\pm 0.26^{+15}_{-20}\, ,\quad
\beta_M^{p-n}=0.80\pm 0.28\pm 0.04\, .
\en
One can combine this with the experimental result for $\alpha_E^{p-n}+\beta_M^{p-n}$ from
Eq.~(\ref{eq:modelA}) in order to get a more accurate estimate:
\eq\label{eq:modelC}
\alpha_E^{p-n}-\beta_M^{p-n}=-2.80\pm 0.72\, ,\quad
\beta_M^{p-n}=0.80\pm 0.28\, .
\en
To summarize, we have now three sets of polarizabilities, referred to as model A~[Eq.~(\ref{eq:modelA})],
model B~[Eq.~(\ref{eq:modelB})]
and model C~[Eq.~(\ref{eq:modelC})]. These correspond to the purely experimental input, the Reggeon dominance
hypothesis and the combination of the lattice results with experimental data. Below, we shall evaluate the
difference of the subtraction functions for the proton and the neutron\footnote{In order to ease the
  notations, we do not attach the superscript $p-n$ to the subtraction functions,
  corresponding to the difference proton minus neutron. From the context it is always clear,
which subtraction function is meant.} using the input from the three distinct models:
\eq\label{eq:ABC}
\mbox{model A:}\quad&&\quad
S_1^{\sf inel}(0)=(0.8\pm 2.7)\, \mbox{GeV}^{-2}\, ,\quad~~
\bar S(0)=(-0.2\pm 2.6)\, \mbox{GeV}^{-2}\, ,
\nonumber\\[2mm]
\mbox{model B:}\quad&&\quad
S_1^{\sf inel}(0)=(-0.7\pm 1.0)\, \mbox{GeV}^{-2}\, ,\quad
\bar S(0)=(-1.7\pm 0.8)\, \mbox{GeV}^{-2}\, ,
\nonumber\\[2mm]
\mbox{model C:}\quad&&\quad
S_1^{\sf inel}(0)=(-1.2\pm 0.5)\, \mbox{GeV}^{-2}\, ,\quad
\bar S(0)=(-2.2\pm 0.6)\, \mbox{GeV}^{-2}\,. 
\en

\subsection{The subtraction function}

\begin{figure}[t]
  \begin{center}
    \includegraphics*[width=0.45\linewidth]{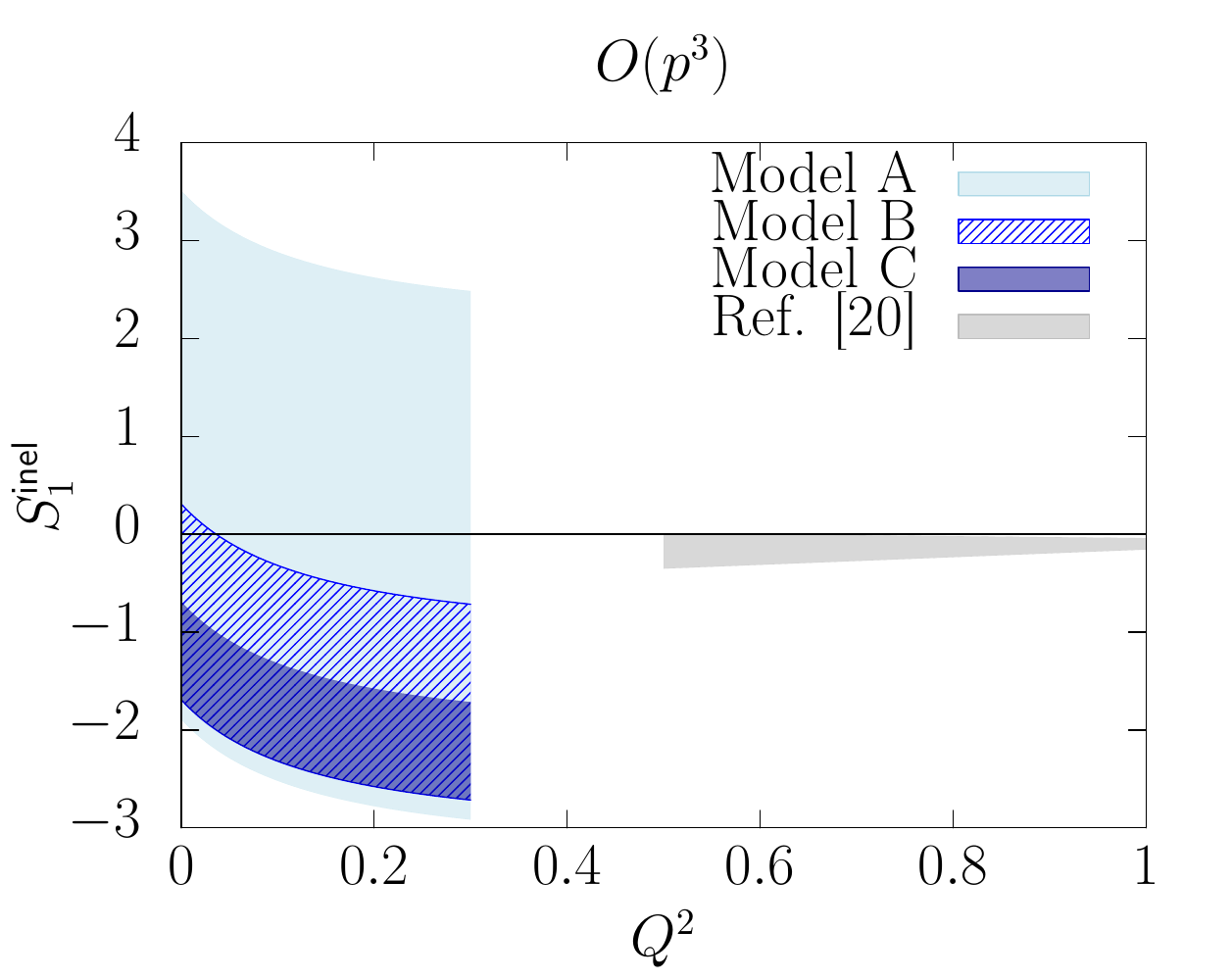}\hspace*{.3cm}
    \includegraphics*[width=0.45\linewidth]{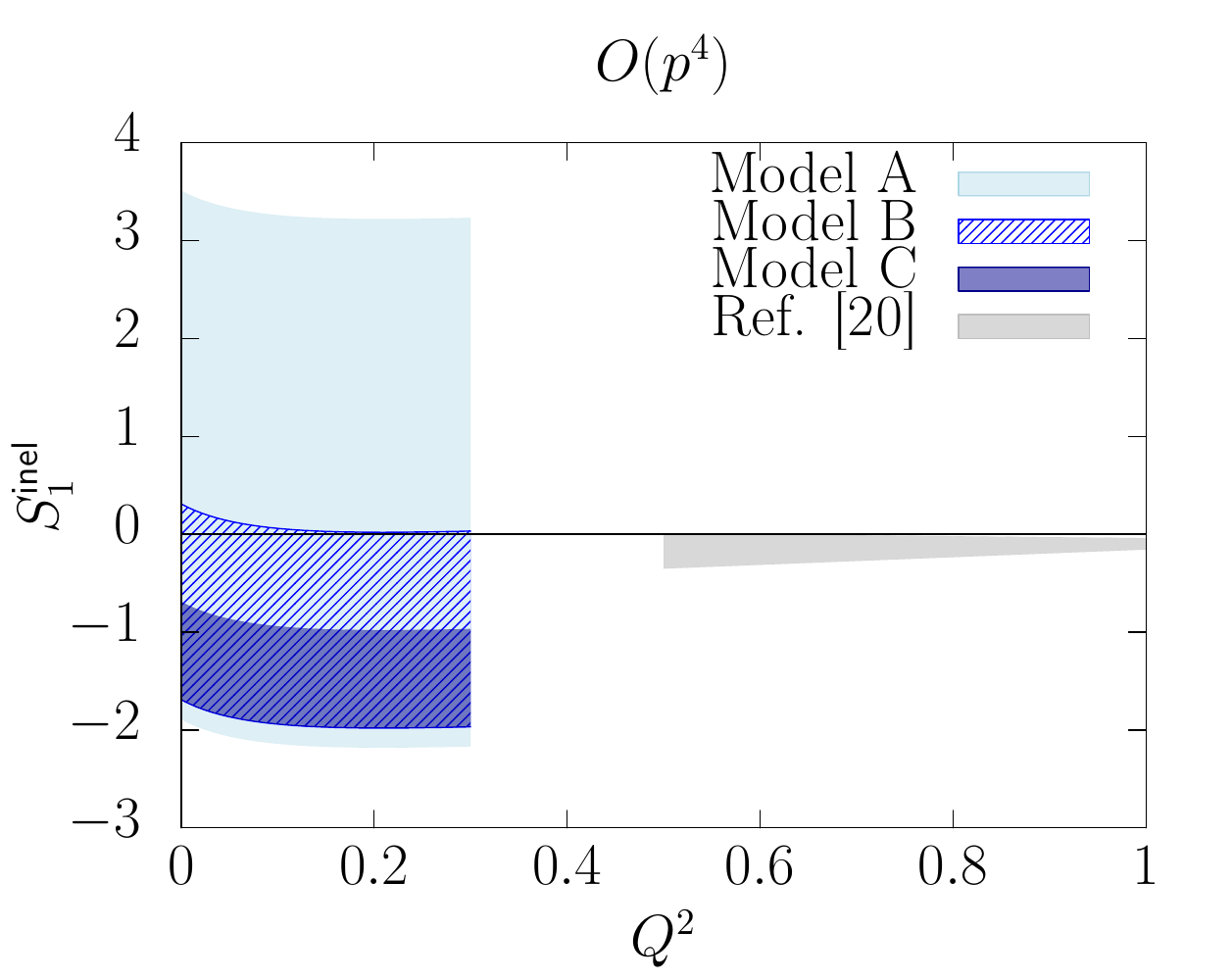}
    \caption{The subtraction function $S_1^{\sf inel}(q^2)$ for proton minus neutron,
      at order $p^3$ and $p^4$
      in the left and right panel, respectively.
      Here,  $Q^2=-q^2$. The light blue, dashed and dark blue bands show the results of
    Models A,B and C, respectively. The result of Ref.~\cite{Gasser:2015dwa},
    which is obtained with the use of the Reggeon dominance hypothesis, is shown
    by the gray band. GeV units are used everywhere.}
    \label{fig:S1}
  \end{center}
\end{figure}

\begin{figure}[t]
  \begin{center}
    \includegraphics*[width=0.45\linewidth]{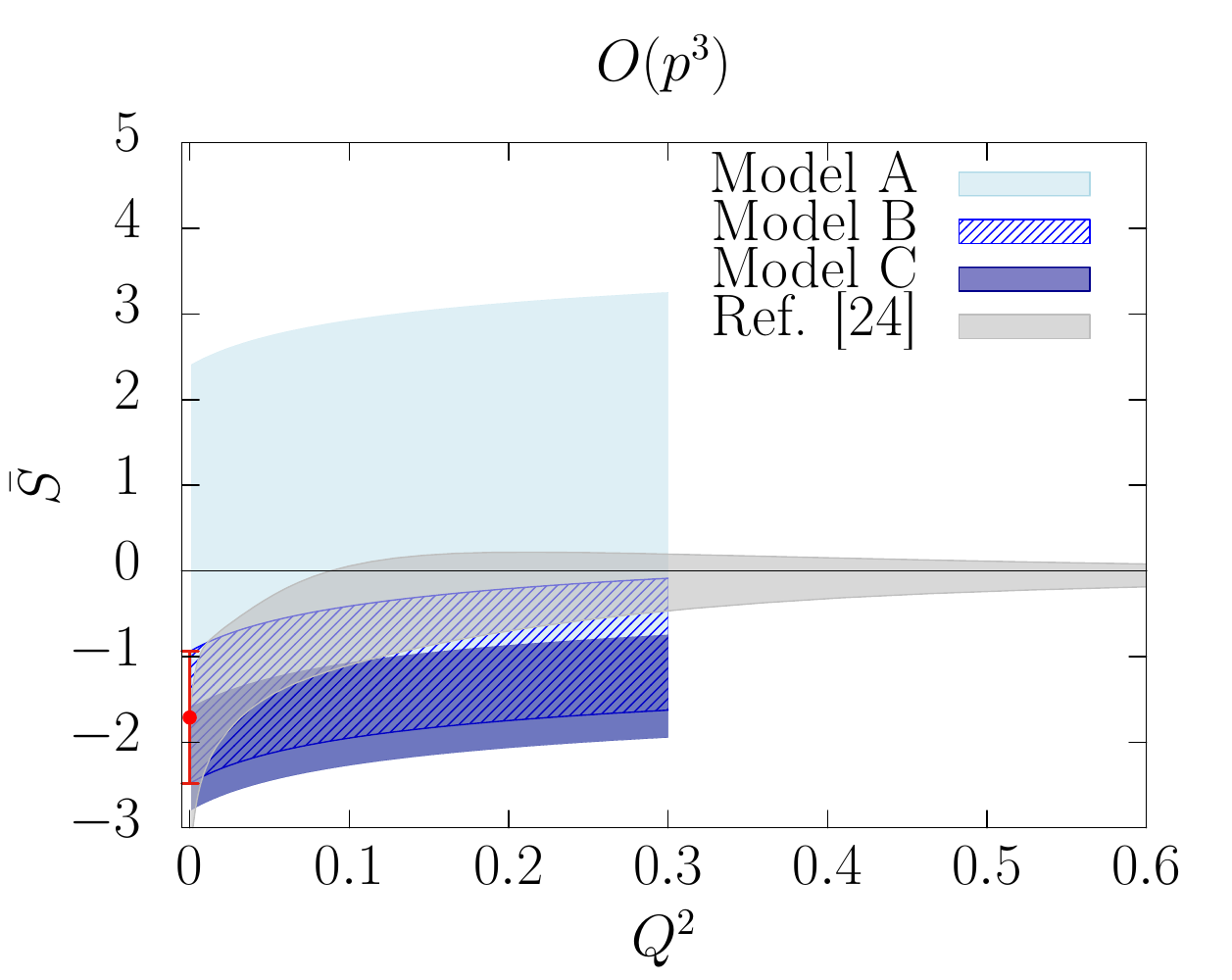}\hspace*{.3cm}
    \includegraphics*[width=0.45\linewidth]{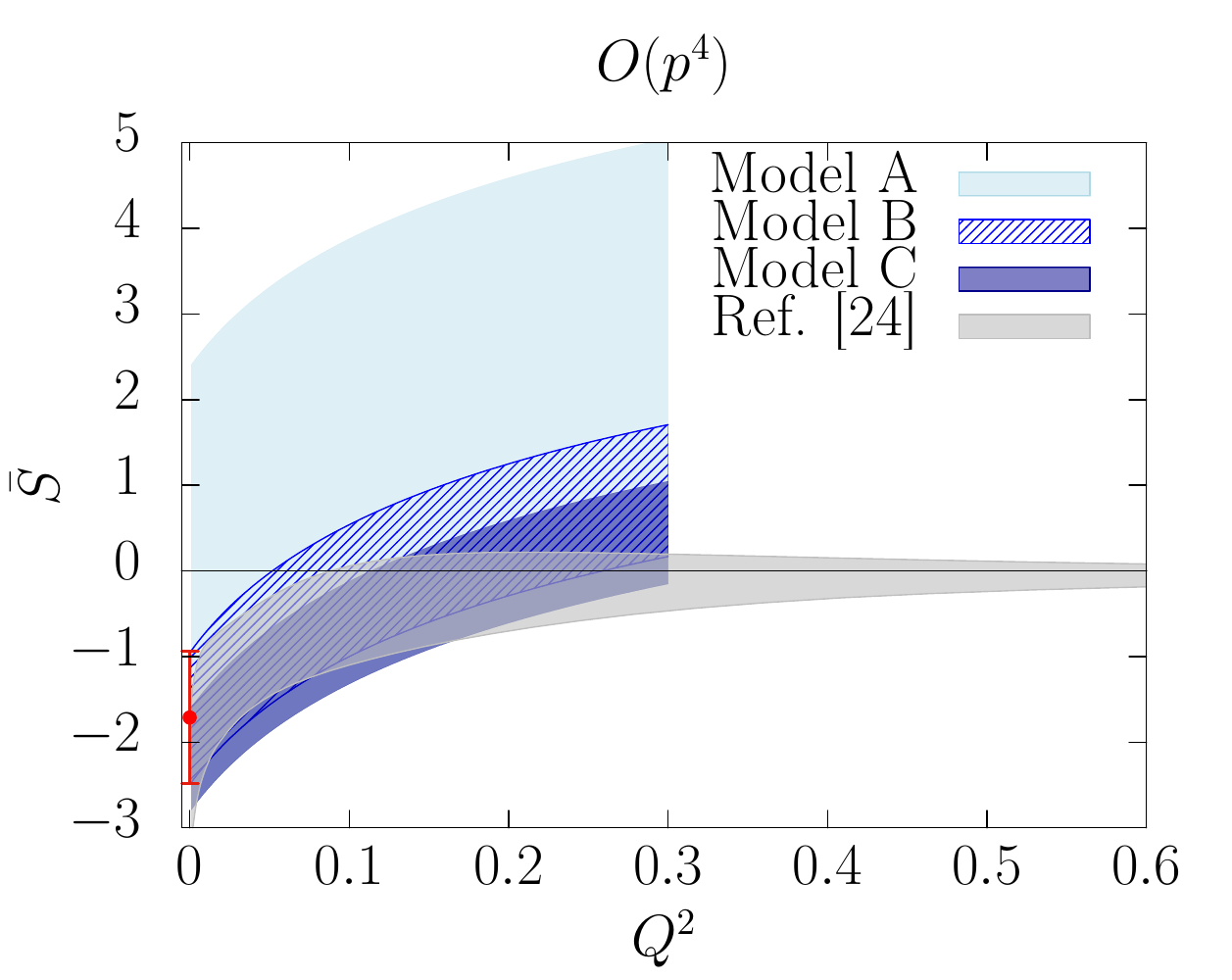}
    \caption{Results for the subtraction function $\bar S(q^2)$ for proton minus neutron.
      The notations are the same as
      in Fig.~\ref{fig:S1}. The gray band shows the result of Ref.~\cite{Gasser:2020hzn}, obtained with
      the use of the Reggeon dominance. GeV units are used everywhere. The data point at the origin shows
      the prediction of the $\bar S(0)$ in model B with Reggeon dominance,
      see Eq.~(\ref{eq:ABC}).}
    \label{fig:Sbar}
  \end{center}
\end{figure}

The main goal of the present paper is to evaluate the finite-volume corrections to the
Compton amplitude. However, having the expression of the infinite-volume
amplitude at hand, one may compare it to the known results from the literature.
For instance, here we shall discuss the comparison to the subtraction
functions $S_1^{\sf inel}(q^2)$ and $\bar S(q^2)$ for proton minus neutron, obtained from the experimental input
by using the Reggeon dominance hypothesis in Refs.~\cite{Gasser:2015dwa}
and \cite{Gasser:2020hzn}, respectively. Note that the experimental input, used
in these papers, leads to a very large uncertainty at small values of $Q^2$, which
comes mainly from the resonance region above the $\Delta$-resonance. On the other
hand, the results of Chiral Perturbation Theory become generally unreliable at higher values of
$Q^2$. Thus, combining both calculations, one can get a coherent picture of the
$Q^2$-dependence of the subtraction function in a wide interval and check the
consistency of the Reggeon dominance hypothesis: Should it turn out that there is an
apparent mismatch between the low-$Q^2$ and high-$Q^2$ regions, this might
cast doubt on the above hypothesis.

\medskip

In order to reduce the error, which stems from the poor knowledge of the higher order
LECs, in our calculations we have attempted to evaluate the $Q^2$-dependence of
the subtraction functions by subtracting
their values at the origin. Eventually, the latter
quantity is expressed in terms of the electric and magnetic polarizabilities, see
Eq.~(\ref{eq:subtr-subtr}). The polarizabilites can be fixed from the different
inputs, leading to what we term the models~A,B and C, see Eq.~(\ref{eq:ABC}). The
corresponding results in the interval $0<Q^2<0.3~\mbox{GeV}^2$
are displayed in Fig.~\ref{fig:S1} for the function
$S_1^{\sf inel}(q^2)$ and in Fig.~\ref{fig:Sbar} for the function $\bar S(q^2)$. Note that
the uncertainty, which is displayed here, comes entirely from the poor knowledge of the polarizabilities.
In other words, we assume that the uncertainties coming from other
LECs are much smaller and do not contribute significantly to the error (it is clear that
even attaching a reasonable uncertainty to other LEcs and adding uncertainties in quadrature,
the changes will  barely be visible due to the huge uncertainty in the polarizabilities).
Note also that, in this scheme,
the subtraction functions at $O(p^3)$ are no more considered as a parameter-free prediction, but
contain polarizabilities as input.

\medskip

From Figs.~\ref{fig:S1} and~\ref{fig:Sbar} one may conclude that the results obtained
in BChPT and by using the Reggeon dominance are reasonably consistent with each other within
the error bars. In case of the subtraction function $\bar S(q^2)$ this can be seen
more clearly, since the results of the calculations, based on the Reggeon dominance,
are available for all values of $Q^2$ down to $Q^2=0$, see Ref.~\cite{Gasser:2020hzn}
(in case of $S_1^{\sf inel}(q^2)$, the calculations extend down to
$Q^2=0.5~\mbox{GeV}^2$, see Ref.~\cite{Gasser:2015dwa}).
Note, however, that the uncertainty in different models, considered in the present paper,
lead to very large error bars in these plots.
What is more important in our opinion, is that
 the amplitudes calculated in the effective field theory show 
a smooth behavior in the vicinity of $Q^2=0$, stated differently, no rapid variations are observed.  Note also
that the convergence is quite poor, and the picture changes significantly when going
from $O(p^3)$ to $O(p^4)$. Still, we do not observe any apparent disagreement to the
Reggeon dominance. Also, it should be noted that both the $O(p^3)$ and $O(p^4)$ contributions
are part of the complete one-loop amplitude, so a true test of convergence could only be achieved by
going to the two-loop level. This, however, is beyond the scope of this paper.

\medskip

We finish this section by briefly mentioning related calculations in the literature. Our result
for the polarizabilities fully agrees with that of Ref.~\cite{Bernard:1991ru}, where the
calculations were done in the relativistic BChPT at $O(p^3)$. Next, our subtraction
function $S_1^{\sf inel}(q^2)$ coincides with the one from Ref.~\cite{Alarcon:2013cba}
at order $p^3$. Moreover, in the recent paper~\cite{Alarcon:2020wjg}, these calculations were
extended up-to-and-including order $p^4$. However, at $O(p^4)$, the contribution
from the $\Delta$-resonance comes into play and this renders the direct comparison more complicated
(note, however, that for proton minus neutron the leading contribution of the
$\Delta$ drops out, as already noted in Ref.~\cite{Bernard:1993bg}).
For this reason, in this paper we restricted ourselves to $O(p^3)$ and verified that the sum of the
polarizabilities $\alpha_E+\beta_M$ is algebraically reproduced in our calculations,
both for the proton and the neutron. Further, expanding our result in inverse
powers of $m$, one should reproduce the non-analytic pieces of the Heavy Baryon ChPT
(HBChPT). The result of Ref.~\cite{Birse:2012eb}, however, cannot be obtained in this way,
whereas the same quantity, evaluated at $O(p^4)$ within HBChPT in Ref.~\cite{Bernard:1993ry}, can be readily
reproduced after the expansion.
Finally, note the calculation of the imaginary part of the subtraction function,
carried out in Ref.~\cite{Nevado:2007dd} in the framework of HBChPT (the dispersion
relations are used to recover the real part). The result in that paper is given in a form
that is not very suitable for a straightforward comparison.

\medskip

Another group of the papers deals with the calculation of the subtraction function
by using dispersion relations and  experimental input (see, e.g., Ref.~\cite{Tomalak:2018dho}), or
modeling it, taking into account the constraints at $Q^2=0$ and at $Q^2$ tending to
infinity~\cite{WalkerLoud:2012bg,Erben:2014hza,Thomas:2014dxa}.
This work has been discussed in great detail in
Refs.~\cite{Gasser:2015dwa,Gasser:2020hzn}, to which the interested reader is referred to.

\section{Finite volume corrections}
\label{sec:finite}

\subsection{Analytic expression for the finite-volume amplitude}

The diagrams that contribute to the Compton amplitude are the same in the
infinite and in a finite volume. The only difference consists in replacing
the three-dimensional integrals with the sums over discrete lattice momenta
(we take that the effects related to a finite size of a lattice in the temporal
direction are already taken into account during the measurement of the energy
levels).
Assuming periodic boundary conditions, in the loop integrals one has to replace:
\eq
\int \frac{d^4k}{(2\pi)^4i}\to\int_V \frac{d^4k}{(2\pi)^4i}\equiv
\int\frac{dk_0}{2\pi i}\frac{1}{L^3}\sum_{\bf k}\, ,\quad\quad {\bf k}=\frac{2\pi}{L}\,{\bf n}\, ,
\quad {\bf n}=\mathbb{Z}^3\, .
\en
Some of the loop integrals in the infinite volume diverge. In a finite volume,
this divergence can be dealt with using dimensional regularization in the same
manner as in the infinite volume. The counterterms that remove the divergences
are the same in both cases. In the following, we shall use this fact and
write down the finite-volume sums in a dimensionally regularized fashion,
without specifying how this is done. These sums will be further split into
the infinite-volume parts and the corrections. The regularization is relevant
only for the first parts, where the standard prescription can be applied.
The finite-volume corrections are ultraviolet-finite and can be calculated
in four dimensions.

\medskip

A closely related question concerns  the presence of the power-counting
breaking terms in the covariant BChPT. It must be stressed that
there are no such terms in the finite-volume part of the amplitude, as all
such terms are suppressed by a factor $\exp(-mL)$ containing the nucleon mass.
On the contrary, in the infinite-volume part these terms are present and
can be dealt with, e.g., by using the EOMS prescription, as described above.
The reason for this is, of course, that the breaking of the power-counting
rules is a high-energy phenomenon that  emerges for loop momenta of the order of
the nucleon mass. On the other hand, the large-$L$ behavior is governed by the
momenta of the order of the pion mass. This region does not contribute to the
breaking of the power counting.

\medskip
In order to carry out the calculations, one needs the expression of the nucleon
$Z$-factor in a finite volume. This quantity is defined in the rest-frame
of the nucleon, where the nucleon propagator is given by:
\eq
S_L(p)=i\int_L d^4x e^{ip_0x_0}\langle 0|T\Psi(x)\bar\Psi(0)|0\rangle\, ,
\quad\quad p^\mu=(p^0,{\bf 0})\, .
\en
In the above expression, the integration is carried out in a finite box. Further,
\eq
S_L(p)=\frac{1}{\mkrig-\gamma^0p^0-\Sigma_L(p)}\, ,
\en
where $\Sigma_L(p)$ denotes the self-energy of the nucleon in a finite volume,
and
\eq
\Sigma_L(p)=A_L(p_0)+\gamma^0p^0B_L(p^0)\, .
\en
The finite-volume mass of the nucleon is implicitly given through the solution
of the equation that contains the scalar functions $A_L(p^0),B_L(p^0)$:
\eq
m_L=\frac{\mkrig-A_L(m_L)}{1+B_L(m_L)}=\mkrig-A_L(m_L)-m_LB_L(m_L)+\cdots\, .
\en
This equation can be solved by iteration, expressing $m_L$ order by order
through the infinite-volume parameters. Further, the $Z$-factor in a finite
volume is given by the residue of the nucleon propagator at the pole $p^0=m_L$.
It can be also expressed in terms of the functions  $A_L(p^0),B_L(p^0)$ and
the derivatives thereof with respect to the variable $p^0$:
\eq
Z_L=\bigl(1+A_L'(m_L)+B_L(m_L)+m_LB'(m_L)\bigr)^{-1}=1-A_L'(m_L)-B_L(m_L)-m_LB'(m_L)+\cdots\, .
\en
The explicit expression for this quantity is given by:
\begin{eqnarray}
Z_L&=& -\frac{3 m g_A^2  \tilde{\mathscr{B}}^{0}_{(1,1)}  (m,M;m)}{2 F^2}+\frac{3 M^2 g_A^2
\left(3 M^2-8 m^2\right) \tilde{\mathscr{B}}_{(1,1)}(m,M;m)}{4 F^2 \left(M^2-4 m^2\right)}
-\frac{3 g_A^2 \tilde{\mathscr{A}}_{(1)}(m)  \left(4 m^2 + 3 M^2\right)}{4 F^2 \left(4 m^2-M^2\right)}
\nonumber\\[2mm]   
&+& \frac{3 g_A^2
\tilde{\mathscr{A}}_{(1)}(M) \left(M^2-2 m^2\right)}{F^2 \left(4 m^2-M^2\right)} +\frac{6 c_2
\tilde{\mathscr{A}}_{(1)}^{00}(M)}{F^2 m} \,.
\label{ZFFV}
\end{eqnarray}
The derivatives of the loop integrals have been reduced again to loop integrals utilizing the
algebraic identities that are specified in Refs.~\cite{Devaraj:1997es,Djukanovic:2008skh}.

\medskip

Finally, note that Lorentz symmetry is broken in a cubic box and one can no
more use the decomposition of the Compton amplitude into two scalar amplitudes.
This is, in fact, not needed, because the energy shift of a nucleon in
the periodic field is directly given by the $11$-component of the Compton
tensor~\cite{Agadjanov:2016cjc,Agadjanov:2018yxh}. Putting together all contributions,
one can write:
\eq
T^{11}_L({\rm neutron})&=&\tilde{\mathscr{T}}_n^{(1)}
+\tilde{\mathscr{T}}_n^{(2)}+\tilde{\mathscr{T}}_n^{(3)}+\tilde{\mathscr{T}}_n^{(4)}+O(p^5)\, ,
\nonumber\\[2mm]
T^{11}_L({\rm proton})&=&\tilde{\mathscr{T}}_p^{(1)}
+\tilde{\mathscr{T}}_p^{(2)}+\tilde{\mathscr{T}}_p^{(3)}+\tilde{\mathscr{T}}_p^{(4)}+O(p^5)\, .
\en
The individual contributions are given by:

\medskip

$O(p)$: {\rm neutron}
\eq\label{eq:chuncho-1}
\tilde{\mathscr{T}}_n^{(1)}=0\, .
\en

\medskip

$O(p)$: {\rm proton}
\eq
\tilde{\mathscr{T}}_p^{(1)}=0\, .
\en

\medskip

$O(p^2)$: {\rm neutron}
\eq
\tilde{\mathscr{T}}_n^{(2)}=0\, .
\en

\medskip

$O(p^2)$: {\rm proton}
\eq
\tilde{\mathscr{T}}_p^{(2)}=2m(2c_6+c_7)\, .
\en

\medskip

$O(p^3)$: {\rm neutron}
\eq
\tilde {\mathscr{T}}_n^{(3)}&=& m^2(2c_6-c_7)^2
\nonumber\\[2mm]
&+&\frac{2g_A^2m^2}{F^2}\,\biggl\{
-4M^2 \tilde{\mathscr{D}}^{11}_{(1,1,1,1)}(m,M,m,M;p,q,p+q)
-4M^2 \tilde{\mathscr{C}}^{11}_{(2,1,1)}(M,m,M;-p,q)
\nonumber\\[2mm]
& -&4M^2 \tilde{\mathscr{C}}^{11}_{(2,1,1)}(m,M,m;-p,q) 
+ M^2q^2\tilde{\mathscr{C}}_{(2,1,1)}(m,M,m;-p,q)
-q^2\tilde{\mathscr{C}}_{(1,1,1)}(m,M,m; -p,q)
\nonumber\\[2mm]
& -&4 \tilde{\mathscr{B}}^{11}_{(2,1)}(m,m;q)
-M^2 \tilde{\mathscr{B}}_{(1,2)}(m,M;p)
-M^2 \tilde{\mathscr{B}}_{(2,1)}(m,M;p)
+ q^2\tilde{\mathscr{B}}_{(2,1)}(m,m;q)
-\tilde{\mathscr{A}}_{(2)}(m)
\biggr\}\, .
\en

$O(p^3)$: {\rm proton}
\eq
\tilde {\mathscr{T}}_p^{(3)} &=&m^2(2c_6+c_7)^2+2q^2(d_6+2d_7)
\nonumber\\[2mm]
&+&\frac{m^2g_A^2}{F^2}\,\biggl\{
4\tilde{\mathscr{C}}^{11}_{(1,1,1)}(m,M,m; - p,q)
-8\tilde{\mathscr{C}}^{11}_{(1,1,1)}(M,m,M;-p,q)
-8M^2\tilde{\mathscr{C}}^{11}_{(2,1,1)}(M,m,M;-p,q) 
\nonumber\\[2mm]
&-&4M^2 \tilde{\mathscr{C}}^{11}_{(2,1,1)}(m,M,m;-p,q) 
+ M^2q^2\tilde{\mathscr{C}}_{(2,1,1)}(m,M,m;-p,q)
-(q^2-2M^2)\tilde{\mathscr{C}}_{(1,1,1)}(m,M,m; - p,q)
\nonumber\\[2mm]
&-&4 \tilde{\mathscr{B}}^{11}_{(2,1)}(m,m;q) 
+\frac{3}{2m^2}\,p_\alpha \tilde{\mathscr{B}}^\alpha_{(1,1)}(M,m;p+q)
-2M^2  \tilde{\mathscr{B}}_{(1,2)}(m,M;p)
-M^2\tilde{\mathscr{B}}_{(2,1)}(m,M;p) 
\nonumber\\[2mm]
&+&q^2 \tilde{\mathscr{B}}_{(2,1)}(m,m;q)
+2\tilde{\mathscr{B}}_{(1,1)}(m,m;q)
-\tilde{\mathscr{B}}_{(1,1)}(m,M;p)
+\tilde{\mathscr{B}}_{(1,1)}(M,m;p+q)
- \tilde{\mathscr{A}}_{(2)}(m)
\biggr\}
\nonumber\\[2mm]
&+&\frac{3g_A^2}{F^2q^2}\,\biggl\{
-p_\alpha q_\beta \tilde{\mathscr{B}}^{\alpha\beta}_{(1,1)}(M,m;p+q)
-p_\alpha p_\beta \tilde{\mathscr{B}}^{\alpha\beta}_{(1,1)}(M,m;p+q)
+\frac{M^2}{2}p_\alpha\tilde{\mathscr{B}}^\alpha_{(1,1)}(M,m;p+q)
\nonumber\\[2mm]
&-&2m^2q_\alpha\tilde{\mathscr{B}}^\alpha_{(1,1)}(M,m;p+q)
-m^2M^2\tilde{\mathscr{B}}_{(1,1)}(m,M;p)
+m^2M^2\tilde{\mathscr{B}}_{(1,1)}(M,m;p+q)
+\frac{m^2}{2}\tilde{\mathscr{A}}_{(1)}(m)
\biggr\}\,.
\en

$O(p^4)$: {\rm neutron}
\eq
\tilde {\mathscr{T}}_n^{(4)}
   &=&8m(2e_{89}+e_{93}+e_{118}-e_{91})q^2
\nonumber\\[2mm]
&+&\frac{m g_A^2}{F^2}\biggl\{
 -  4(2c_6+c_7) q^2m^2 \tilde{\mathscr{D}}^{11}_{(1,1,1,1)}(m,M,m,M;p,-q,p-q)-8m^2(2c_6-c_7) \tilde{\mathscr{C}}^{11}_{(1,1,1)}(m,M,m;p,q)  
 \nonumber\\[2mm]
   & -&8m^2(2c_6-c_7) \tilde{\mathscr{C}}^{11}_{(1,1,1)}(M,m,M; - p,q) +4(2c_6+c_7) q^2M^2m^2 \tilde{\mathscr{C}}_{(2,1,1)}(m,M,m;p,q) \nonumber\\[2mm]
   &-&2  (4c_6m^2M^2 -c_7(q^2(M^2-2m^2) +2m^2M^2)) \tilde{\mathscr{C}}_{(1,1,1)}(m,M,m;p,q)    \nonumber\\[2mm]
&+&(2c_6-c_7)q_\alpha \left( \tilde{\mathscr{B}}^\alpha_{(1,1)}(M,m;p+q) -\mathscr{B}^\alpha_{(1,1)}(M,m;p-q) \right)   + 4m^2q^2 (2c_6+c_7)\tilde{\mathscr{B}}_{(2,1)}(m,m;q) 
\nonumber\\[2mm]
   &-&  \left(c_7(q^2+4m^2-M^2) + 2c_6 (q^2+M^2)\right) \tilde{\mathscr{B}}_{(1,1)}(M,m;p+q)
   \nonumber\\[2mm]
   &+& 4c_7m^2\tilde{\mathscr{B}}_{(1,1)}(m,M;p)
   - 2 (4c_6m^2-c_7(q^2+2m^2))\tilde{\mathscr{B}}_{(1,1)}(m,m;q)
   \nonumber\\[2mm]
   &- & 2(2c_6-c_7) \tilde{\mathscr{B}}_{(1,1)}^{11} (M,M;q)
   +  2 (2c_6-c_7)  p_\alpha \tilde{\mathscr{B}}_{(1,1)}^\alpha(M,m;p+q) 
- (2c_6 - c_7)\tilde{\mathscr{A}}_{(1)}(m)
   \biggr\}
   \nonumber\\[2mm]
   &+& \frac{4}{mF^2}\,\biggl\{
    (2c_6 -c_7)m^2(\tilde{\mathscr{B}}^{11}_{(1,1)}(M,M;q)
   + \frac{1}{2}\tilde{\mathscr{A}}_{(1)}(M)) - c_2 p_\alpha p_\beta \left( 4 \tilde{\mathscr{B}}_{(2,1)}^{11\alpha\beta} (M,M;q) + \tilde{\mathscr{A}}^{\alpha\beta}_{(2)}(M) \right) 
    \nonumber\\[2mm]
   &+ &
    (2c_1  - c_3) M^2 m^2 \left( 4 \tilde{\mathscr{B}}_{(2,1)}^{11} (M,M;q) +  \tilde{\mathscr{A}}_{(2)}(M) \right)
   \biggr\}\, .
\en

$O(p^4)$: {\rm proton}
\eq\label{eq:chuncho-6}
\tilde {\mathscr{T}}_p^{(4)} &=& 8m(2e_{89}+e_{93}+e_{118}+e_{91})q^2
+4m((2e_{54}+e_{74})q^2-4(2e_{105}+e_{106})M^2)
\nonumber\\[2mm]
   &+&
\frac{m g_A^2}{F^2}\biggl\{
4(c_7-2c_6)q^2m^2 \tilde{\mathscr{D}}^{11}_{(1,1,1,1)} (m,M,m,M;p,-q,p-q) +16c_7m^2\tilde{\mathscr{C}}^{11}_{(1,1,1)} 
(m,M,m;p,q) \nonumber\\[2mm]
& - & 8m^2 (2c_6+c_7)\tilde{\mathscr{C}}^{11}_{(1,1,1)}(M,m,M;-p,q) +2 \left(2 c_6+c_7\right) m^2 M^2 q^2 \tilde{\mathscr{C}}_{(2,1,1)}(m,M,m;p,q) 
\nonumber\\[2mm]
&+& \left(2 c_6 q^2 \left(M^2-2 m^2\right)+c_7 \left(q^2 \left(M^2-2 m^2\right)+8 m^2 M^2\right)\right)  \tilde{\mathscr{C}}_{(1,1,1)}(m,M,m;p,q)
\nonumber\\[2mm]
&-& \frac{1}{4} \left(2 c_6+c_7\right) q_{\alpha } \left(\tilde{\mathscr{B}}^{\alpha}_{(1,1)}(M,m;p-q)-\tilde{\mathscr{B}}^{\alpha}_{(1,1)}(M,m;p+q)\right) -3 \left(2 c_6+c_7\right) p_{\alpha } \tilde{\mathscr{B}}^{\alpha}_{(1,1)}(M,m;p)
\nonumber\\[2mm]
&+&
8 \left(c_6+c_7\right) p_{\alpha } \tilde{\mathscr{B}}^{\alpha}_{(1,1)}(M,m;p+q) -6 \left(2 c_6+c_7\right) M^2 p_{\alpha } \tilde{\mathscr{B}}^{\alpha}_{(2,1)}(M,m;p)
\nonumber\\[2mm]
&+& \left(4 c_6 \left(m^2-M^2\right)+2 c_7 m^2\right) \tilde{\mathscr{B}}_{(1,1)}(m,M;p)+\left(c_7 \left(8
   m^2+q^2\right)+2 c_6 q^2\right)  \tilde{\mathscr{B}}_{(1,1)}(m,m;q)
   \nonumber\\[2mm]
   &-& 2 \left(2 c_6+c_7\right)  \tilde{\mathscr{B}}^{11}_{(1,1)}(M,M;q)-\frac{1}{2} \left(2 c_6+c_7\right) \left(4
   m^2-M^2+q^2\right)  \tilde{\mathscr{B}}_{(1,1)}(M,m;p+q) 
\nonumber\\[2mm]   
   &+& 2 \left(2 c_6+c_7\right) m^2 q^2
    \tilde{\mathscr{B}}_{(2,1)}(m,m;q)+\frac{3}{2} \left(2 c_6+c_7\right) M^2  \tilde{\mathscr{B}}_{(1,1)}(M,m;p)
+ \left(3 c_6+\frac{7 c_7}{2}\right)  \tilde{\mathscr{A}}_{(1)}(m) 
 \biggr\}\nonumber\\[2mm]
   &+&
\frac{m g_A^2}{F^2 q^2}\biggl\{ 
\frac{9}{4}(2 c_6 + c_7)   p_\alpha q_\beta   \tilde{\mathscr{B}}^{\alpha\beta }_{(1,1)} 
(M,m;p-q) 
- \frac{3}{4}(2 c_6 + c_7) (8 p_\alpha p_\beta +9  p_\alpha q_\beta + 4 q_\alpha q_\beta)  \tilde{\mathscr{B}}^{\alpha\beta }_{(1,1)} 
(M,m;p+q) 
\nonumber\\[2mm]
&+& \frac{1}{8}(c_6 (32 m^2-6M^2)+3 c_7 (16 m^2 - M^2)) q_\alpha \tilde{\mathscr{B}}^{\alpha}_{(1,1)} 
(M,m;p-q) \nonumber\\[2mm]
&+& \left( c_6 \left( \left( \frac{ 9M^2}{4}-4m^2 \right) q_\alpha + 6 M^2 p_\alpha\right) +\frac{3}{8} c_7 (8 M^2 p_\alpha  +( 3 M^2-16 m^2) q_\alpha \right)  \tilde{\mathscr{B}}^{\alpha}_{(1,1)} 
(M,m;p+q) \nonumber\\[2mm]
& + & 3 (2c_6+c_7) 
m^2 \left( 2 M^2 ( \tilde{\mathscr{B}}_{(1,1)} 
(M,m;p+q) -  \tilde{\mathscr{B}}_{(1,1)} (M,m;p)  ) + \tilde{\mathscr{A}}_{(1)} (m)  \right)
  \biggr\}
\nonumber\\[2mm]
&+& \frac{4 }{mF^2}\,\biggl\{
 (2c_6 -c_7)m^2(\tilde{\mathscr{B}}^{11}_{(1,1)}(M,M;q)
   + \frac{1}{2}\tilde{\mathscr{A}}_{(1)}(M)) - c_2 p_\alpha p_\beta \left( 4 \tilde{\mathscr{B}}_{(2,1)}^{11\alpha\beta} (M,M;q) + \tilde{\mathscr{A}}^{\alpha\beta}_{(2)}(M) \right) 
    \nonumber\\[2mm]
   &+ &
    (2c_1  - c_3) M^2 m^2 \left( 4 \tilde{\mathscr{B}}_{(2,1)}^{11} (M,M;q) +  \tilde{\mathscr{A}}_{(2)}(M) \right)
 +  \frac{3}{q^2}\, c_2  q_\alpha q_\beta \, \tilde{\mathscr{A}}_{(1)}^{\alpha\beta}(M)  
+m^2 c_6 \tilde{\mathscr{A}}_{(1)}(M)
\nonumber\\[2mm]
&+&c_7 m^2 \left( \tilde{\mathscr{B}}^{11}_{(1,1)}(M,M;q) + \tilde{\mathscr{A}}_{(1)} (M) \right)  -c_4m^2 
\tilde{\mathscr{B}}^{11}_{(1,1)}(M,M;q)\biggr\}\, .
\en
Note that in the above expressions, we have replaced $m_L$, emerging from the
kinematics, by the infinite-volume nucleon mass, $m$. Up to the chiral order
we are working, this is a perfectly valid procedure. Finally,
the expressions for the various finite-volume sums, which enter the above
formulae, are listed in appendix~\ref{app:FV-integrals}.

\subsection{Numerical results}
\label{sec:numerics}

In this section, we evaluate the finite-volume corrections to the $11$-component of
the Compton tensor, which enters the expression of the energy shift in the
periodic background field. The quantity, which will be calculated here, is given by:
\eq
\Delta=\frac{T_L^{11}(p,q)-T^{11}(p,q)}{T^{11}(p,q)}\, .
\en
We calculate this quantity for the physical value of the pion mass
and for several different values of $q^2$, separately for the proton and the neutron.
Having explicit expressions for the amplitude, it is straightforward to carry
out the calculations for unphysical quark masses as well, if needed. It should be stressed
that we are mainly interested in the order-of-magnitude estimate of the correction,
which is needed to answer the following question: How large the box size $L$ should be so that one can
safely neglect the finite-volume artifacts?

\medskip

The results at $O(p^3)$ can be obtained directly from Eqs.~(\ref{eq:chuncho-1})-(\ref{eq:chuncho-6}), since
these contain no unknown LECs. At $O(p^4)$, however, the LECs
$e_y^\pm$, defined in Eq.~(\ref{eq:exy}), appear. We fit these 
to the magnetic polarizability $\beta_M$:
\eq
\mbox{model A:}&&\quad\quad e_y^+=(0.42\pm 0.10)\,\mbox{GeV}^{-3}\, ,\quad\quad
e_y^-=(0.56\pm 0.34)\,\mbox{GeV}^{-3}\, ,
\nonumber\\[2mm]
\mbox{model C:}&&\quad\quad e_y^+=(0.53\pm 0.06)\,\mbox{GeV}^{-3}\, ,\quad\quad
e_y^-=(0.91\pm 0.07)\,\mbox{GeV}^{-3}\,,
\en
(note that we do not have separate inputs for the proton and neutron for model B).
For the further calculations, we use the following values of the $O(p^4)$ LECs:
\eq
e_y^+=(0.46\pm 0.14)\,\mbox{GeV}^{-3}\, ,\quad\quad
e_y^-=(0.60\pm 0.38)\,\mbox{GeV}^{-3}\, .
\en
This choice covers model A, as well as model C.

\medskip

\begin{figure}[t]
  \begin{center}

   \includegraphics[width=0.35\linewidth]{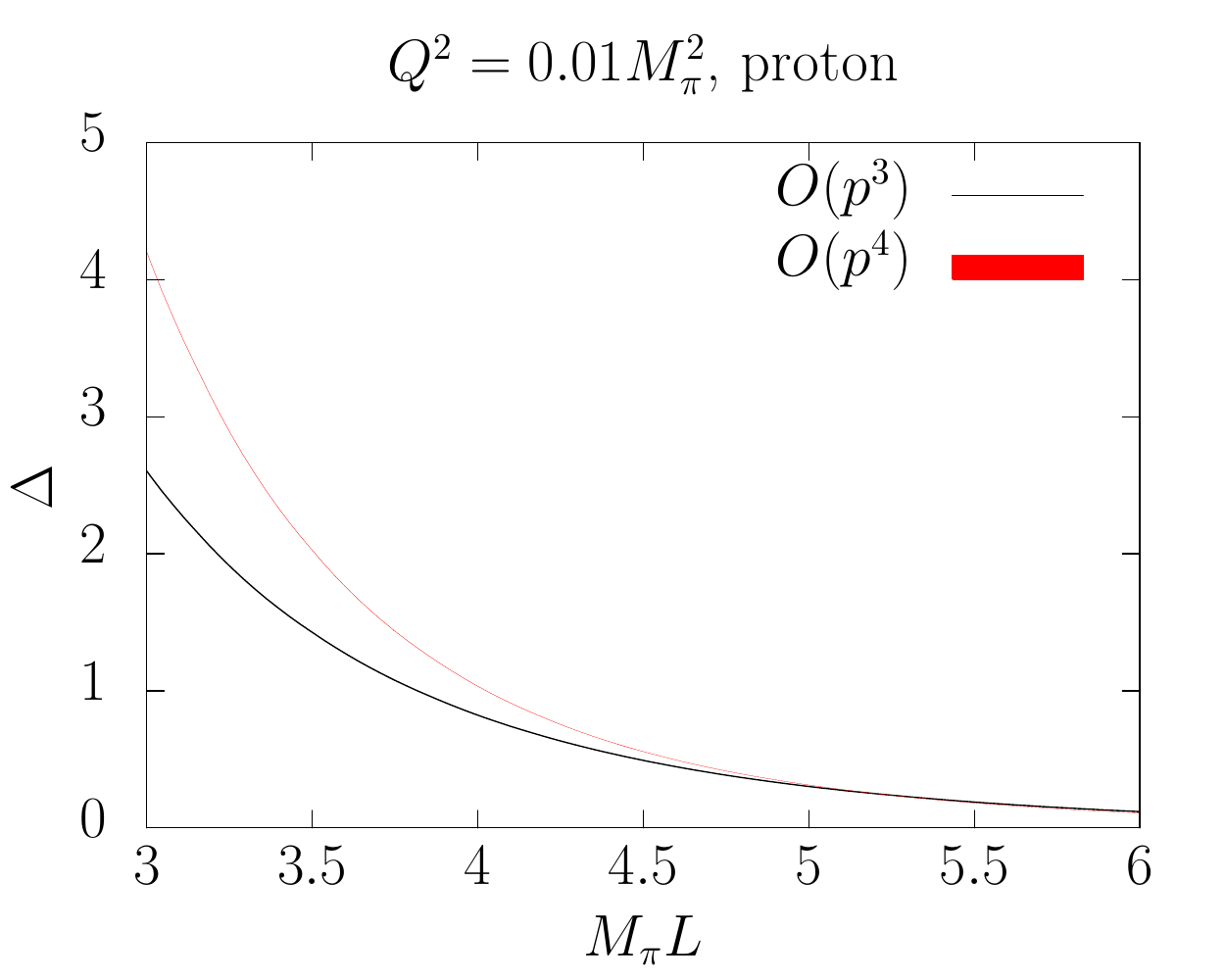}\hspace*{.4cm}
    \includegraphics[width=0.35\linewidth]{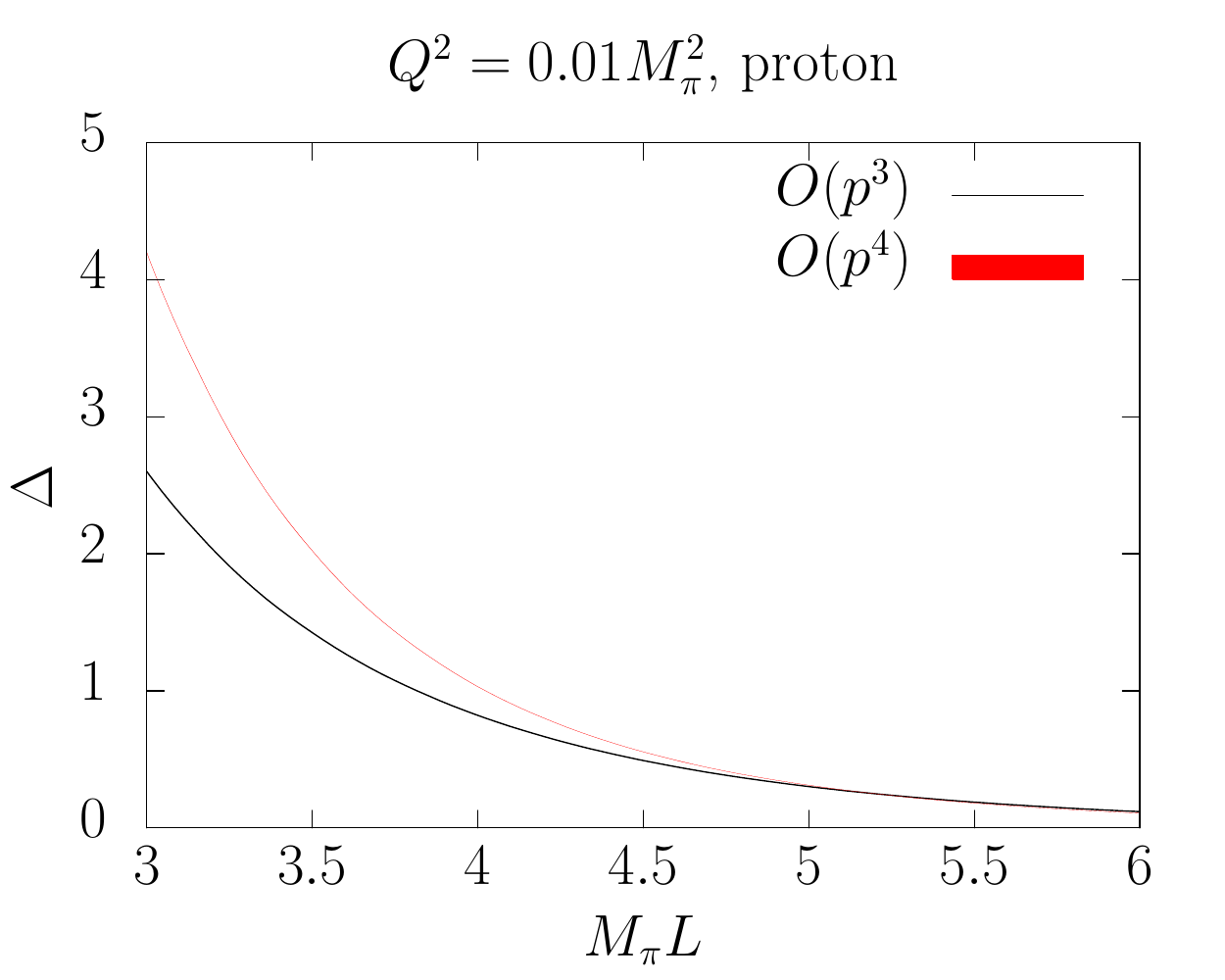}

    \vspace*{.5cm}
 
    \includegraphics[width=0.35\linewidth]{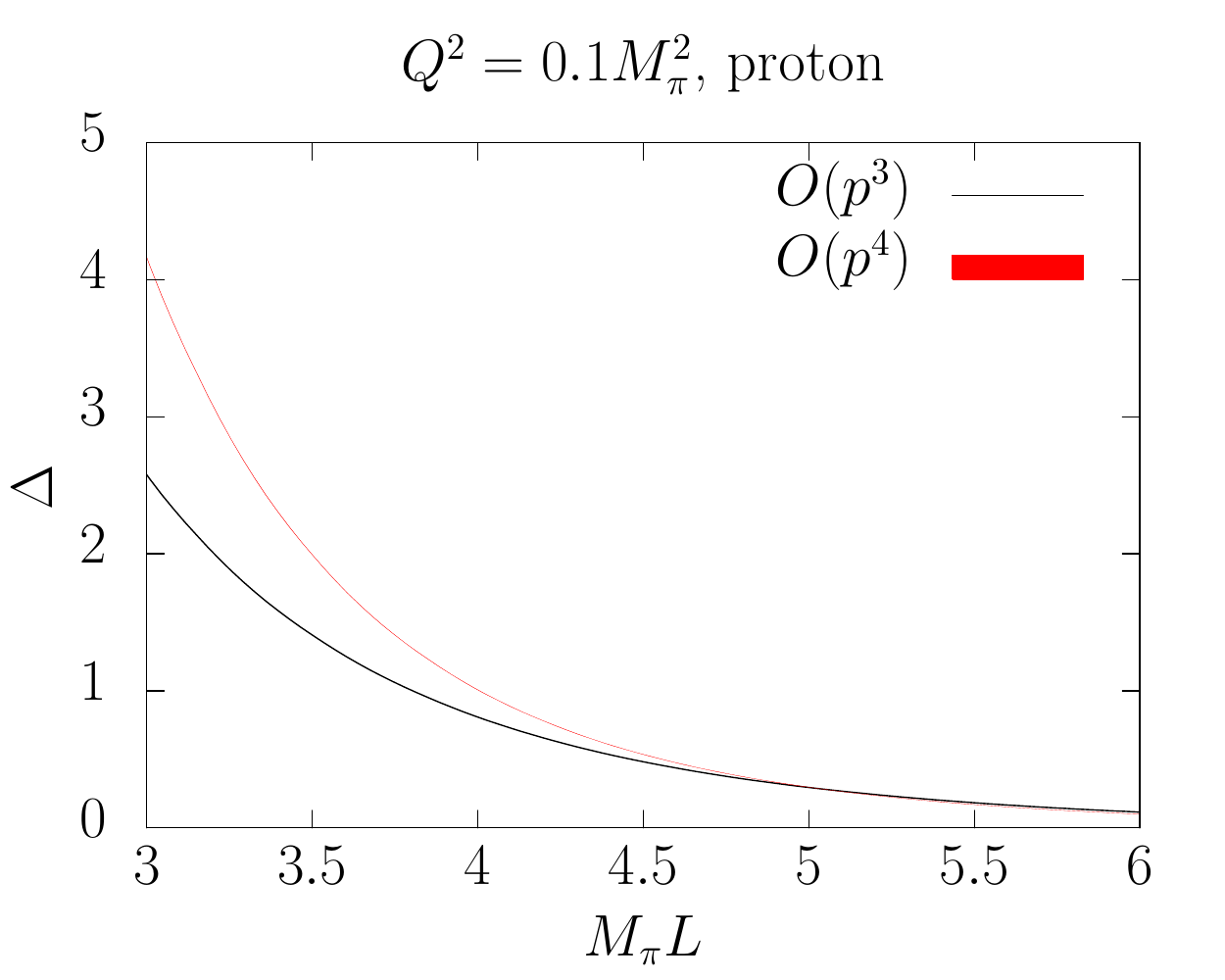}\hspace*{.4cm}
    \includegraphics[width=0.35\linewidth]{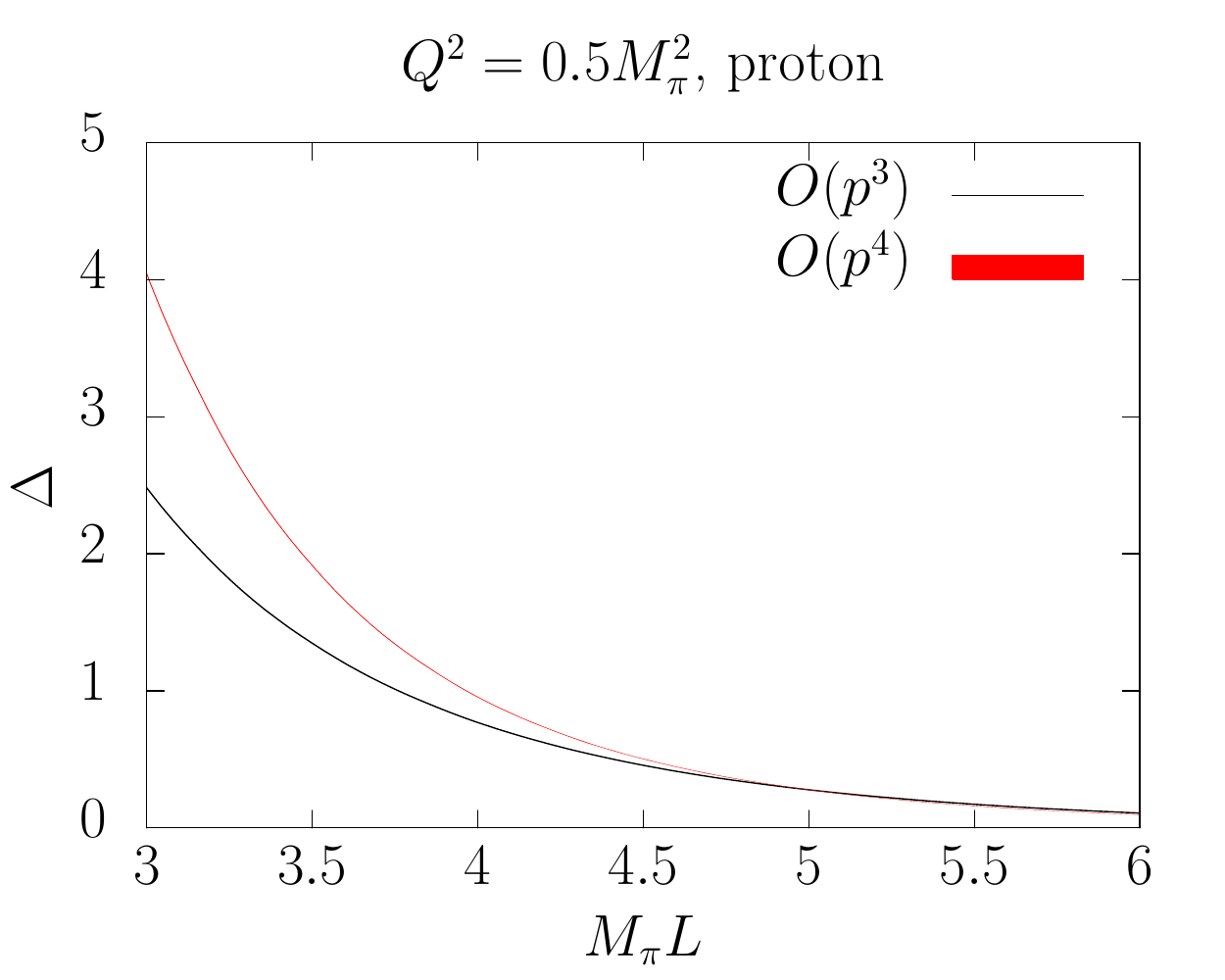}

    \vspace*{.5cm}

    \includegraphics[width=0.35\linewidth]{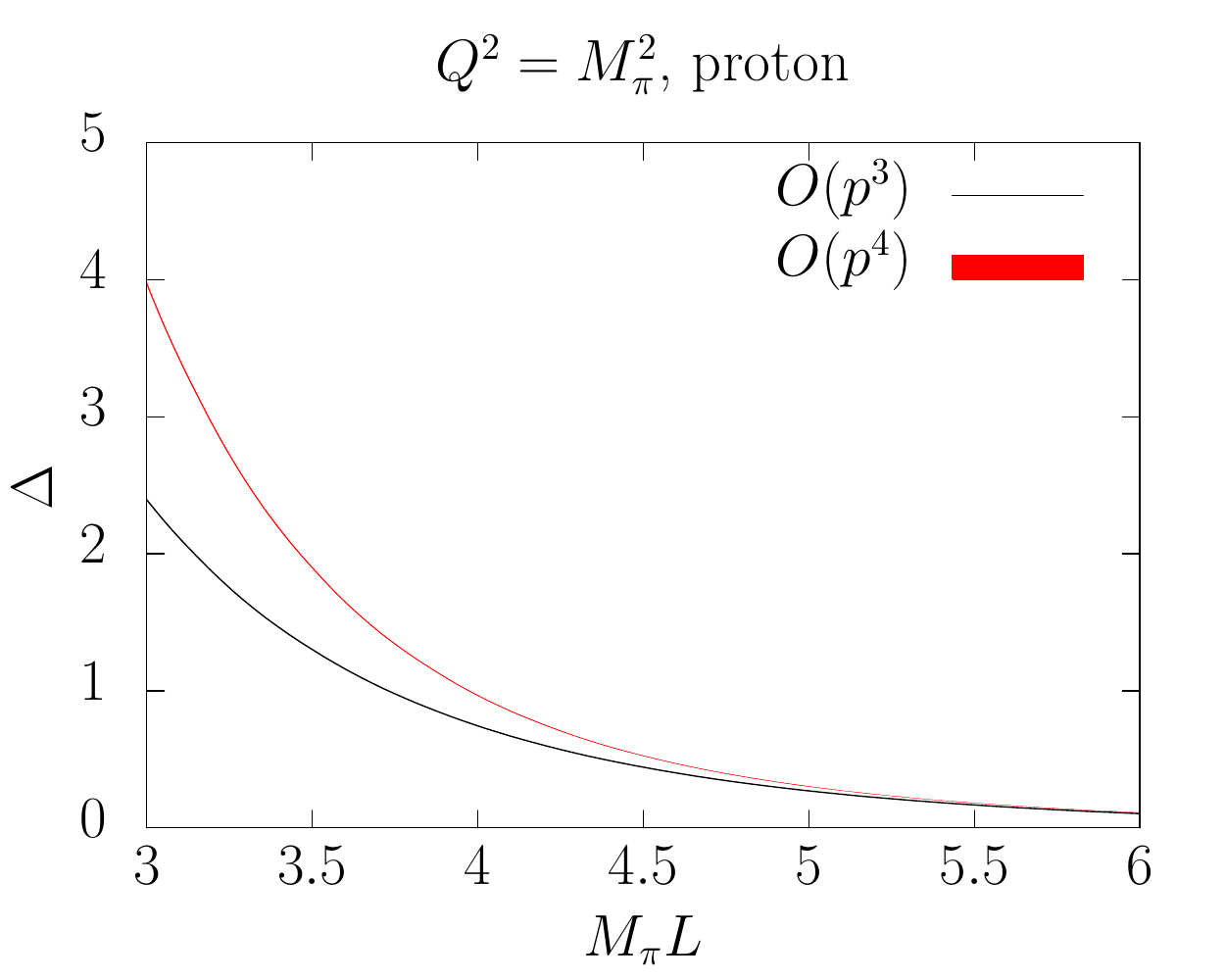}\hspace*{.4cm}
    \includegraphics[width=0.35\linewidth]{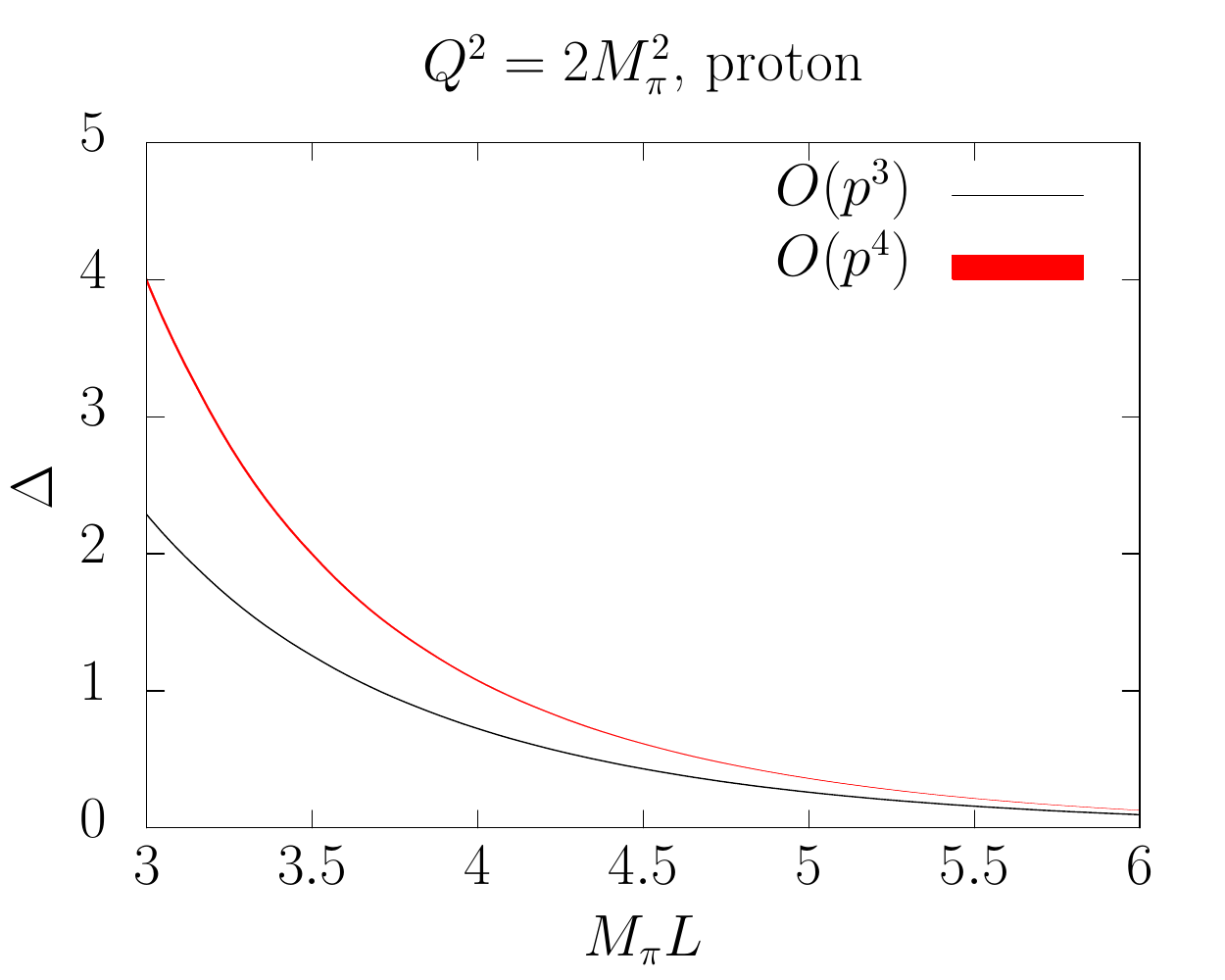}

    \caption{The finite-volume effect in the proton amplitude versus the dimensionless variable $M_\pi L$.
      The quantity $\Delta$ is shown in percent. The uncertainty in the knowledge of $e_y^+$ does not translate
      into a large uncertainty in the final results, the width of the red band is barely visible by eye.
      }
    \label{fig:FV-proton}
  \end{center}
\end{figure}

\begin{figure}[t]
  \begin{center}
   \includegraphics[width=0.35\linewidth]{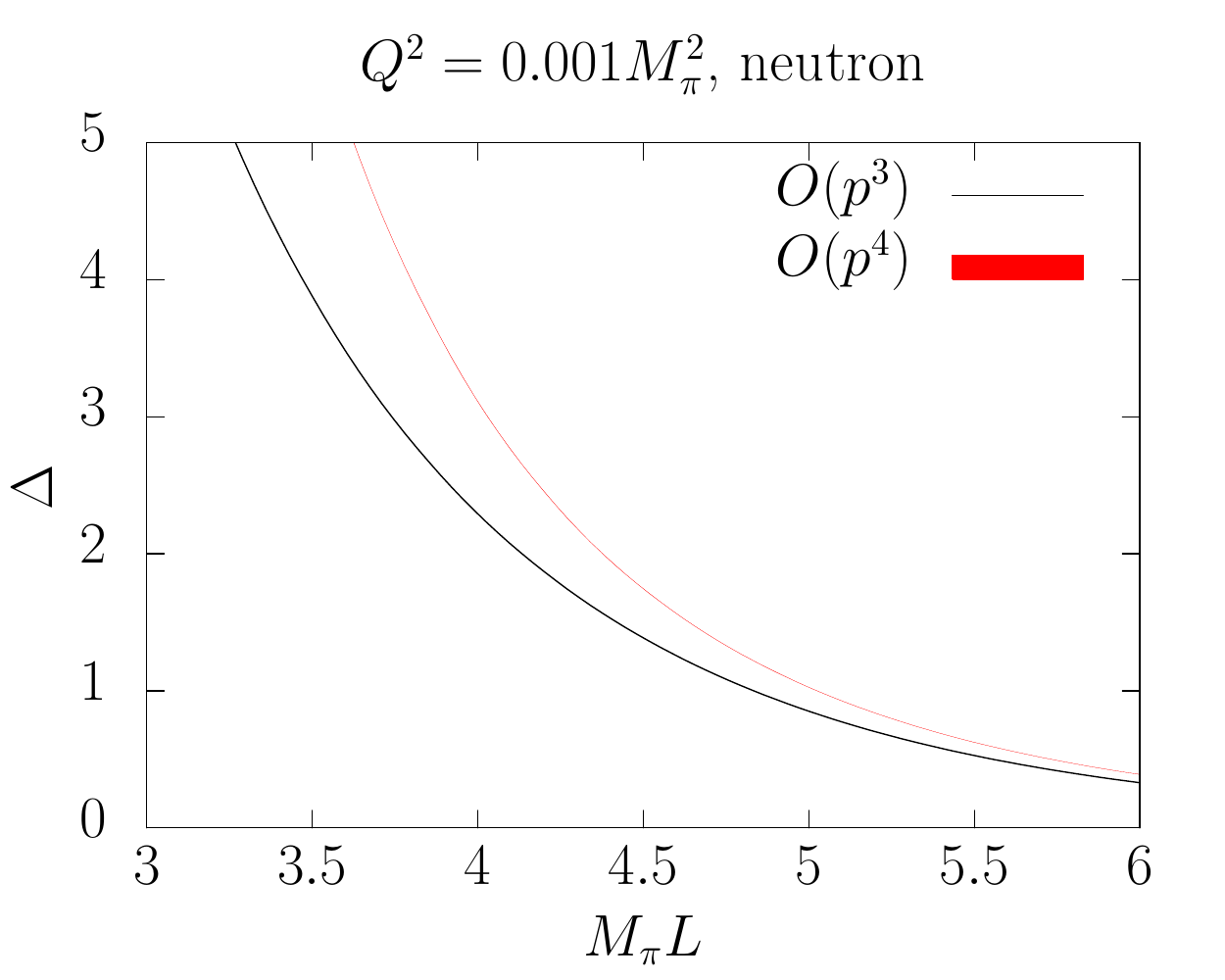}\hspace*{.4cm}
    \includegraphics[width=0.35\linewidth]{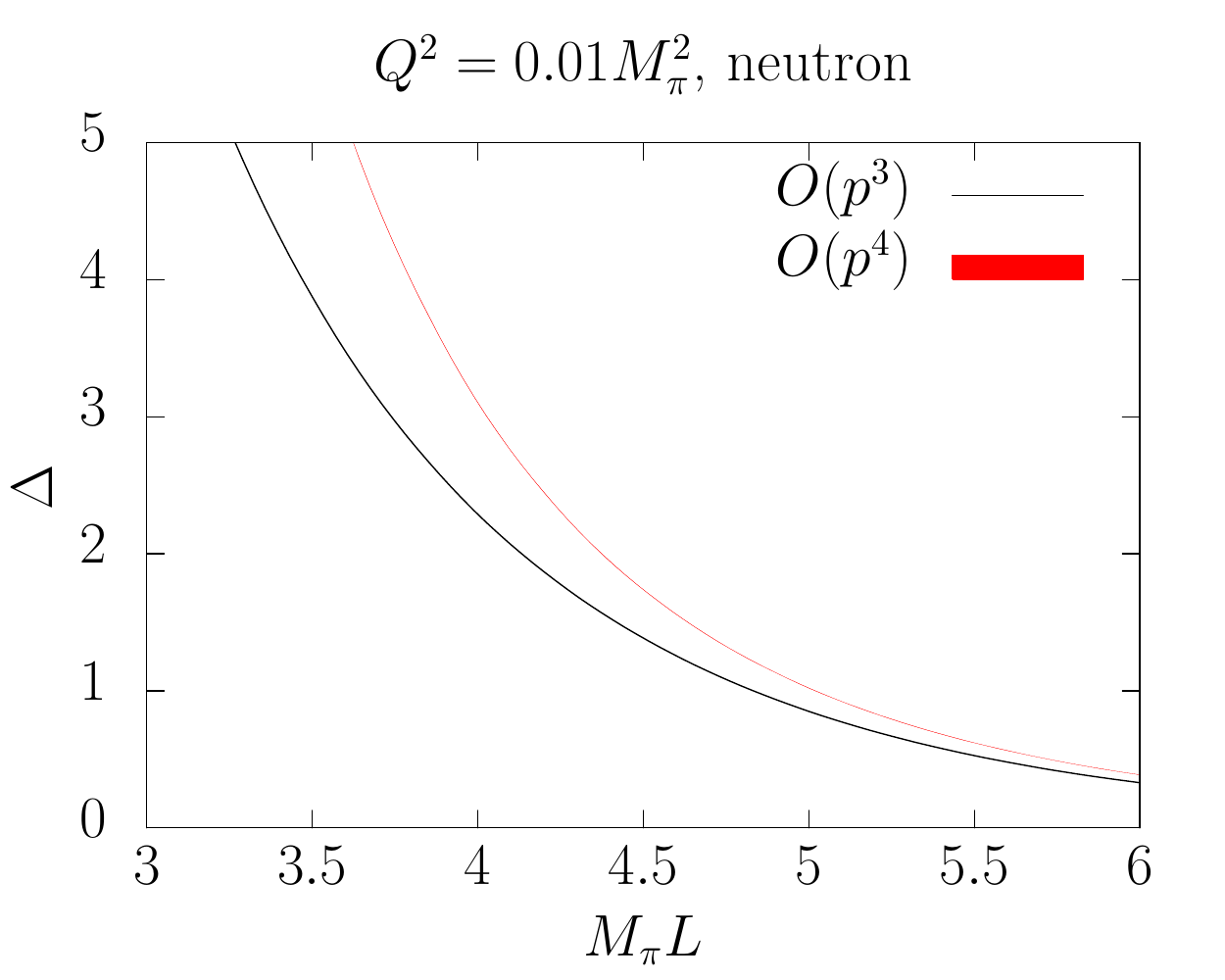}

    \vspace*{.5cm}

    \includegraphics[width=0.35\linewidth]{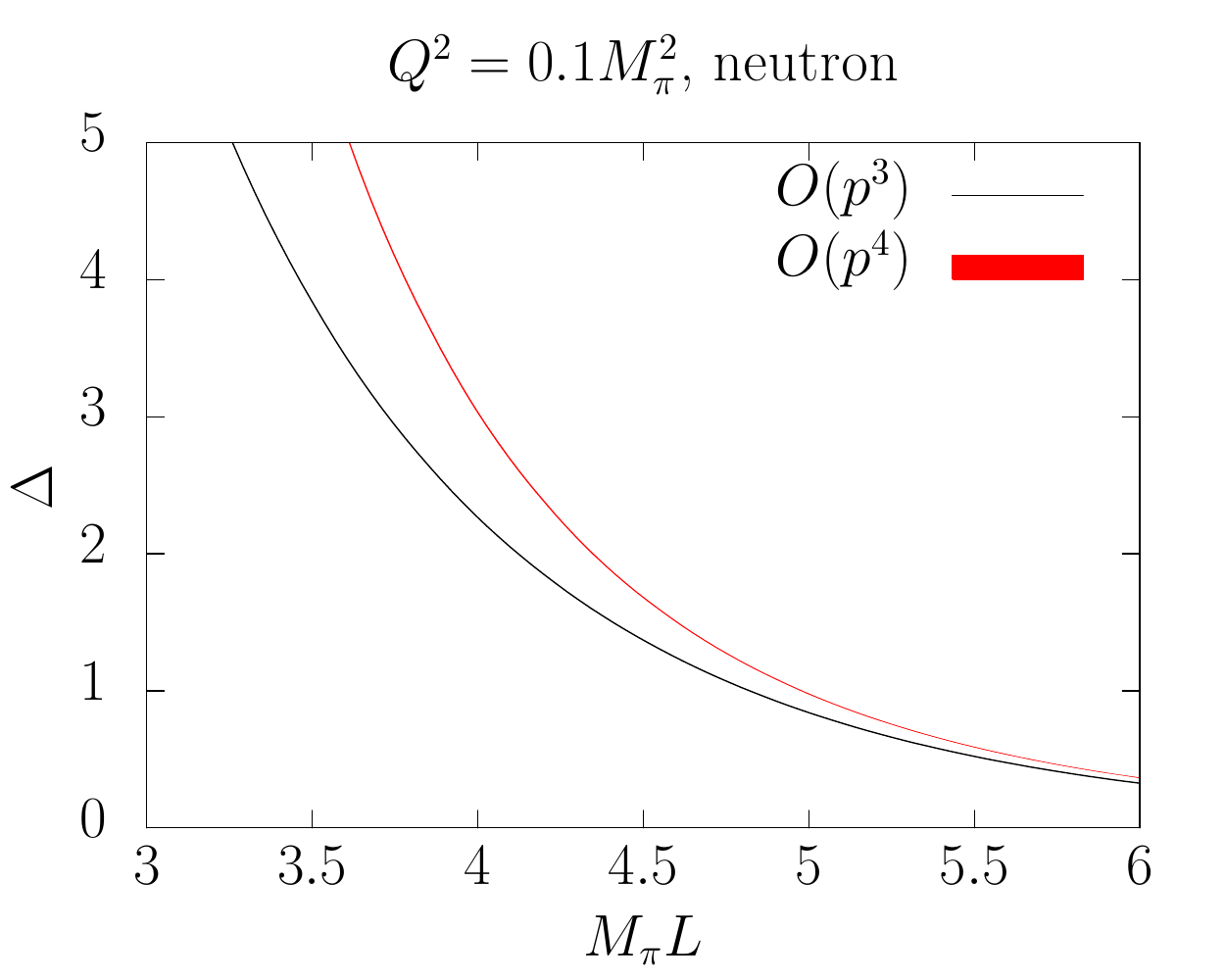}\hspace*{.4cm}
    \includegraphics[width=0.35\linewidth]{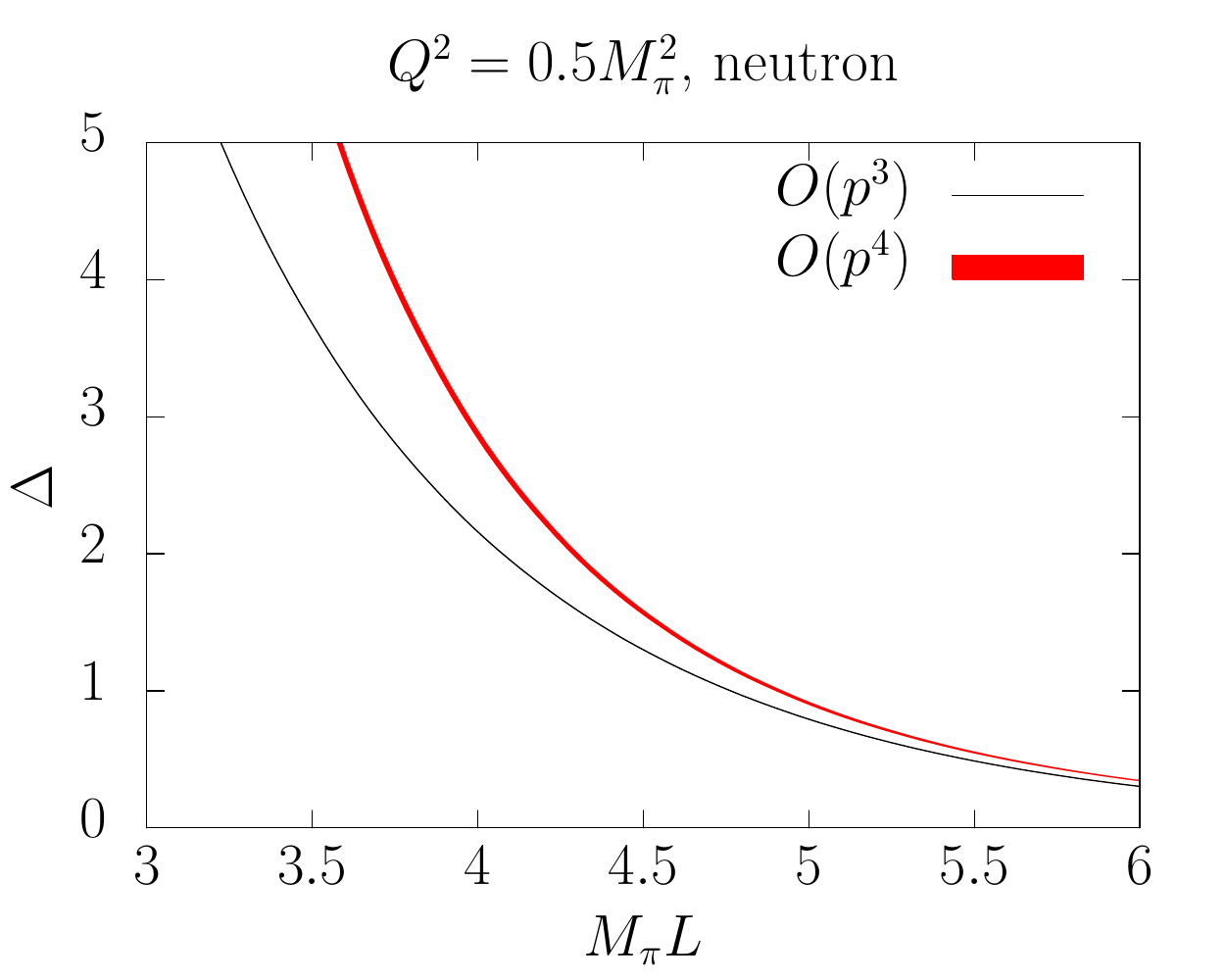}

    \vspace*{.5cm}

    \includegraphics[width=0.35\linewidth]{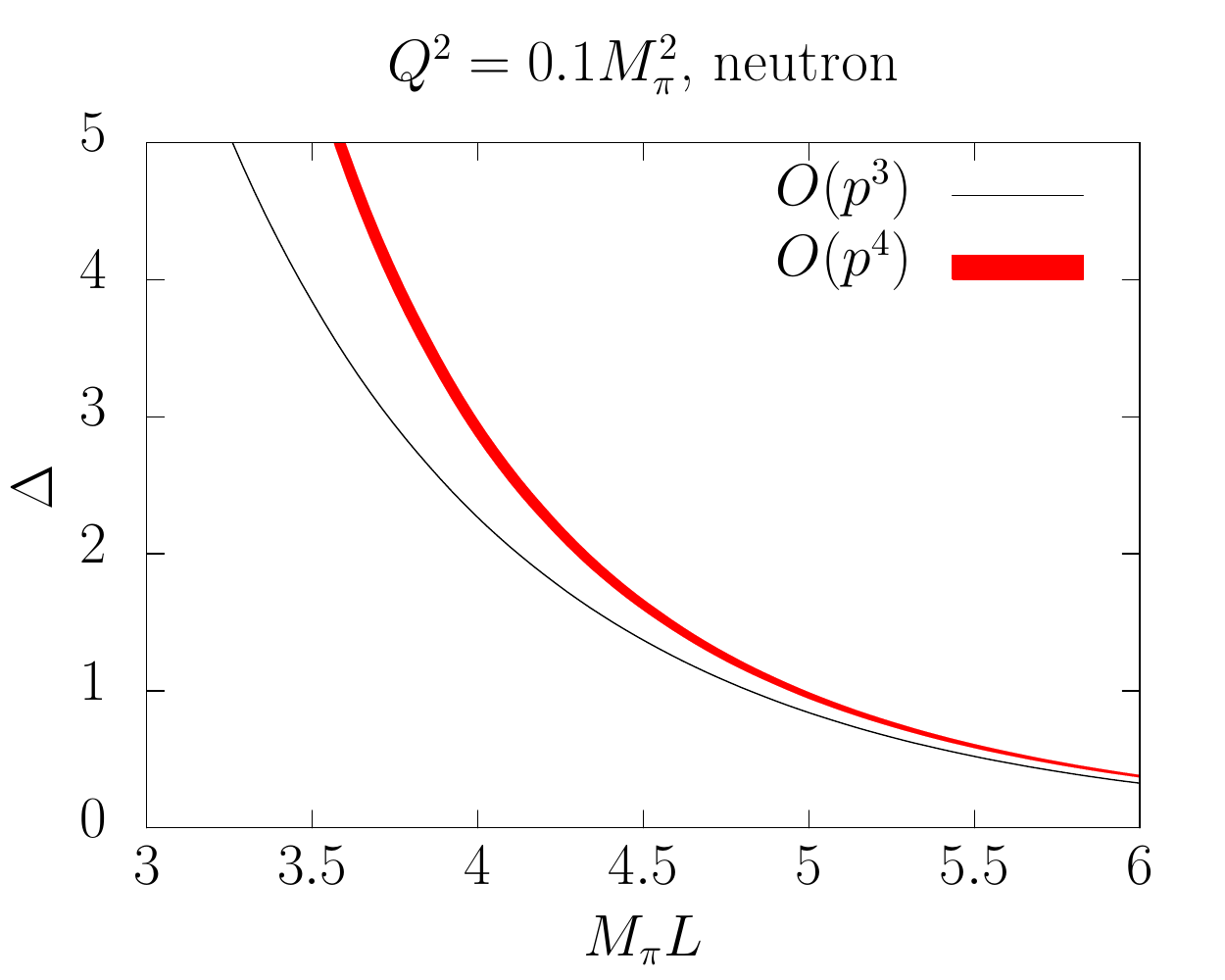}\hspace*{.4cm}
    \includegraphics[width=0.35\linewidth]{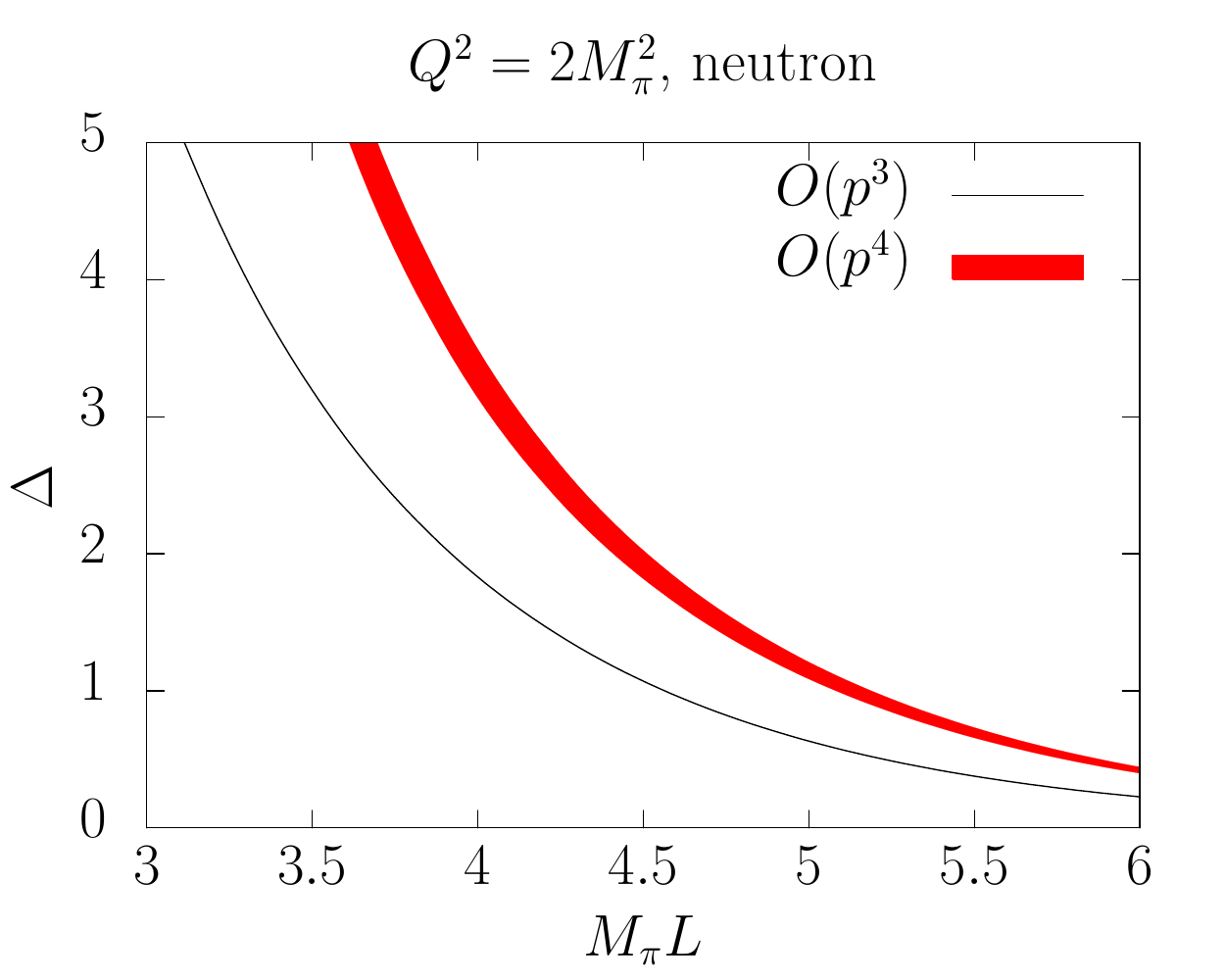}

    \caption{The same as in Fig.~\ref{fig:FV-proton} for the neutron.}
    \label{fig:FV-neutron}
  \end{center}
\end{figure}

The finite-volume corrections to the Compton amplitude are shown in
Figs.~\ref{fig:FV-proton} and \ref{fig:FV-neutron} for the following values of the variable $Q^2$:
\eq
Q^2=0.001M_\pi^2\, ,~0.01M_\pi^2\, ,~0.1M_\pi^2\, ,~0.5 M_\pi^2\, ,~M_\pi^2\, ,~2M_\pi^2\, .
\en
These figures contain our main result, answering the question about the feasibility of the extraction of
the subtraction function on the lattice. It is seen that, both for the proton and the neutron,
the finite-volume corrections are encouragingly small for $M_\pi L\geq 4$ (the fact that they are
slightly smaller for the proton than for the neutron stems from the presence of the large second-order
pole term proportional to the combination $2c_6+c_7$,
in the infinite-volume proton amplitude). Further, it is very comforting to
see that the convergence of the result at fourth order is reasonable. Moreover,
the uncertainty caused by the poor knowledge of the $O(p^4)$ LECs is indeed
moderate in the final results (for example, in case of the proton,
it is hardly visible by the bare eye). Taking this fact into account, one might wonder whether the uncertainty
in the LECs at lower orders might play a more significant role.
Following our expectation, a 20-30\,\% error in a final result  generously covers the effect coming from
these LECs. Another easy way to estimate the uncertainty of calculations (not limited
necessarily to the poorly determined LECs) is to compare the results at $O(p^3)$ and $O(p^4)$. Taking into
account the fact that the present study was primarily intended to serve
as a rough estimate of the size of the exponentially suppressed corrections to the amplitude,
we did not try to investigate this question further.

\medskip

Finally, note that the relative
correction stays almost constant from $Q^2\simeq 0$ to $Q^2\simeq 2M_\pi^2$ and,
possibly, even for higher values of $Q^2$. In other words, the finite-volume
artifacts do not hinder an accurate extraction of the amplitude at large $Q^2$.
Here we remind a reader  that an accurate measurement of the inelastic part on the lattice
becomes more difficult as $Q^2\to 0$, because the elastic contribution dominates
in this limit~\cite{Agadjanov:2016cjc,Agadjanov:2018yxh}. Hence, the finite-volume
artifacts do not further restrict an interval in $Q^2$, where an accurate calculation
of the subtraction function is possible.

\section{Summary}
\label{sec:concl}

\begin{itemize}
\item[i)]
  Using Baryon Chiral Perturbation Theory and the EOMS renormalization
  scheme of Refs.~\cite{Gegelia:1999gf,Fuchs:2003qc},
  which guarantees that the renormalized expressions satisfy the standard
  power counting, we have evaluated the doubly virtual spin-averaged
  Compton scattering amplitude off nucleons up to $O(p^4)$ in the infinite
  volume. We have further calculated finite-volume corrections to the so-called
  subtraction function at the same chiral order.

\item[ii)]
  The inelastic parts of the infinite-volume subtraction functions $S^{\sf inel}_1(q^2)$
  and $\bar S(q^2)$ show a rather regular behavior
  at small values of $q^2$, that is of the order of the pion mass squared. None of the loop
  diagrams up-to-and-including order $p^4$ leads to a rapid variation of the
  calculated subtraction functions at small scales. Note that in these
  calculations we have fixed the numerical values of some of the $O(p^4)$
  LECs through the proton and neutron magnetic
  polarizabilities which, at present, are
  not known to high accuracy (especially, the one for the neutron).

\item[iii)]
  The main result of this work is the calculation of the finite-volume
  corrections in the Compton scattering amplitude. These
  calculations are interesting,
  first and foremost, in  view of the perspective of a measurement
  of the subtraction function on the lattice using  periodic external
  fields. Note also that, since the cubic lattice does not preserve
  Lorentz invariance, the definition of the subtraction function in a finite
  volume is ambiguous. On the other hand, what is extracted from the nucleon
  mass shift, measured on the lattice,
  is a particular component of the second-rank Compton tensor $T^{11}_L$, which
  is a well-defined quantity and which tends to 
  $S_1(q^2)$ (modulo an overall kinematic factor) in the infinite-volume limit.
  The numerical results quoted in this work refer to this quantity.

\item[iv)]
  At this stage, we do not know,  on the one hand,
  how the other subtraction function $\bar S(q^2)$
  can be measured on the lattice. On the other hand, the two subtraction
  functions are related to each other: Their difference is a convergent
  integral containing experimentally measured electroproduction cross
  sections.

\item[v)]
  Our results show that the exponentially vanishing
  finite-volume corrections to the quantity 
  $T^{11}_L$ amount up to 2-3\% percent or less at $M_\pi L\simeq 4$ both for
  the proton and the neutron.
  This means that one can extract the infinite-volume subtraction function
  with a good accuracy already using reasonably large lattices. We also note that
  the convergence of our results is rather good, and the poor knowledge of the
  $O(p^4)$ LECs does not pose  a real obstacle as the resulting uncertainty is
  very small.

\item[vi)]
  As pointed out in Refs.~\cite{Agadjanov:2016cjc,Agadjanov:2018yxh}, the
  large elastic part renders the extraction of the subtraction function
  at low $q^2$ problematic. Having the finite-volume corrections well under
  control might help one to carry out the analysis at lower  values of $q^2$.  
  On the other hand, the observed $q^2$-behavior of the quantity $\Delta$
  is very mild up to $-q^2=2M_\pi^2$, and one may try to push up the upper limit
  in $q^2$ higher, allowing for the extraction of the subtraction function
  in a larger interval of $q^2$.

  \end{itemize}

\section*{Acknowledgments}

The authors would like to thank D.~Djukanovic, J.~Gasser, H.~Leutwyler, V.~Pascalutsa and
G.~Schierholz for interesting discussions.
J.~L. would like to acknowledge the financial support from the fellowship ``Regierungsstipendiaten
CONACYT-DAAD mit Mexiko'' under the grant number 2016 (57265507).
A.~A. was supported by the European Research Council (ERC) under the European Union’s
Horizon 2020 research and innovation programme through grant agreement
771971-SIMDAMA.
J.~G.  was supported by BMBF (Grant No. 05P18PCFP1) and by the 
Georgian Shota Rustaveli National Science Foundation (Grant No. FR17-354). 
The work of U.-G.~M. and  A.~R.  was  supported in part by the DFG (CRC 110 ``Symmetries 
and the Emergence of Structure in QCD'', grant no. TRR110), Volkswagenstiftung 
(grant no. 93562) and the Chinese Academy of Sciences (CAS) President's
International Fellowship Initiative (PIFI) (grant nos. 2018DM0034 and 2021VMB0007).

\appendix

\section{Definition of the loop integrals}
\label{app:integrals}

The one-loop integrals appearing in our calculations are defined as follows:

\medskip
{\em One factor in the denominator:}

\eq
\int\frac{d^n k}{k^2-m^2+i \epsilon}  &=&  A_0(m)\,,\nonumber\\[2mm]
\int\frac{d^n k}{[k^2-m^2+i \epsilon]^N}  &=&  A_0^{(N)}(m) \,,\nonumber\\[2mm]
\int\frac{d^n k k^\mu k^\nu}{k^2-m^2+i \epsilon}  &=&  g^{\mu\nu}A_{00}(m) \,.
\label{ints1}
\
\en

\medskip
{\em Two factors in the denominator:}

\eq
\int\frac{d^nk}{[k^2-M^2+i \epsilon] [(p+k)^2-m^2+i \epsilon]} &=& B_0\left(p^2;M,m\right)\,, \nonumber\\[2mm]
\int\frac{d^nk}{[k^2-M^2+i \epsilon]^M [(p+k)^2-m^2+i \epsilon]^N} &=& B^{(M,N)}_0\left(p^2;M,m\right)\,,\nonumber\\[2mm]
\int\frac{d^nk \, k^\mu}{[k^2-M^2+i \epsilon] [(p+k)^2-m^2+i \epsilon]}  &=&  p^{\mu } B_1\left(p^2;M,m\right) \,,
\nonumber\\[2mm]
\int\frac{d^nk \, k^\mu}{[k^2-M^2+i \epsilon]^M [(p+k)^2-m^2+i \epsilon]^N } & =&  p^{\mu } B^{(M,N)}_1\left(p^2;M,m\right) \,,
\nonumber\\[2mm]
  \int\frac{d^nk \, k^\mu k^\nu }{[k^2-M^2+i \epsilon] [(p+k)^2-m^2+i \epsilon]} &=& g^{\mu \nu } B_{00}\left(p^2;M,m\right)+p^{\mu } p^{\nu }
   B_{11}\left(p^2;M,m\right)\,,\nonumber\\[2mm]
    \int\frac{d^nk \, k^\mu k^\nu }{[k^2-M^2+i \epsilon]^M [(p+k)^2-m^2+i \epsilon]^N} &=& g^{\mu \nu } B^{(M,N)}_{00}\left(p^2;M,m\right)
                                        +\,p^{\mu } p^{\nu }
   B^{(M,N)}_{11}\left(p^2;M,m\right)\,,\nonumber\\[2mm]
\int \frac{d^nk\, k^\mu k^\nu k^\alpha}{(k^2-M^2+i \epsilon)^2[(k-q)^2-M^2+i \epsilon]}
&=& -q^\alpha q^\mu q^\nu B_{111}^{(2,1)} (q^2;M,M)
\nonumber\\[2mm]
&-&( q^\nu g^{\alpha\nu}   + q^\mu g^{\alpha\mu}  + q^\alpha g^{\mu\nu})B_{001}^{(2,1)} (q^2;M,M)\, ,
\nonumber\\[2mm]
\int \frac{d^nk\, k^\mu k^\nu k^\alpha k^\beta}{(k^2-M^2+i \epsilon)^2[(k-q)^2-M^2+i \epsilon]}
&=& (g^{\alpha \nu} g^{\beta\mu} + g^{\alpha \mu} g^{\beta\nu}+g^{\alpha \beta} g^{\mu\mu}  )B_{0000}^{(2,1)} (q^2;M,M)
\nonumber\\[2mm]
&+&\, q^\mu q^\nu q^\alpha q^\beta B_{1111}^{(2,1)}(q^2;M,M)
\nonumber\\[2mm]&&\hspace*{-4.cm}
 +\,  ( q^\beta q^\nu g^{\alpha\mu} + q^\mu q^\nu g^{\alpha\beta} +q^\mu q^\beta g^{\alpha \nu}
+\, q^\alpha q^\beta g^{\mu\nu} +q^\mu q^\alpha g^{\beta\nu} + q^\nu q^\alpha g^{\mu\beta}) B_{0011}^{(2,1)} (q^2;M,M)\, .  \label{ints2}
\
 \en

\medskip
{\em Three factors in the denominator:}

 \eq
 \int\frac{d^nk}{[k^2-M^2+i \epsilon] [(p+k)^2-m_1^2+i \epsilon] [(q+k)^2-m_2^2+i \epsilon]} &=&\,  C_0\left(p^2,(p-q)^2,q^2;M,m_1,m_2\right) \,,
\nonumber\\[2mm]
\int\frac{d^nk}{[k^2-M^2+i \epsilon]^M [(p+k)^2-m_1^2+i \epsilon]^N [(q+k)^2-m_2^2+i \epsilon]^K } &=&\,  C^{(M,N,K)}_0\left(p^2,(p-q)^2,q^2;M,m_1,m_2\right) \,,
\nonumber\\[2mm]
\int\frac{d^nk \, k^\mu}{[k^2-M^2+i \epsilon] [(p+k)^2-m_1^2+i \epsilon] [(q+k)^2-m_2^2+i \epsilon]}  &=&\,  q^{\mu } C_2\left(p^2,(p-q)^2,q^2;M,m_1,m_2\right)
\nonumber\\[2mm]
&+& p^{\mu } C_1\left(p^2,(p-q)^2,q^2;M,m_1,m_2\right)\,,\nonumber\\[2mm]
\int\frac{d^nk \, k^\mu}{[k^2-M^2+i \epsilon]^M [(p+k)^2-m_1^2+i \epsilon]^N [(q+k)^2-m_2^2+i \epsilon]^K}
&=&\,  q^{\mu } C^{(M,N,K)}_2\left(p^2,(p-q)^2,q^2;M,m_1,m_2\right)
\nonumber\\[2mm]
&+& p^{\mu } C^{(M,N,K)}_1\left(p^2,(p-q)^2,q^2;M,m_1,m_2\right)\,,
\nonumber\\[2mm]
\int\frac{d^nk \, k^\mu k^\nu}{[k^2-M^2+i \epsilon] [(p+k)^2-m_1^2+i \epsilon] [(q+k)^2-m_2^2+i \epsilon]}
&=&\,  g^{\mu \nu } C_{00}\left(p^2,(p-q)^2,q^2;M,m_1,m_2\right)
\nonumber\\[2mm]
&+& q^{\mu } q^{\nu }
C_{22}\left(p^2,(p-q)^2,q^2;M,m_1,m_2\right)
\nonumber\\[2mm]
 &+&\, p^{\mu} p^{\nu } C_{11}\left(p^2,(p-q)^2,q^2;M,m_1,m_2\right)
\nonumber\\[2mm]
 &&\hspace*{-2.5cm}+\,  \left( p^{\nu } q^{\mu }+ p^{\mu
   } q^{\nu } \right) C_{12}\left(p^2,(p-q)^2,q^2;M,m_1,m_2\right)\,,\nonumber\\[2mm]
    \int\frac{d^nk \, k^\mu k^\nu}{[k^2-M^2+i \epsilon]^M [(p+k)^2-m_1^2+i \epsilon]^N [(q+k)^2-m_2^2+i \epsilon]^K }  &=&\,  g^{\mu \nu } C^{(M,N,K)}_{00}\left(p^2,(p-q)^2,q^2;M,m_1,m_2\right)
\nonumber\\[2mm]
    &+& q^{\mu } q^{\nu }
    C^{(M,N,K)}_{22}\left(p^2,(p-q)^2,q^2;M,m_1,m_2\right)
    \nonumber\\[2mm]
    &+&\, p^{\mu} p^{\nu } C^{(M,N,K)}_{11}\left(p^2,(p-q)^2,q^2;M,m_1,m_2\right) 
\nonumber\\[2mm]&&\hspace*{-2.5cm}
   +\,\left( p^{\mu
   } q^{\nu } +p^{\nu } q^{\mu }\right) C^{(M,N,K)}_{12}\left(p^2,(p-q)^2,q^2;M,m_1,m_2\right) \, .
   \label{ints3}
\end{eqnarray}

\section{Tree-level expressions -- individual diagrams}
\label{app:infinite-tree}

Below we list the tree-order contributions to
the quantities $T_1(0,q^2)$ and $T_2(0,q^2)$, coming from the
individual diagrams in Figs.~\ref{fig:O1}-\ref{fig:O4}.
At the order we are working, one may safely replace $\mkrig$ by $m$, $M$ by $M_\pi$ and
$F$ by $F_\pi$ everywhere.

\medskip

{\em $O(p)$ contributions:}

\eq
T_1^1&=&0\, ,
\nonumber\\
T_2^1&=&2m^2(\tau^3+1)\frac{1}{q^4}\, .
\en

{\em $O(p^2)$ contributions:}

\begin{eqnarray}
T_1^{2,3} & = & \frac{  m }{ q^2} \left(2 c_6+c_7\right) \left(\tau ^3+1\right) 
\,,\nonumber\\
T_2^{2,3} & = & 0
\,.
\label{TreeKs2}
\end{eqnarray}

{\em $O(p^3)$ contributions:}

  \eq
T_1^{4,5} & = &  0\,,
\nonumber\\
T_2^{4,5} & = & - 4\frac{m^2}{q^2} \left(d_6+2 d_7\right)  \left(\tau ^3+1\right)\,,
\nonumber\\
T_1^{6} & = &    \frac{ m^2}{q^2} \left(4 c_7 c_6 \tau ^3+4 c_6^2+c_7^2\right)\,,
\nonumber\\
T_2^{6} & = &  -\frac{m^2}{q^2} \left(4 c_7 c_6 \tau ^3+4 c_6^2+c_7^2\right)\, ,
\nonumber\\
T_1^{7} & = &  \left(d_6+2 d_7\right)  \left(\tau ^3+1\right)\,,
\nonumber\\
T_2^{7} & = & 0\,.
\label{TreeKs3}
\en

{\em $O(p^4)$ contributions:}
\begin{eqnarray}
T_1^{33} & = &  
8m\left(2e_{89}+e_{93}+e_{118} +\tau^3 e_{91}\right) 
\,,\nonumber\\
T_2^{33} & = & 4m \left(2e_{90}+e_{94}+e_{117}+\tau^3e_{92}\right)
\,,\nonumber\\
T_1^{34,35} & = & \frac{2 m }{
	 q^2} \left(\tau ^3+1\right) \left(\left(2 e_{54}+e_{74}\right) q^2-4 \left(2 e_{105}+e_{106}\right) M^2\right)
\,,\nonumber\\
T_2^{34,35} & = &  0
\,,\nonumber\\
T_1^{36,37} & = &  0
\,,\nonumber\\
T_2^{36,37} & = &  -2 m \left(c_7 \left(d_6 \tau ^3+2 d_7\right)+2 c_6 \left(2 d_7 \tau ^3+d_6\right)\right)
\,.
\label{TreeKs4}
\end{eqnarray}

\section{One-loop expressions -- individual diagrams}
\label{app:infinite-loop}

Below one finds the one-loop contributions to the amplitudes $T_1(0,q^2)$ and $T_2(0,q^2)$:

\medskip

{\em $O(p^3)$ contributions: } 
\begin{eqnarray}
  T_1^{8} & = &  -\frac{ m^2 g_A^2 }{8 \pi ^2 F^2}
\biggl\{ M^2 \left(4
                \left({C}_{22}^{(2,1,1)}\left(m^2,m^2+q^2,q^2;M,m,M\right)\right. \right.
+\left. {C}_2^{(2,1,1)}\left(m^2,m^2+q^2,q^2;M,m,M\right)\right)\\
            &+& \left. {C}_0^{(2,1,1)}\left(m^2,m^2+q^2,q^2;M,m,M\right)\right)
+4
                {C}_{22}\left(m^2,m^2+q^2,q^2;M,m,M\right)\nonumber\\
                &+&{C}_0\left(m^2,m^2+q^2,q^2;M,m,M\right)
+  4{C}_2\left(m^2,m^2+q^2,q^2;M,m,M\right)\biggr\}\,,
    \nonumber\\
  T_2^{8} & = &    \frac{ m^4 g_A^2 }{2 \pi ^2 F^2 q^2}
                \left(M^2 C_{11}^{(2,1,1)}\left(m^2,m^2+q^2,q^2;M,m,M\right)\right.
        +\left. C_{11}\left(m^2,m^2+q^2,q^2;M,m,M\right)\right)\,,
            \nonumber\\  
T_1^{9} & = &  \frac{ m^2 \left(\tau ^3-1\right) g_A^2 }{16 \pi ^2 F^2}
\biggl\{  M^2 \left(4
              \left({C}_{22}^{(2,1,1)}\left(m^2,q^2,m^2+q^2;M,m,m\right)\right.\right.
        -\left. {C}
_{22}^{(2,1,1)}\left(m^2,m^2+q^2,q^2;M,m,M\right) \right.   
\nonumber\\
            &+& \left. \left. 
{C}_{22}^{(2,1,1)
}\left(m^2+q^2,m^2,q^2;M,m,M\right)  \right. \right.  
+  \left. \left. 2
{C}_{12}^{(2,1,1)}\left(m^2+q^2,m^2,q^2;M,m,M\right)
\right. \right. 
\nonumber\\ 
&+& \left. \left. 
{C}_{11}^
{(2,1,1)}\left(m^2+q^2,m^2,q^2;M,m,M\right)+{C}_2^{(2,1,1)}\left(m^
2,q^2,m^2+q^2;M,m,m\right)
\right. \right. 
\nonumber\\ 
&-& \left. \left. 
{C}_2^{(2,1,1)}\left(m^2,m^2+q^2,q^2;M,m,M
\right)+{C}_2^{(2,1,1)}\left(m^2+q^2,m^2,q^2;M,m,M\right)
\right. \right. 
\nonumber\\ 
&+& \left. \left. 
{C}_1^{(2,1,1)}\left(m^2+q^2,m^2,q^2;M,m,M\right)\right)+{C}_0^{(2,1,1)
}\left(m^2,q^2,m^2+q^2;M,m,m\right)
\right. 
\nonumber\\ 
&-& \left. 
{C}_0^{(2,1,1)}\left(m^2,m^2+q^2,
q^2;M,m,M\right)+{C}_0^{(2,1,1)}\left(m^2+q^2,m^2,q^2;M,m,M\right)\right)
\nonumber\\ 
&+& 
4 {C}_{22}\left(m^2,q^2,m^2+q^2;M,m,m\right)-2
{C}_{22}\left(m^2,m^2+q^2,q^2;M,m,M\right)+2
{C}_{22}\left(m^2+q^2,m^2,q^2;M,m,M\right)
\nonumber\\
&+&4
{C}_{12}\left(m^2+q^2,m^2,q^2;M,m,M\right)+2
{C}_{11}\left(m^2+q^2,m^2,q^2;M,m,M\right)+{C}_0\left(m^2,q^2,m^
2+q^2;M,m,m\right)
\nonumber\\
& - & {C}_0\left(m^2,m^2+q^2,q^2;M,m,M\right)+{C}_0
\left(m^2+q^2,m^2,q^2;M,m,M\right)+4
{C}_2\left(m^2,q^2,m^2+q^2;M,m,m\right)
\nonumber\\
&-& 3
{C}_2\left(m^2,m^2+q^2,q^2;M,m,M\right)+3
{C}_2\left(m^2+q^2,m^2,q^2;M,m,M\right)+3
{C}_1\left(m^2+q^2,m^2,q^2;M,m,M\right)\biggr\}
\,,\nonumber\\
T_2^{9} & = &  - \frac{ m^4 \left(\tau ^3-1\right) g_A^2 }{8 \pi ^2 F^2 q^2} \left(2 M^2
C_{22}^{(2,1,1)}\left(m^2,q^2,m^2+q^2;M,m,m\right)+4 M^2
C_{12}^{(2,1,1)}\left(m^2,q^2,m^2+q^2;M,m,m\right)
\right.  \nonumber\\
&+& \left. 
2 M^2
C_{11}^{(2,1,1)}\left(m^2,q^2,m^2+q^2;M,m,m\right)-2 M^2
C_{11}^{(2,1,1)}\left(m^2,m^2+q^2,q^2;M,m,M\right)
\right.  \nonumber\\
&+& \left. 
2 M^2
C_{11}^{(2,1,1)}\left(m^2+q^2,m^2,q^2;M,m,M\right)+2 M^2
C_1^{(2,1,1)}\left(m^2,q^2,m^2+q^2;M,m,m\right)
\right.  \nonumber\\
&-& \left. 
2 M^2
C_1^{(2,1,1)}\left(m^2,m^2+q^2,q^2;M,m,M\right)+2 M^2
C_1^{(2,1,1)}\left(m^2+q^2,m^2,q^2;M,m,M\right)
\right.  \nonumber\\
&+& \left. 
\left(2 M^2-q^2\right)
C_2^{(2,1,1)}\left(m^2,q^2,m^2+q^2;M,m,m\right)-q^2
C_1^{(2,1,1)}\left(m^2,q^2,m^2+q^2;M,m,m\right)
\right.  \nonumber\\
&+& \left. 
q^2
C_1^{(2,1,1)}\left(m^2,m^2+q^2,q^2;M,m,M\right)-q^2
C_1^{(2,1,1)}\left(m^2+q^2,m^2,q^2;M,m,M\right)
\right.  \nonumber\\
&+& \left. 
2
C_{22}\left(m^2,q^2,m^2+q^2;M,m,m\right)+4
C_{12}\left(m^2,q^2,m^2+q^2;M,m,m\right)+2
C_{11}\left(m^2,q^2,m^2+q^2;M,m,m\right)
\right.  \nonumber\\
&-& \left. 
C_{11}\left(m^2,m^2+q^2,q^2;M,m,M\right)+C_{11}\left(m^2+q^2,m^2,q^2;M,m,M\right)+2
C_2\left(m^2,q^2,m^2+q^2;M,m,m\right)
\right.  \nonumber\\
&+& \left. 
2
C_1\left(m^2,q^2,m^2+q^2;M,m,m\right)-C_1\left(m^2,m^2+q^2,q^2;M,m,M\right)+C_
1\left(m^2+q^2,m^2,q^2;M,m,M\right)\right)
\,,\nonumber\\   
T_1^{10} & = & \frac{ m^2 \left(\tau ^3-3\right) g_A^2}{8 \pi ^2 F^2}
\biggl\{  {B}_{11}^{(2,1)}\left(q^2;m,m\right)+{B}_1^{(2,1)}\left(q^2;m,m\right)+M^2
\left({C}_{22}^{(1,2,1)}\left(m^2,q^2,m^2+q^2;M,m,m\right) \right. \nonumber\\
&+& \left. {C}
_2^{(1,2,1)}\left(m^2,q^2,m^2+q^2;M,m,m\right)\right)\biggr\} 
\,,\nonumber\\
T_2^{10} & = &  \frac{  m^2 \left(\tau ^3-3\right) g_A^2}{32 \pi ^2 F^2 q^2} 
\left( -4 m^2 M^2
\left(C_{22}^{(1,2,1)}\left(m^2,q^2,m^2+q^2;M,m,m\right) 
\right.\right. \nonumber\\
&+& \left. \left.  2
\left(C_{12}^{(1,2,1)}\left(m^2,q^2,m^2+q^2;M,m,m\right)+C_1^{(1,2,1)}\left(m^2,q^2,m^2+q^2;M,m,m\right)\right) 
\right.\right. \nonumber\\
&+& \left. \left. C_{11}^{(1,2,1)}\left(m^2,q^2,m^2+q^2;
M,m,m\right)+C_0^{(1,2,1)}\left(m^2,q^2,m^2+q^2;M,m,m\right) 
\right.\right. \nonumber\\
&+& \left. \left.  2
C_2^{(1,2,1)}\left(m^2,q^2,m^2+q^2;M,m,m\right)\right)+B_0\left(m^2+q^2;M,m\right)+B_1\left(m^2+q^2;M,m\right)\right)
\,,\nonumber\\   
T_1^{11,12} & = &  -\frac{ m^2 \left(\tau ^3-1\right) g_A^2
	\left({C}_{22}\left(m^2,q^2,m^2+q^2;M,m,m\right)+{C}_2\left(m^2,
	q^2,m^2+q^2;M,m,m\right)\right)}{4 \pi ^2 F^2}
\,,\nonumber\\
T_2^{11,12} & = & - \frac{ m^2 \left(\tau ^3-1\right) g_A^2}{8 \pi ^2 F^2 q^2}
\left(B_0\left(m^2+q^2;M,m\right)+B_1\left(m^2+q^2;M,m\right)-2 m^2
\left(C_{22}\left(m^2,q^2,m^2+q^2;M,m,m\right)
\right.\right. \nonumber\\
&+& \left. \left. 
2
C_{12}\left(m^2,q^2,m^2+q^2;M,m,m\right)+C_{11}\left(m^2,q^2,m^2+q^2;M,m,m\right)+C_2\left(m^2,q^2,m^2+q^2;M,m,m\right) 
\right.\right. \nonumber\\
&+& \left. \left.  C_1\left(m^2,q^2,m^2+q^2;M,m,m\right
)\right)\right)
\,,\nonumber\\
T_1^{13,14} & = &  \frac{ m^2 g_A^2 \left(2
	{C}_{22}\left(m^2,m^2+q^2,q^2;M,m,M\right)+{C}_2\left(m^2,m^2+q^
	2,q^2;M,m,M\right)\right)}{4 \pi ^2 F^2}
\,,\nonumber\\
T_2^{13,14} & = &  -\frac{ m^4 g_A^2}{2 \pi
	^2 F^2 q^2}\,  C_{11}\left(m^2,m^2+q^2,q^2;M,m,M\right)
\,,\nonumber\\
T_1^{15} & = &  0
\,,\nonumber\\
T_2^{15} & = &  0
\,,\nonumber\\
T_1^{16} & = &  0 \,,\nonumber\\
T_2^{16} & = & - \frac{ m^2g_A^2}{
	8 \pi ^2 F^2 q^2}
\left(B_0\left(m^2+q^2;M,m\right)+B_1\left(m^2+q^2;M,m\right)\right)
\,,\nonumber\\
T_1^{17} & = & 0
\,,\nonumber\\
T_2^{17} & = &  - \frac{3  m^2 \left(\tau ^3+1\right) g_A^2 }{32
	\pi ^2 F^2 \left(q^2\right)^3}
\left(\left(2
m^2+q^2\right) \left(M^2 B_1\left(m^2+q^2;M,m\right)+2
\left(\left(m^2+q^2\right)
B_{11}\left(m^2+q^2;M,m\right)  \right. \right. \right. \nonumber\\
&+& \left. \left. \left.  B_{00}\left(m^2+q^2;M,m\right)\right)\right)+M^2 \left(-\left(4 m^2+q^2\right)\right)
B_0\left(m^2+q^2;M,m\right)-2 \left(3 m^2+q^2\right) A_0(m)\right) 
\,,
\nonumber\\
T_1^{18,19} & = & -\frac{ m^2 \left(\tau ^3+1\right) g_A^2 }{8 \pi ^2 F^2
	q^2}
\biggl\{ {B}_0\left(m^2+q^2;m,M\right)+2
{B}_1\left(m^2+q^2;m,M\right)+M^2
{C}_0\left(m^2,m^2+q^2,q^2;M,m,M\right) \nonumber\\
&+& \left(2 M^2+q^2\right)
{C}_2\left(m^2,m^2+q^2,q^2;M,m,M\right)+2 q^2
{C}_{22}\left(m^2,m^2+q^2,q^2;M,m,M\right)\biggr\}
\,,\nonumber\\
T_2^{18,19} & = &  \frac{ m^2 \left(\tau ^3+1\right) g_A^2}{4 \pi ^2 F^2 (q^2)^2}  \left(m^2
\left(B_0\left(m^2+q^2;m,M\right)+B_1\left(m^2+q^2;m,M\right)-B_1\left
(m^2;M,m\right) 
\right.\right. \nonumber\\
&+& \left. \left. 
\left(q^2-2 M^2\right)
C_1\left(m^2,m^2+q^2,q^2;M,m,M\right)+q^2
C_{11}\left(m^2,m^2+q^2,q^2;M,m,M\right)\right)+B_{00}\left(q^2;M,M\right)\right)
\,,\nonumber\\
T_1^{20-23} & = &  0
\,,\nonumber\\
T_2^{20-23} & = & - \frac{ m^2 \left(\tau ^3+1\right) g_A^2 }{8 \pi ^2 F^2 (q^2)^2}  \left(M^2
B_0\left(m^2+q^2;M,m\right)+q^2 B_1\left(m^2+q^2;M,m\right)-2 m^2
B_1\left(m^2;M,m\right) \right. \nonumber\\
&+& \left. A_0(m)+A_0(M)\right)
\,,\nonumber\\
T_1^{24,25} & = & \frac{ m^2 \left(\tau ^3+1\right) g_A^2 }{8 \pi ^2 F^2
	q^2} 
\left( {B}_1\left(m^2+q^2;M,m\right)-{B}_1\left(q^2;m,m\right)+\left(q^2-M^2\right)
{C}_2\left(m^2,q^2,m^2+q^2;M,m,m\right) \right.  \nonumber\\
&+& \left. q^2
{C}_{22}\left(m^2,q^2,m^2+q^2;M,m,m\right)\right) 
\,,\nonumber\\
T_2^{24,25} & = & - \frac{ m^2\left(\tau ^3+1\right) g_A^2 }{16 \pi ^2 F^2 (q^2)^2} \left(-q^2
\left(B_0\left(m^2+q^2;M,m\right)+B_1\left(m^2+q^2;M,m\right)  \right.\right. \nonumber\\
&-& \left. \left. 2 m^2
\left(C_{22}\left(m^2,q^2,m^2+q^2;M,m,m\right)+2
C_{12}\left(m^2,q^2,m^2+q^2;M,m,m\right)
\right.\right. \right. \nonumber\\
&+& \left. \left. \left. C_{11}\left(m^2,q^2,m^2+q^2;M
,m,m\right)\right)\right)-4 m^2 M^2
\left(C_0\left(m^2,q^2,m^2+q^2;M,m,m\right) 
\right.\right. \nonumber\\
&+& \left. \left.  C_2\left(m^2,q^2,m^2+q^2;M
,m,m\right)+C_1\left(m^2,q^2,m^2+q^2;M,m,m\right)\right)+A_0(M)\right)
\,,\nonumber\\
T_1^{29-32} & = &  0\,,\nonumber\\ 
T_2^{29-32} & = &  \frac{ m^2 \left(\tau ^3+1\right)  \left({A}_0(M)-2
	{B}_{00}\left(q^2;M,M\right)\right)}{4 \pi ^2 F^2 (q^2)^2
} \,.
\label{DiagRess3}
\end{eqnarray}

\medskip

{\em $O(p^4)$ contributions:}  

\begin{eqnarray}
T_1^{38,39} & = &  -\frac{ \left(2 c_6+c_7\right)  g^2 m \left(\tau ^3-3\right)  }{64 \pi ^2 F^2}
\biggl\{ 4 m^2
\left({B}_0^{(2,1)}\left(q^2;m,m\right)+M^2
{C}_0^{(1,2,1)}\left(m^2,q^2,m^2+q^2;M,m,m\right)\right)  
\nonumber\\
&-& 
{B}_0\left(m^2+q^2;M,m\right)-{B}_1\left(m^2+q^2;M,m\right)+2
{B}_0\left(q^2;m,m\right)+2 {B}_1\left(q^2;m,m\right) \nonumber\\
&+& 2 M^2
\left({C}_0\left(m^2,q^2,m^2+q^2;M,m,m\right)+{C}_2\left(m^2,q^2,m
^2+q^2;M,m,m\right)\right)\biggr\} 
\,,\nonumber\\
T_2^{38,39} & = &  0
\,,\nonumber\\   
T_1^{40} & = &  -\frac{  m^3 g_A^2 \left(2 c_6-c_7 \tau ^3\right) }{16 \pi ^2 F^2}
\biggl\{ 4 q^2
{C}_{22}^{(2,1,1)}\left(q^2,m^2,m^2+q^2;m,m,M\right)+8 q^2
{C}_{12}^{(2,1,1)}\left(q^2,m^2,m^2+q^2;m,m,M\right) 
\nonumber\\
&+& 4 q^2
{C}_{11}^{(2,1,1)}\left(q^2,m^2,m^2+q^2;m,m,M\right)-4 q^2
{C}_{11}^{(2,1,1)}\left(q^2,m^2+q^2,m^2;m,m,M\right)\nonumber\\
&+& 4 q^2
{C}_{11}^{(2,1,1)}\left(m^2+q^2,q^2,m^2;m,M,M\right)+q^2
{C}_0^{(2,1,1)}\left(q^2,m^2,m^2+q^2;m,m,M\right)
\nonumber\\
& - & q^2
{C}_0^{(2,1,1)}\left(q^2,m^2+q^2,m^2;m,m,M\right)+q^2
{C}_0^{(2,1,1)}\left(m^2+q^2,q^2,m^2;m,M,M\right)\nonumber\\
&+& 4 q^2
{C}_2^{(2,1,1)}\left(q^2,m^2,m^2+q^2;m,m,M\right)+4 q^2
{C}_1^{(2,1,1)}\left(q^2,m^2,m^2+q^2;m,m,M\right) \nonumber\\
&-& 4 q^2
{C}_1^{(2,1,1)}\left(q^2,m^2+q^2,m^2;m,m,M\right)+4 q^2
{C}_1^{(2,1,1)}\left(m^2+q^2,q^2,m^2;m,M,M\right)
\nonumber\\
&+& {C}_0\left(q^2
,m^2,m^2+q^2;m,m,M\right)-{C}_0\left(q^2,m^2+q^2,m^2;m,m,M\right)+2
{C}_2\left(q^2,m^2,m^2+q^2;m,m,M\right) \nonumber\\
&+& 2
{C}_1\left(q^2,m^2,m^2+q^2;m,m,M\right)-2
{C}_1\left(q^2,m^2+q^2,m^2;m,m,M\right)\biggr\}
\,,\nonumber\\
T_2^{40} & = &  \frac{  m^3 g_A^2 \left(2 c_6-c_7 \tau ^3\right)  }{16 F^2 \pi ^2 q^2}
\left(-4
{B}_1^{(2,1)}\left(m^2;m,M\right) m^2-4
{B}_1^{(2,1)}\left(m^2+q^2;m,M\right) m^2-2
{B}_{11}^{(2,1)}\left(m^2;m,M\right) m^2
\right.  \nonumber\\
&-&  \left.         
2
{B}_{11}^{(2,1)}\left(m^2+q^2;m,M\right) m^2+4 q^2
{C}_0^{(2,1,1)}\left(q^2,m^2,m^2+q^2;m,m,M\right) m^2
\right.  \nonumber\\
&-&  \left.      
4 q^2
{C}_0^{(2,1,1)}\left(q^2,m^2+q^2,m^2;m,m,M\right) m^2+4 q^2
{C}_0^{(2,1,1)}\left(m^2+q^2,q^2,m^2;m,M,M\right) m^2
\right.  \nonumber\\
&+&  \left.      
8 q^2
{C}_2^{(2,1,1)}\left(q^2,m^2,m^2+q^2;m,m,M\right) m^2-8 q^2
{C}_2^{(2,1,1)}\left(q^2,m^2+q^2,m^2;m,m,M\right) m^2
\right.  \nonumber\\
&+&  \left.      
8 q^2
{C}_2^{(2,1,1)}\left(m^2+q^2,q^2,m^2;m,M,M\right) m^2+4 q^2
{C}_{22}^{(2,1,1)}\left(q^2,m^2,m^2+q^2;m,m,M\right) m^2
\right.  \nonumber\\
&-&  \left.      
4 q^2
{C}_{22}^{(2,1,1)}\left(q^2,m^2+q^2,m^2;m,m,M\right) m^2+4 q^2
{C}_{22}^{(2,1,1)}\left(m^2+q^2,q^2,m^2;m,M,M\right) m^2
\right.  \nonumber\\
&+ &  \left.      
8 q^2
{C}_1^{(2,1,1)}\left(m^2+q^2,q^2,m^2;m,M,M\right) m^2+8 q^2
{C}_{12}^{(2,1,1)}\left(m^2+q^2,q^2,m^2;m,M,M\right) m^2
\right.  \nonumber\\
&+&  \left.      
4 q^2
{C}_{11}^{(2,1,1)}\left(m^2+q^2,q^2,m^2;m,M,M\right) m^2+2
{A}_0^{(2)}(m)-3
{B}_0\left(m^2;m,M\right)+{B}_0\left(m^2+q^2;m,M\right)
\right.  \nonumber\\
&-&  \left.      
3 {B}_1\left(m^2;m,M\right)+{B}_1\left(m^2+q^2;m,M\right)+\left(M^2-
2 m^2\right) {B}_0^{(2,1)}\left(m^2;m,M\right)
\right.  \nonumber\\
&+&  \left.      
\left(-2
m^2+M^2-q^2\right) {B}_0^{(2,1)}\left(m^2+q^2;m,M\right)+M^2
{B}_1^{(2,1)}\left(m^2;m,M\right)+M^2
{B}_1^{(2,1)}\left(m^2+q^2;m,M\right)
\right.  \nonumber\\
&-&  \left.      
3 q^2
{B}_1^{(2,1)}\left(m^2+q^2;m,M\right)-2 q^2
{B}_{11}^{(2,1)}\left(m^2+q^2;m,M\right)-2
\left({B}_{00}^{(2,1)}\left(m^2;m,M\right)
\right.  \right. \nonumber\\
&+&  \left. \left.     
{B}_{00}^{(2,1)}\left(m^2+q^2;m,M\right)\right)+q^2
{C}_0\left(q^2,m^2,m^2+q^2;m,m,M\right)-q^2
{C}_0\left(q^2,m^2+q^2,m^2;m,m,M\right)
\right.  \nonumber\\
&+&  \left.      
2 q^2
{C}_0\left(m^2+q^2,q^2,m^2;m,M,M\right)+q^2
{C}_2\left(q^2,m^2,m^2+q^2;m,m,M\right)-q^2
{C}_2\left(q^2,m^2+q^2,m^2;m,m,M\right)
\right.  \nonumber\\
&+&  \left.      
2 q^2
{C}_2\left(m^2+q^2,q^2,m^2;m,M,M\right)+2 q^2
{C}_1\left(m^2+q^2,q^2,m^2;m,M,M\right)\right) 
\,,\nonumber\\   
T_1^{41,42} & = &   -\frac{  m g_A^2 \left(2 c_6-c_7 \tau ^3\right) }{16 \pi ^2 F^2}
\biggl\{  -{B}_1\left(m^2+q^2;M,m\right)+{B}_0\left(q^2;m,m\right)+2
{B}_1\left(q^2;m,m\right) \nonumber\\
&+& M^2
{C}_0\left(m^2,q^2,m^2+q^2;M,m,m\right)+2 \left(M^2-2 m^2\right)
{C}_2\left(m^2,q^2,m^2+q^2;M,m,m\right)\biggr\}
\,,\nonumber\\
T_2^{41,42} & = &  - \frac{  m^3 g_A^2 \left(2 c_6-c_7 \tau ^3\right) }{8 \pi ^2 F^2 q^2}
\left(-{B}_0\left(m^2+q^2;M,m\right)-{B}_1\left(m^2+q^2;M,m\right)
+{B}_0\left(m^2;M,m\right)+{B}_1\left(m^2;M,m\right)    \right.  \nonumber\\
&-&  \left.      
q^2
\left({C}_0\left(m^2,q^2,m^2+q^2;M,m,m\right)+{C}_2\left(m^2,q^2,m
^2+q^2;M,m,m\right)+{C}_1\left(m^2,q^2,m^2+q^2;M,m,m\right)\right)\right)
\,,\nonumber\\
T_1^{43,44} & = &  - \frac{  g_A^2 m \left(2 c_6 \left(\tau ^3+3\right)
	+c_7 \left(3 \tau ^3+1\right)\right)}{128 \pi ^2
	F^2 q^2}
\biggl\{ 
4 m^2 \left({B}_0\left(m^2+q^2;m,M\right)+2
{B}_1\left(m^2+q^2;m,M\right) \right. \nonumber\\
&-& \left. {B}_0\left(m^2;M,m\right)+q^2
\left(4
\left({C}_{22}\left(m^2,m^2+q^2,q^2;M,m,M\right)+{C}_2\left(m^2,m^
2+q^2,q^2;M,m,M\right)\right) \right. \right. \nonumber\\
&+& \left. \left. {C}_0\left(m^2,m^2+q^2,q^2;M,m,M\right)\right)\right)+q^2 \left(4
    \left({B}_{11}\left(q^2;M,M\right)+{B}_1\left(q^2;M,M\right)\right)
    +{B}_0\left(q^2;M,M\right)\right)  \nonumber\\
    & - & 2 {A}_0(M)\biggr\} 
    \,,
    \nonumber\\
T_2^{43,44} & = & - \frac{  m^3 g_A^2 \left(2 c_6 \left(\tau ^3+3\right)+c_7 \left(3
	\tau ^3+1\right)\right) }{32 \pi ^2 F^2q^2}
\left({B}_0\left(m^2+q^2;m,M\right)+{B}_1\left(m^2+q^2;m,M\right)
+{B}_1\left(m^2;M,m\right)                  \right. \nonumber\\
&-&  \left. 
4 m^2
{C}_{11}\left(m^2,m^2+q^2,q^2;M,m,M\right)-q^2
{C}_1\left(m^2,m^2+q^2,q^2;M,m,M\right)\right) 
\,,\nonumber\\   
T_1^{45,47} & = &   \frac{  g_A^2 m \left(2 c_6 \left(3 \tau ^3-1\right)
	+c_7 \left(3-\tau ^3\right)\right) }{64 \pi ^2 F^2 q^2}
\biggl\{ -M^2 {B}_0\left(m^2;M,m\right)+\left(6 m^2+q^2\right)
{B}_1\left(m^2+q^2;M,m\right)
\nonumber\\
&+& \left(4 m^2+q^2\right)
{B}_0\left(q^2;m,m\right)+M^2 \left(4 m^2+q^2\right)
{C}_0\left(m^2,q^2,m^2+q^2;M,m,m\right) \nonumber\\
&+& 8 m^2 q^2
\left({C}_{22}\left(m^2,q^2,m^2+q^2;M,m,m\right)+{C}_2\left(m^2,q^
2,m^2+q^2;M,m,m\right)\right)-{A}_0(m)\biggr\}
\,,\nonumber\\
T_2^{45,47} & = &  \frac{  m^3 g_A^2}{8 \pi ^2 F^2 (q^2)^2} \left(c_6 \left(\tau ^3-1\right)+c_7\right)
\left(q^2 \left(3
{B}_1\left(m^2+q^2;M,m\right)-{B}_1\left(m^2;M,m\right)
\right. \right. \nonumber\\
&-&  \left. \left.      
4 m^2
\left({C}_{22}\left(m^2,q^2,m^2+q^2;M,m,m\right)+2
{C}_{12}\left(m^2,q^2,m^2+q^2;M,m,m\right) \right. \right.\right.
\nonumber\\
&+& \left. \left.  \left. {C}_{11}\left(m^2,q^2,m
^2+q^2;M,m,m\right)\right)
+  q^2
\left({C}_0\left(m^2,q^2,m^2+q^2;M,m,m\right) \right. \right. \right.
\nonumber\\
&+& \left.\left. \left.  {C}_2\left(m^2,q^2,m
^2+q^2;M,m,m\right)+{C}_1\left(m^2,q^2,m^2+q^2;M,m,m\right)\right)\right) +    
\left(M^2-q^2\right) {B}_0\left(m^2;M,m\right) \right. \nonumber\\
&+& \left. \left(M^2+2 q^2\right)
{B}_0\left(m^2+q^2;M,m\right) +       
2 m^2
\left({B}_1\left(m^2+q^2;M,m\right)+{B}_1\left(m^2;M,m\right)\right)+2 {A}_0(m)-2 {A}_0(M)\right)
\,,\nonumber\\   
T_1^{46,48} & = &   -\frac{ \left(2 c_6-3 c_7\right)  g_A^2 m \left(\tau ^3+1\right) }{64 \pi ^2
	F^2 q^2}  \biggl\{ M^2
{B}_0\left(m^2;M,m\right)+\left(6 m^2+q^2\right)
{B}_1\left(m^2+q^2;M,m\right) \nonumber\\
&+& \left(4 m^2-q^2\right)
{B}_0\left(q^2;m,m\right)-2 q^2 {B}_1\left(q^2;m,m\right)+2 q^2
\left(\left(4 m^2-M^2\right) {C}_2\left(m^2,q^2,m^2+q^2;M,m,m\right) \right.
\nonumber\\
& + &  \left. 4
m^2 {C}_{22}\left(m^2,q^2,m^2+q^2;M,m,m\right)\right)+M^2 \left(4
m^2-q^2\right)
{C}_0\left(m^2,q^2,m^2+q^2;M,m,m\right)+{A}_0(m)\biggr\}
\,,\nonumber\\
T_2^{46,48} & = &  -\frac{ \left(2 c_6-3 c_7\right)  m^3 \left(\tau ^3+1\right) g_A^2 }{32 \pi ^2 F^2 (q^2)^2}
\left(q^2 \left(3
{B}_1\left(m^2+q^2;M,m\right)-{B}_1\left(m^2;M,m\right)
\right. \right. \nonumber\\
&-&  \left. \left.     
4 m^2
\left({C}_{22}\left(m^2,q^2,m^2+q^2;M,m,m\right)+2
{C}_{12}\left(m^2,q^2,m^2+q^2;M,m,m\right)  \right. \right. \right. \nonumber\\
&+& \left. \left. \left. {C}_{11}\left(m^2,q^2,
m^2+q^2;M,m,m\right)\right)
+ q^2
\left({C}_0\left(m^2,q^2,m^2+q^2;M,m,m\right)+{C}_2\left(m^2,q^2,
m^2+q^2;M,m,m\right) \right. \right. \right. \nonumber\\
&+& \left. \left. \left. {C}_1\left(m^2,q^2,m^2+q^2;M,m,m\right)\right)\right) +    
\left(M^2-q^2\right) {B}_0\left(m^2;M,m\right)+\left(M^2+2 q^2\right)
{B}_0\left(m^2+q^2;M,m\right)
\right.  \nonumber\\
&+&  \left.   
2 m^2
\left({B}_1\left(m^2+q^2;M,m\right)+{B}_1\left(m^2;M,m\right)\right)+2 {A}_0(m)-2 {A}_0(M)\right) 
\,,\nonumber\\   
&&\hspace{-1.6cm}T_1^{49-52}  =  -\frac{  m g_A^2 }{16 \pi ^2
	F^2 q^2}  \left(2 c_6+c_7 \tau ^3\right)  \left(M^2
\left({B}_0\left(m^2+q^2;M,m\right)+{B}_0\left(m^2;M,m\right)\right)  \right.  \nonumber\\
&+& \left.  q^2 {B}_1\left(m^2+q^2;M,m\right)+2 {A}_0(m)\right) 
\,,\nonumber\\
&& \hspace{-1.6cm} T_2^{49-52}  =   0
\,,\nonumber\\   
T_1^{53,62} & = &  \frac{  m \left(c_7+2 c_6\tau_3\right) {B}_{00}\left(q^2;M,M\right)}{4
	\pi ^2 F^2 q^2}
\,,\nonumber\\
T_2^{53,62} & = &  0
\,,\nonumber\\   
T_1^{54,61} & = &  -\frac{3  \left(2 c_6+c_7\right)  g^2 m \left(\tau ^3+1\right) }{64 \pi ^2 F^2 \left(q^2\right)^2}
\biggl\{
\left(2
m^2+q^2\right) \left(M^2 {B}_1\left(m^2+q^2;M,m\right)+2
\left(\left(m^2+q^2\right)
{B}_{11}\left(m^2+q^2;M,m\right) \right. \right. \nonumber\\
&+& \left. \left.   {B}_{00}\left(m^2+q^2;M,m\right)\right)\right)+M^2 \left(-\left(4 m^2+q^2\right)\right)
{B}_0\left(m^2+q^2;M,m\right)-2 \left(3 m^2+q^2\right)
{A}_0(m)\biggr\}
\,,\nonumber\\
T_2^{54,61} & = &  0
\,,
\nonumber\\
T_1^{55,56} & = &  \frac{ c_4  m \left(\tau ^3+1\right) {B}_{00}\left(q^2;M,M\right)}{8 \pi
	^2 F^2 q^2}
\,,\nonumber\\
T_2^{55,56} & = &  0
\,,\nonumber\\   
T_1^{57,58} & = &  \frac{  \left(\tau ^3+1\right) \left(c_6 m^2 {A}_0(M)+3 c_2
	{A}_{00}(M)\right)}{8 \pi ^2 F^2 m q^2}
\,,\nonumber\\
T_2^{57,58} & = &  -\frac{3 c_2  m \left(\tau ^3+1\right) A_{00}(M)}{2 \pi ^2 F^2 \left(q^2\right)^2}\,,\nonumber\\
T_1^{59,60} & = &  -\frac{  m \left(c_7+2 c_6\tau_3\right) {A}_0(M)}{8 \pi ^2 F^2 q^2}
\,,\nonumber\\
T_2^{59,60} & = &  0
\,,\nonumber\\   
T_1^{63} & = &   0
\,,\nonumber\\
T_2^{63} & = &  -\frac{3   m^2\left(\tau ^3+1\right) }{4 \pi ^2 F^2 m \left(q^2\right)^3}  \left(2 c_1 m^2 M^2
{A}_0(M)-{A}_{00}(M) \left(c_3 \mathit{d} m^2+c_2
\left(m^2+q^2\right)\right)\right)  
\,,\nonumber\\   
T_1^{64} & = &   -\frac{  m  }{4 \pi ^2 F^2}  
\biggl\{ -2 c_1 M^2 \left(4
\left({B}_{11}^{(2,1)}\left(q^2;M,M\right)+{B}_1^{(2,1)}\left(
q^2;M,M\right)\right)+{B}_0^{(2,1)}\left(q^2;M,M\right)\right) 
\nonumber\\
& + & c_3
\left(M^2 \left(4
\left({B}_{11}^{(2,1)}\left(q^2;M,M\right)+{B}_1^{(2,1)}\left(
q^2;M,M\right)\right)+{B}_0^{(2,1)}\left(q^2;M,M\right)\right)+4
{B}_{11}\left(q^2;M,M\right)
\right. \nonumber\\
& + & \left. {B}_0\left(q^2;M,M\right)+4
{B}_1\left(q^2;M,M\right)\right)+c_2
\left({B}_{00}^{(2,1)}\left(q^2;M,M\right)+4
\left({B}_{001}^{(2,1)}\left(q^2;M,M\right)+{B}_{0011}^{(2,1)}
\left(q^2;M,M\right)\right)\right) \biggr\} 
\,,\nonumber\\
T_2^{64} & = &  \frac{2  c_2  m}{\pi ^2 F^2 q^2} \, {B}_{0000}^{(2,1)}\left(q^2;M,M\right) 
\,,\nonumber\\   
T_1^{65,66} & = & -\frac{ c_3  m \left({B}_0\left(q^2;M,M\right)-4
	{B}_{11}\left(q^2;M,M\right)\right)}{4 \pi ^2 F^2}   
\,,\nonumber\\
T_2^{65,66} & = & - \frac{ c_2  m}{\pi ^2 F^2 q^2} \, {B}_{00}\left(q^2;M,M\right)
\,,\nonumber\\   
T_1^{67} & = &  0
\,,\nonumber\\
T_2^{67} & = & \frac{ c_2  m}{4 \pi ^2 F^2 q^2} \, {A}_0(M)
\,.   
\label{DiagResNeutron4}
\end{eqnarray}

\medskip
\medskip

{\em Contributions of tree diagrams to the elastic parts of $T_1(0,q^2)$ and $T_2(0,q^2)$:}

\begin{eqnarray}
T_{1,{\sf el}}^{1} & = & 0\,,\nonumber\\
T_{2,{\sf el}}^{1} & = &  2m^2  (\tau^3+1)	\frac{ 1}{q^4}\,, \nonumber\\
T_{1,{\sf el}}^{2-3} & = &  m (2c_6  + c_7)(\tau^3+1)  \frac{1}{ q^2}  \,,\nonumber\\
T_{2,{\sf el}}^{2-3} & = &   0 \,, \nonumber\\
T_{1,{\sf el}}^{4-5} & = &  (\tau^3+1)(2d_7  + d_6)    \,,\nonumber\\
T_{2,{\sf el}}^{4-5} & = &   - 4 m^2  (\tau^3+1)(2d_7  + d_6) \frac{1} {q^2}   \,, \nonumber\\
T_{1,{\sf el}}^{6} & = &  (2c_6\tau^3 + c_7)^2  \frac{q^2 + 4m^2}{4 q^2  }  \,,\nonumber\\
T_{2,{\sf el}}^{6} & = &  -m^2 (2c_6\tau^3 + c_7)^2  \frac{1}{ q^2  }  \,, \nonumber\\
T_{1,{\sf el}}^{34-35} & = & 2 m  (\tau^3+1) \frac{\left((2e_{54} + e_{74})q^2 - 4(2e_{105}+e_{106})M^2\right)}{q^2 }   \,,\nonumber\\
T_{2,{\sf el}}^{34-35} & = &  0  \,, \nonumber\\
T_{1,{\sf el}}^{36-37} & = &  \frac{q^2 }{2m}(2c^6\tau^3 + c_7 )(d_6 \tau^3 + 2d_7)  \,,\nonumber\\
T_{2,{\sf el}}^{36-37} & = & - 2 m  (2c^6\tau^3 + c_7 )(d_6 \tau^3 + 2d_7)   \,.
\end{eqnarray}

\medskip

{\em Contributions of $O(p^3)$ one-loop diagrams to the elastic parts of $T_1(0,q^2)$ and $T_2(0,q^2)$:}

\begin{eqnarray}
T_{1,{\sf el}}^{18,19} & = &    -\frac{ m^2 \left(\tau ^3+1\right) g_A^2}{8 \pi ^2 F^2 q^2}
\biggl\{
{B}_0\left(m^2;m,M\right)+2
{B}_1\left(m^2;m,M\right)+M^2
{C}_0\left(m^2,m^2,q^2;M,m,M\right)
\nonumber\\
&+& \left(2 M^2+q^2\right)
{C}_2\left(m^2,m^2,q^2;M,m,M\right)+2 q^2
{C}_{22}\left(m^2,m^2,q^2;M,m,M\right)\biggr\} \,,\nonumber\\
T_{2,{\sf el}}^{18,19} & = &    \frac{ m^2 \left(\tau ^3+1\right) g_A^2 }{4 \pi ^2 F^2 \left(q^2\right)^2}
\biggl\{  m^2
\left({B}_0\left(m^2;m,M\right)+{B}_1\left(m^2;m,M\right)-{B}_1\left(m^2;M,m\right)  \right. \nonumber\\
&+& \left. \left(q^2-2 M^2\right)
{C}_1\left(m^2,m^2,q^2;M,m,M\right)+q^2
{C}_{11}\left(m^2,m^2,q^2;M,m,M\right)\right)+{B}_{00}\left
(q^2;M,M\right)\biggr\}\,,\nonumber\\
T_{1,{\sf el}}^{20-23} & = &    0\,,\nonumber\\
T_{2,{\sf el}}^{20-23} & = &     
-\frac{ m^2 \left(\tau ^3+1\right) g_A^2 \left(M^2
	{B}_0\left(m^2;M,m\right)-2 m^2
	{B}_1\left(m^2;M,m\right)+{A}_0(m)+{A}_0(M)\right)}{
	8 \pi ^2 F^2 \left(q^2\right)^2}\,,\nonumber\\
T_{1,{\sf el}}^{24,25} & = & \frac{ m^2 \left(\tau ^3+1\right) g_A^2}{8 \pi ^2 F^2 q^2}
\biggl\{ {B}_1\left(m^2;M,m\right)-{B}_1\left(q^2;m,m\right)+
M^2 \left(-{C}_2\left(m^2,q^2,m^2;M,m,m\right)\right)  \nonumber\\
&+&  q^2
{C}_{22}\left(m^2,q^2,m^2;M,m,m\right)\biggr\}\,,\nonumber\\
T_{2,{\sf el}}^{24,25} & = & -\frac{ m^2 \left(\tau ^3+1\right) g_A^2 }{16 \pi ^2 F^2 \left(q^2\right)^2}
\biggl\{ 2 m^2
\left(q^2 \left({C}_{22}\left(m^2,q^2,m^2;M,m,m\right)+2
{C}_{12}\left(m^2,q^2,m^2;M,m,m\right) \right.\right. \nonumber\\
&+& \left. \left. {C}_{11}\left(m^2,q^
2,m^2;M,m,m\right)\right)-2 M^2
\left({C}_0\left(m^2,q^2,m^2;M,m,m\right)+{C}_2\left(m^2,q^
2,m^2;M,m,m\right) \right.\right. \nonumber\\
&+& \left. \left. {C}_1\left(m^2,q^2,m^2;M,m,m\right)\right)\right)+{A}_0(M)\biggr\} \,,\nonumber\\   
  T_{1,{\sf el}}^{29-32} & = &    0\,,\nonumber\\
  T_{2,{\sf el}}^{29-32} & = &    
\frac{m^2 \left(\tau ^3+1\right) \left(A_0(M)-2
   B_{00}\left(q^2;M,M\right)\right)}{4 \pi ^2 F^2 \left(q^2\right)^2}\,.
\label{Born3}
\end{eqnarray}

\medskip

{\em Contributions of $O(p^4)$ one-loop diagrams to the elastic parts of $T_1(0,q^2)$ and $T_2(0,q^2)$:}

\begin{eqnarray}
T_{1,{\sf el}}^{43,44} & = & -\frac{ m g_A^2 \left(c_7 \tau ^3+2 c_6\right) }{32 \pi ^2 F^2 q^2}
\biggl\{
4 m^2
\left({B}_0\left(m^2;m,M\right)-{B}_0\left(m^2;M,m\right)+2
{B}_1\left(m^2;m,M\right) \right.
\nonumber\\
&+& \left. q^2 \left(4
\left({C}_{22}\left(m^2,m^2,q^2;M,m,M\right)+{C}_2\left(m^2,m^2,q^2;M,m,
M\right)\right)+{C}_0\left(m^2,m^2,q^2;M,m,M\right)\right)\right)
\nonumber\\
&+& q^2 \left(4
{B}_{11}\left(q^2;M,M\right)+3 {B}_0\left(q^2;M,M\right)+8
{B}_1\left(q^2;M,M\right)\right)-2 {A}_0(M)
\biggr\}\,,\nonumber\\
T_{2,{\sf el}}^{43,44} & = & \frac{ m^5 g_A^2 \left(c_7 \tau ^3+2 c_6\right)
	C_{11}\left(m^2,m^2,q^2;M,m,M\right)}{2 \pi ^2 F^2 q^2} \,,\nonumber\\
T_{1,{\sf el}}^{45,47} & = & \frac{ m g_A^2 }{64 \pi ^2 F^2 q^2}
\biggl\{ c_7 \left(8 m^2 \left(3
{B}_1\left(m^2;M,m\right)+2 {B}_0\left(q^2;m,m\right)+2 M^2
{C}_0\left(m^2,q^2,m^2;M,m,m\right)  \right. \right. \nonumber\\
& + &  \left. \left.  
4 q^2
{C}_{22}\left(m^2,q^2,m^2;M,m,m\right)\right)+M^2 \left(3 \tau ^3-1\right)
{B}_0\left(m^2;M,m\right)+\left(3 \tau ^3-1\right) {A}_0(m)\right)  \nonumber\\
&+ &  2 c_6
\left(4 m^2 \left(\tau ^3-1\right) \left(3 {B}_1\left(m^2;M,m\right)+2
{B}_0\left(q^2;m,m\right)+2 M^2 {C}_0\left(m^2,q^2,m^2;M,m,m\right)
\right. \right. \nonumber\\
& + &  \left. \left.  
+4
q^2 {C}_{22}\left(m^2,q^2,m^2;M,m,m\right)\right)+M^2 \left(1-3 \tau ^3\right)
{B}_0\left(m^2;M,m\right)+\left(1-3 \tau ^3\right)
{A}_0(m)\right)\biggr\}
\,,\nonumber\\
T_{2,{\sf el}}^{45,47} & = &  \frac{ m^3 g_A^2 \left(c_6 \left(\tau ^3-1\right)+c_7\right) }{4 \pi ^2 F^2 \left(q^2\right)^2} 
\biggl\{ 2 m^2
\left(B_1\left(m^2;M,m\right)-q^2 \left(C_{22}\left(m^2,q^2,m^2;M,m,m\right) \right.\right. \nonumber\\
&+& \left. \left. 2
C_{12}\left(m^2,q^2,m^2;M,m,m\right)+C_{11}\left(m^2,q^2,m^2;M,m,m\right)\right)\right
)+M^2 B_0\left(m^2;M,m\right)+A_0(m)-A_0(M)\biggr\} \,,\nonumber\\
T_{1,{\sf el}}^{46,48} & = & -\frac{\left(2 c_6-3 c_7\right)  m \left(\tau ^3+1\right) g_A^2 }{64 \pi ^2 F^2 q^2}
\biggl\{ 2
m^2 \left(3 {B}_1\left(m^2;M,m\right)+2 {B}_0\left(q^2;m,m\right)+2 M^2
{C}_0\left(m^2,q^2,m^2;M,m,m\right)  \right. \nonumber\\
&+& \left. 4 q^2
{C}_{22}\left(m^2,q^2,m^2;M,m,m\right)\right)+M^2
{B}_0\left(m^2;M,m\right)+{A}_0(m)\biggr\} \,,\nonumber\\
T_{2,{\sf el}}^{46,48} & = &  \frac{\left(2 c_6-3 c_7\right)  m^3 \left(\tau ^3+1\right) g_A^2 }{16 \pi ^2
	F^2 \left(q^2\right)^2}
\biggl\{  2
m^2 \left(q^2 \left(C_{22}\left(m^2,q^2,m^2;M,m,m\right)+2
C_{12}\left(m^2,q^2,m^2;M,m,m\right) \right.\right.\nonumber\\
&+& \left.\left. C_{11}\left(m^2,q^2,m^2;M,m,m\right)\right)-B_1\left(m^2;M,m\right)\right)-M^2 B_0\left(m^2;M,m\right)-A_0(m)+A_0(M)\biggr\} \,,\nonumber\\
T_{1,{\sf el}}^{49-52} & = & -\frac{ m g_A^2 \left(c_7 \tau ^3+2 c_6\right) \left(M^2
	{B}_0\left(m^2;M,m\right)+{A}_0(m)\right)}{8 \pi ^2 F^2 q^2}\,,\nonumber\\
T_{2,{\sf el}}^{49-52} & = &  0\,,\nonumber\\
T_{1,{\sf el}}^{53,62} & = & \frac{ m \left(c_7+2 c_6\tau_3\right) {B}_{00}\left(q^2;M,M\right)}{4 \pi ^2
	F^2 q^2} \,,\nonumber\\
T_{2,{\sf el}}^{53,62} & = & 0 \,,\nonumber\\
T_{1,{\sf el}}^{55,56} & = & \frac{c_4  m \left(\tau ^3+1\right) {B}_{00}\left(q^2;M,M\right)}{8 \pi ^2 F^2
	q^2} \,,\nonumber\\
T_{2,{\sf el}}^{55,56} & = & 0 \,,\nonumber\\
T_{1,{\sf el}}^{57,58} & = & \frac{ \left(\tau ^3+1\right) \left(c_6 m^2 {A}_0(M)+3 c_2
	{A}_{00}(M)\right)}{8 \pi ^2 F^2 m q^2}\,,\nonumber\\
T_{2,{\sf el}}^{57,58} & = &  -\frac{3 c_2  m \left(\tau ^3+1\right) A_{00}(M)}{2 \pi ^2 F^2 \left(q^2\right)^2}\,,\nonumber\\
T_{1,{\sf el}}^{59,60} & = & -\frac{m \left(c_7+2 c_6\tau_3\right) {A}_0(M)}{8 \pi ^2 F^2 q^2}\,,\nonumber\\
T_{2,{\sf el}}^{59,60} & = & 0 \,.
\label{Born4}
\end{eqnarray}

\section{Finite-volume sums}
\label{app:FV-integrals}

\eq
\int_V \frac{d^n k}{(2\pi)^n} \frac{\{1,k^\mu,k^\mu k^\nu\}}{(k^2-m^2)^M}
&\equiv& \tilde{\mathscr{A}}_{(M)}^{\{ \hphantom{1},\mu,\mu\nu\}}(m^2) \,,
\nonumber\\[2mm]
\int_V \frac{d^n k}{(2\pi)^n} \frac{\{1,k^\mu,k^\mu k^\nu, k^\mu k^\nu k^\alpha,k^\mu k^\nu k^\alpha k^\beta\}}{(k^2-m_1^2)^M((k-p)^2-m_2^2)^N}&\equiv& \tilde{\mathscr{B}}_{(M,N)}^{\{ \hphantom{1},\mu,\mu\nu,\mu\nu\alpha,\mu\nu\alpha\beta\}}(m_1,m_2;p) \,,
\nonumber\\[2mm]
\int_V \frac{d^n k}{(2\pi)^n} \frac{\{1,k^\mu,k^\mu k^\nu\}}{(k^2-m_1^2)^M((k-p)^2-m_2^2)^N((k-q)^2-m_3^2)^R}&\equiv& \tilde{\mathscr{C}}_{(M,N,R)}^{\{ \hphantom{1},\mu,\mu\nu\}}(m_1,m_2,m_3;p,q) \,,
\nonumber\\[2mm]
\int_V \frac{d^n k}{(2\pi)^n} \frac{\{1,k^\mu,k^\mu k^\nu\}}{(k^2-m_1^2)^M((k-p)^2-m_2^2)^N((k-q)^2-m_3^2)^R((k-r)^2-m_4^2)^S}
&\equiv& \tilde{\mathscr{D}}_{(M,N,R,S)}^{\{ \hphantom{1},\mu,\mu\nu\}}(m_1,m_2,m_3,m_4;p,q,r) \,.
\nonumber\\
\en
Here, $\int_V$ denotes an integral in the finite volume, which really is a sum.
The calculation of these sums by using the Poisson formula is considered
in appendix~\ref{app:Poisson}.

\section{Evaluation of the finite-volume sums using Poisson's formula}
\label{app:Poisson}

The calculation of  finite-volume sums with the use of Poisson's formula is nowadays
a standard procedure. For a detailed introduction, we refer the reader, e.g.,
to Ref.~\cite{Bijnens:2013doa}, and list only the final results here.

\medskip

The following notations are used:
\eq
\dxn&=&\delta(1-x_1-\cdots-x_n)dx_1\cdots dx_n\, ,
\nonumber\\[2mm]
\int\dxn f(x_1,\cdots,x_n)
&=&\int_0^1dx_1\int_0^{1-x_1}dx_2\cdots
\int_0^{1-\cdots - x_{n-2}}dx_{n-1}f(x_1,\cdots,x_{n-1},1-x_1-\cdots -x_{n-1})\, ,
\en
where $n^\mu=(0,{\bf n})$ is a unit spacelike vector, whose components take
integer values. Further, $K_\nu(z)$ denote the modified Bessel functions of
the second kind.

\medskip

The finite-volume sums, which are displayed in appendix~\ref{app:FV-integrals},
contain an infinite-volume piece and finite-volume correction.
The ultraviolet divergences are contained only in the former, while the latter
is ultraviolet-convergent and vanishes exponentially for large values of $L$.
In order to ease the notation, we list only the finite-volume corrections.
The following notation is used:
\eq
\dintL=\intL-\int\frac{d^nk}{(2\pi)^ni}\, .
\en
The full list of the finite-volume sums entering the amplitude
at $O(p^4)$ is given below. Note that in the expressions, which contain only
nucleon propagators, the finite-volume corrections are extremely small (proportional
to the factor $\exp(-mL)$) and can therefore be neglected. We shall indicate these
quantities by writing $\approx 0$ at the end. Also, note that the
structure of the integrands,
which appear in the infinite and in a finite volume, is in general different. This is related
to the fact that  Lorentz-invariance is used in the infinite volume to reduce tensor
integrals to  scalar ones. Some factors in the denominator get canceled during this procedure.
One cannot apply the same trick in a finite volume.

\medskip

{\em One factor in the denominator:}

\eq
I_1&=&\dintL \frac{1}{k^2-m^2}
=-\frac{m}{4\pisq L}\sumn \frac{1}{|{\bf n}|}\, K_1(|{\bf n}|mL)\approx 0\, ,
\\[3mm]
I_2&=&\dintL \frac{1}{(k^2-m^2)^2}
=\frac{1}{8\pisq}\,\sumn K_0(|{\bf n}|mL)\approx 0\, ,
\\[3mm]
I_3&=&\dintL \frac{1}{k^2-M^2}
=-\frac{M}{4\pisq L}\sumn \frac{1}{|{\bf n}|}\, K_1(|{\bf n}|ML)\, ,
\\[3mm]
I_4&=&\dintL \frac{1}{(k^2-M^2)^2}
=\frac{1}{8\pisq}\,\sumn K_0(|{\bf n}|ML)\, ,
\\[3mm]
I_5^{\mu\nu}&=&\dintL \frac{k^\mu k^\nu}{k^2-M^2}
=\frac{M^2}{4\pisq L^2}\,\sumn\biggl\{\frac{1}{{\bf n}^2}\,
K_2(|{\bf n}|ML)g^{\mu\nu}+\frac{ML}{|{\bf n}|^3}\,K_3(|{\bf n}|ML)n^\mu n^\nu\biggr\}\, ,
\\[3mm]
I_6^{\mu\nu}&=&\dintL \frac{k^\mu k^\nu}{(k^2-M^2)^2}=-\frac{M}{8\pisq L}\,\sumn\biggl\{\frac{1}{|{\bf n}|}\,
K_1(|{\bf n}|ML)g^{\mu\nu}+\frac{ML}{{\bf n}^2}\,K_2(|{\bf n}|ML)n^\mu n^\nu\biggr\}\, .
\en

\medskip

{\em Two factors in the denominator:}

\eq
I_7&=&\dintL \frac{1}{(k^2-m^2)((k-q)^2-m^2)}
=\frac{1}{8\pisq}\,\sumn\int_0^1 dx e^{iLx{\bf n}{\bf q}}
K_0(|{\bf n}|L\sqrt{g})\approx 0\, ,
\nonumber\\[2mm]
&&g=m^2-x(1-x)q^2\, .
\\[3mm]
I_8&=&\dintL \frac{1}{(k^2-m^2)((k-p)^2-M^2)^2}
=-\frac{L}{16\pisq}\,\sumn\int_0^1 dx x e^{iLx{\bf p}{\bf n}}
\frac{|\bf n|}{\sqrt{g}}\,K_1(|{\bf n}|L\sqrt{g})\, ,
\nonumber\\[2mm]
&&g=(1-x)m^2+xM^2-x(1-x)p^2\, .
\\[3mm]
I_9&=&\dintL\frac{1}{(k^2-m^2)^2((k-p)^2-M^2)}
=-\frac{L}{16\pisq}\,\sumn\int_0^1 dx(1-x)e^{iLx{\bf n}{\bf p}}
\frac{|{\bf n}|}{\sqrt{g}}\,K_1(|{\bf n}|L\sqrt{g})\, ,
\nonumber\\[2mm]
&&g=(1-x)m^2+xM^2-x(1-x)p^2\,.
\\[3mm]
I_{10}&=&\dintL
\frac{1}{(k^2-m^2)^2((k-q)^2-m^2)}
=-\frac{L}{16\pisq}\,\sumn\int_0^1 dx(1-x)e^{iLx{\bf n}{\bf q}}
\frac{|{\bf n}|}{\sqrt{g}}\,K_1(|{\bf n}|L\sqrt{g})\approx 0\, ,
\nonumber\\[2mm]
&&g=m^2-x(1-x)q^2\, .
\\[3mm]
I_{11}&=&\dintL\frac{1}{(k^2-M^2)((k-p-q)^2-m^2)}=\frac{1}{8\pisq}\,
\sumn\int_0^1 dx e^{iL(1-x){\bf n}({\bf p}+{\bf q})}
K_0(|{\bf n}|L\sqrt{g})\, ,
\nonumber\\[2mm]
&&g=m^2(1-x)+M^2x-x(1-x)(p+q)^2\, .
\\[3mm]
I_{12}&=&\dintL \frac{1}{(k^2-M^2)((k-q)^2-M^2)}=\frac{1}{8\pisq}\,\sumn\int_0^1 dx e^{iLx{\bf n}{\bf q}}\Ka\, ,
\nonumber\\[2mm]
&&g=M^2-x(1-x)q^2\, .
\\[3mm]
I_{13}&=&\dintL\frac{1}{(k^2-m^2)((k-p)^2-M^2)}
=\frac{1}{8\pisq}\sumn\int_0^1dx e^{iLx{\bf n}{\bf p}}
K_0(|{\bf n}|L\sqrt{g})\, ,
\nonumber\\[2mm]
&&g=m^2(1-x)+M^2x-x(1-x)p^2\, .
\\[3mm]
I_{14}^\mu&=&\dintL\frac{k^\mu}{(k^2-m^2)((k-p)^2-M^2)}
\nonumber\\[2mm]
&=&\frac{1}{8\pisq}\sumn\int_0^1dx e^{iLx{\bf n}{\bf p}}
\biggl\{xp^\mu K_0(|{\bf n}|L\sqrt{g})
+in^\mu \frac{\sqrt{g}}{|{\bf n}|}\,K_1(|{\bf n}|L\sqrt{g})
\biggr\}\, ,
\nonumber\\[2mm]
&&g=m^2(1-x)+M^2x-x(1-x)p^2\, .
\\[3mm]
I_{15}^\mu&=&\dintL\frac{k^\mu}{(k^2-M^2)((k-p-q)^2-m^2)}
\nonumber\\[2mm]
&=&\frac{1}{8\pisq}\,\sumn\int_0^1 dx e^{iL(1-x){\bf n}({\bf p}+{\bf q})}
\biggl\{(1-x)(p+q)^\mu K_0(|{\bf n}|L\sqrt{g})
+in^\mu \frac{\sqrt{g}}{|{\bf n}|}\,K_1(|{\bf n}|L\sqrt{g})\biggr\}\, ,
\nonumber\\[2mm]
&&g=m^2(1-x)+M^2x-x(1-x)(p+q)^2\, .
\\[3mm]
I_{16}^\mu&=&\dintL\frac{k^\mu}{(k^2-M^2)^2((k-p)^2-m^2)}
\nonumber\\[2mm]
&=&-\frac{L}{16\pisq}\,\sumn \int_0^1 dx xe^{-iL(1-x){\bf n}{\bf p}}
\biggl\{(1-x)p^\mu \frac{|{\bf n}|}{\sqrt{g}}\,K_1(|{\bf n}|L\sqrt{g})
-in^\mu K_0(|{\bf n}|L\sqrt{g})\biggr\}\, ,
\nonumber\\[2mm]
&&g=m^2(1-x)+M^2x-x(1-x)p^2\, .
\\[3mm]
I_{17}^{\mu\nu}&=&\dintL
\frac{k^\mu k^\nu}{(k^2-m^2)^2((k-q)^2-m^2)}
\nonumber\\[2mm]
&=&-\frac{1}{16\pisq}\,\sumn \int_0^1 dx(1-x)e^{iLx{\bf n}{\bf q}}
\biggl\{
x^2q_\mu q_\nu \frac{|{\bf n}|L}{\sqrt{g}}\,
K_1(|{\bf n}|L\sqrt{g})
+ixL(q_\mu n_\nu+n_\mu q_\nu)K_0(|{\bf n}|L\sqrt{g})
\nonumber\\[2mm]
&-&n_\mu n_\nu \frac{L\sqrt{g}}{|{\bf n}|}\,K_1(|{\bf n}|L\sqrt{g})
-g_{\mu\nu}K_0(|{\bf n}|L\sqrt{g})\biggr\}\, ,
\nonumber\\[2mm]
&&g=m^2-x(1-x)q^2\, .
\\[3mm]
I_{18}^{\mu\nu}&=&\dintL\frac{k^\mu k^\nu}{(k^2-M^2)((k-p-q)^2-m^2)}
\nonumber\\[2mm]
&=&\frac{1}{8\pisq}\,\sumn\int_0^1 dx e^{iL(1-x){\bf n}({\bf p}+{\bf q})}
\biggl\{(1-x)^2(p+q)^\mu(p+q)^\nu K_0(|{\bf n}|L\sqrt{g})
-g^{\mu\nu}\frac{\sqrt{g}}{|{\bf n}|L}\,K_1(|{\bf n}|L\sqrt{g})
\nonumber\\[2mm]
&+&i((p+q)^\mu n^\nu+n^\mu(p+q)^\nu)\frac{(1-x)\sqrt{g}}{|{\bf n}|}\,
K_1(|{\bf n}|L\sqrt{g})
-n^\mu n^\nu\frac{g}{|{\bf n}|^2}\,K_2(|{\bf n}|L\sqrt{g})\biggr\}\, ,
\nonumber\\[2mm]
&&g=m^2(1-x)+M^2x-x(1-x)(p+q)^2\, .
\\[3mm]
I_{19}^{\mu\nu}&=&\dintL \frac{k^\mu k^\nu}{(k^2-M^2)((k-q)^2-M^2)}
\nonumber\\[2mm]
&=&\frac{1}{8\pisq}\,\sumn\int_0^1 dx e^{iLx{\bf n}{\bf q}}
\biggl\{x^2q^\mu q^\nu \Ka +\frac{ix\sqrt{g}}{|{\bf n}|}\,(q^\mu n^\nu+q^\nu n^\mu)\Kb
\nonumber\\[2mm]
&-&g^{\mu\nu}\frac{\sqrt{g}}{L|{\bf n}|}\,\Kb
-\frac{g}{|{\bf n}|^2}\,n^\mu n^\nu \Kc\biggr\}\, ,
\nonumber\\[2mm]
&&g=M^2-x(1-x)q^2\, .
\\[3mm]
I_{20}^{\mu\nu}&=&\dintL \frac{k^\mu k^\nu}{(k^2-M^2)^2((k-q)^2-M^2)}
\nonumber\\[2mm]
&=&-\frac{1}{16\pisq}\,\sumn\int_0^1 dx xe^{-iL(1-x){\bf n}{\bf q}}
\biggl\{\biggl((1-x)^2q^\mu q^\nu\frac{|{\bf n}|}{\sqrt{g}}-\frac{\sqrt{g}}{|{\bf n}|}\,n^\mu n^\nu\biggr) L K_1(|{\bf n}|L\sqrt{g})
\nonumber\\[2mm]
&-&(g^{\mu\nu}+iL(1-x)(q^\mu n^\nu+n^\mu q^\nu)) K_0(|{\bf n}|L\sqrt{g})\biggr\}\,,
\nonumber\\[2mm]
&&g=M^2-x(1-x)q^2\, .
\\[3mm]
I_{21}^{\mu\nu\alpha\beta}&=&\dintL \frac{k^\mu k^\nu k^\alpha k^\beta}{(k^2-M^2)^2((k-q)^2-M^2)}
\nonumber\\[2mm]
&=&-\frac{1}{16\pisq}\,\sumn\int_0^1 dx xe^{-iL(1-x){\bf n}{\bf q}}
\biggl\{(1-x)^4J_0^{\mu\nu\alpha\beta}+(1-x)^3J_1^{\mu\nu\alpha\beta}+(1-x)^2J_2^{\mu\nu\alpha\beta}
\nonumber\\[2mm]
&+&(1-x)J_3^{\mu\nu\alpha\beta}+J_4^{\mu\nu\alpha\beta}\biggr\}\, ,
\nonumber\\[2mm]
J_0^{\mu\nu\alpha\beta}&=&
q^\mu q^\nu q^\alpha q^\beta \frac{L|{\bf n}|}{\sqrt{g}}\,K_1(|{\bf n}|L\sqrt{g})\, ,
\nonumber\\[2mm]
J_1^{\mu\nu\alpha\beta}&=&
(q^\mu q^\nu q^\alpha n^\beta+\mbox{perm})(-iL)K_0(|{\bf n}|L\sqrt{g})\, ,
\nonumber\\[2mm]
J_2^{\mu\nu\alpha\beta}&=&
(q^\mu q^\nu g^{\alpha\beta}+\mbox{perm})(-K_0(|{\bf n}|L\sqrt{g}))
+(q^\mu q^\nu n^\alpha n^\beta+\mbox{perm})\biggl(-\frac{L\sqrt{g}}{|{\bf n}|}\,K_1(|{\bf n}|L\sqrt{g})\biggr)\, ,
\nonumber\\[2mm]
J_3^{\mu\nu\alpha\beta}&=&
(q^\mu n^\nu g^{\alpha\beta}+\mbox{perm})
\biggl(\frac{i\sqrt{g}}{|{\bf n}|}\,K_1(|{\bf n}|L\sqrt{g})\biggr)
+(q^\mu n^\nu n^\alpha n^\beta+\mbox{perm})\biggl(\frac{iLg}{{\bf n}^2}\,K_2(|{\bf n}|L\sqrt{g})\biggr)\, ,
\nonumber\\[2mm]
J_4^{\mu\nu\alpha\beta}&=&
(g^{\mu\nu}g^{\alpha\beta}+\mbox{perm})
\biggl(\frac{\sqrt{g}}{L|{\bf n}|}\,K_1(|{\bf n}|L\sqrt{g})\biggr)
+(g^{\alpha\beta}n^\mu n^\nu+\mbox{perm})
\biggl(\frac{g}{{\bf n}^2}\,K_2(|{\bf n}|L\sqrt{g})\biggr)
\nonumber\\[2mm]
&+&n^\mu n^\nu n^\alpha n^\beta
\biggl(\frac{Lg^{3/2}}{|{\bf n}|^{3/2}}\,K_3(|{\bf n}|L\sqrt{g})\biggr)\, ,
\nonumber\\[2mm]
&&g=M^2-x(1-x)q^2\, .
\en
In the above equations, ``perm'' stands for all permutations of the indices $\mu,\nu,\alpha,\beta$.

\medskip

{\em Three factors in the denominator:}

\eq
I_{22}&=&\dintL\frac{1}{(k^2-m^2)^2((k+p)^2-M^2)((k-q)^2-m^2)}
\nonumber\\[2mm]
&=&\frac{L^2}{32\pisq}\,\sumn\int\dxthree\ei
\frac{{\bf n}^2}{g}\, K_2(|{\bf n}|L\sqrt{g}) x_1\, ,
\nonumber\\[2mm]
&&g=(x_1+x_3)m^2+x_2M^2-x_1x_2p^2-x_1x_3q^2-x_2x_3(p+q)^2\, .
\\[3mm]
I_{23}&=&\dintL\frac{1}{(k^2-m^2)((k+p)^2-M^2)((k-q)^2-m^2)}
\nonumber\\[2mm]
&=&-\frac{1}{16\pisq}\,\sumn \int \dxthree\ei
\frac{|{\bf n}|L}{\sqrt{g}}\,K_1(|{\bf n}|L\sqrt{g})\, ,
\nonumber\\[2mm]
&&g=(x_1+x_3)m^2+x_2M^2-x_1x_2p^2-x_1x_3q^2-x_2x_3(p+q)^2\, .
\\[3mm]
I_{24}^{\mu\nu}&=&\dintL \frac{k^\mu k^\nu}{(k^2-m^2)((k-p)^2-M^2)((k-q)^2-m^2)}
\nonumber\\[2mm]
&=&-\frac{1}{16\pisq}\,\sumn\int\dxthree\ei
\biggl\{
\frac{|{\bf n}|L}{\sqrt{g}}\,r^\mu r^\nu K_1(|{\bf n}|L\sqrt{g})
+i L (r^\mu n^\nu+n^\mu r^\nu)K_0(|{\bf n}|L\sqrt{g})
\nonumber\\[2mm]
&-&g^{\mu\nu}K_0(|{\bf n}|L\sqrt{g})
- \frac{L\sqrt{g}}{|{\bf n}|}\,
n^\mu n^\nu K_1(|{\bf n}|L\sqrt{g})\biggr\}\, ,
\nonumber\\[2mm]
&&r^\mu=x_2q^\mu+x_3p^\mu\, ,
\nonumber\\[2mm]
&&g=(x_1+x_2)m^2+x_3M^2-x_3x_1p^2-x_2x_1q^2-x_2x_3(p-q)^2\, .
\\[3mm]
I_{25}^{\mu\nu}&=&\dintL\frac{k^\mu k^\nu}{(k^2-M^2)^2((k+p)^2-m^2)((k-q)^2-M^2)}
\nonumber\\[2mm]
&=&\frac{L^2}{32\pisq}\,\sumn\int\dxthree x_1\ei
\biggl\{ 
r^\mu r^\nu \frac{{\bf n}^2}{g}\,K_2(|{\bf n}|L\sqrt{g})
+i(r^\mu n^\nu+n^\mu r^\mu)\frac{|{\bf n}|}{ \sqrt{g}}\,K_1(|{\bf n}|L\sqrt{g})
\nonumber\\[2mm]
&-&g^{\mu\nu}\frac{|{\bf n}|}{L\sqrt{g}}\,K_1(|{\bf n}|L\sqrt{g})
-n^\mu n^\nu K_0(|{\bf n}|L\sqrt{g})\biggr\}\, ,
\nonumber\\[2mm]
&&r_\mu=x_3 q_\mu-x_2p_\mu\, ,
\nonumber\\[2mm]
&&g=(x_1+x_3) M^2+x_2m^2-x_1x_2p^2-x_1x_3q^2-x_2x_3(p+q)^2\, .
\\[3mm]
I_{26}^{\mu\nu}&=&\dintL\frac{k^\mu k^\nu}{(k^2-m^2)^2((k+p)^2-M^2)((k-q)^2-m^2)}
\nonumber\\[2mm]
&=&\frac{L^2}{32\pisq}\,\sumn\int\dxthree\ei x_1
\biggl\{
\frac{{\bf n}^2}{g}\,r_\mu r_\nu K_2(|{\bf n}|L\sqrt{g})
+i(r_\mu n_\nu+n_\mu r_\nu)\frac{|{\bf n}|}{\sqrt{g}}\,
K_1(|{\bf n}|L\sqrt{g})
\nonumber\\[2mm]
&-&g_{\mu\nu}\frac{|{\bf n}|}{L\sqrt{g}}\,K_1(|{\bf n}|L\sqrt{g})
-n_\mu n_\nu K_0(|{\bf n}|L\sqrt{g})\biggr\}\, ,
\nonumber\\[2mm]
&&r^\mu=x_3q^\mu-x_2p^\mu\, ,
\nonumber\\[2mm]
&&g=(x_1+x_3)m^2+x_2M^2-x_1x_2p^2-x_1x_3q^2-x_2x_3(p+q)^2\, .
\\[3mm]
I_{27}^{\mu\nu}&=&\dintL\frac{k^\mu k^\nu}{(k^2-M^2)((k+p)^2-m^2)((k-q)^2-M^2)}
\nonumber\\[2mm]
&=&-\frac{1}{16\pisq}\,\sumn\int \dxthree\ei
\biggl\{r^\mu r^\nu\frac{|{\bf n}|L}{\sqrt{g}}\,
K_1(|{\bf n}|L\sqrt{g})
+iL(r^\mu n^\nu+n^\mu r^\nu)K_0(|{\bf n}|L\sqrt{g})
\nonumber\\[2mm]
&-&g^{\mu\nu}K_0(|{\bf n}|L\sqrt{g})
-n^\mu n^\nu\frac{L\sqrt{g}}{|{\bf n}|}\,K_1(|{\bf n}|L\sqrt{g})\biggr\}\, ,
\nonumber\\[2mm]
&&r^\mu=x_3 q^\mu-x_2p^\mu\,,
\nonumber\\[2mm]
&&g=(x_1+x_3) M^2+x_2m^2-x_1x_2p^2-x_1x_3q^2-x_2x_3(p+q)^2\, .
\\[3mm]
I_{28}^{\mu\nu}&=&\dintL\frac{k_\mu k_\nu}{(k^2-m^2)((k+p)^2-M^2)((k-q)^2-m^2)}
\nonumber\\[2mm]
&=&-\frac{1}{16\pisq}\,\sumn\int \dxthree\ei
\biggl\{
r^\mu r^\nu\frac{|{\bf n}|L}{\sqrt{g}}\,K_1(|{\bf n}|L\sqrt{g})
+iL(r^\mu n^\nu+n^\mu r^\nu)K_0(|{\bf n}|L\sqrt{g})
\nonumber\\[2mm]
&-&g^{\mu\nu}K_0(|{\bf n}|L\sqrt{g})
-n^\mu n^\nu\frac{\sqrt{g}L}{|{\bf n}|}\,K_1(|{\bf n}|L\sqrt{g})
\biggr\}\, ,
\nonumber\\[2mm]
&&r^\mu=x_3q^\mu-x_2p^\mu\, ,
\nonumber\\[2mm]
&&g=(x_1+x_3)m^2+x_2M^2-x_1x_2p^2-x_1x_3q^2-x_2x_3(p+q)^2\, .
\en

\medskip

{\em Four factors in the denominator:}
\eq
I_{29}^{\mu\nu}&=&\dintL\frac{k^\mu k^\nu}{(k^2-m^2)((k-p)^2-M^2)((k-q)^2-m^2)((k-p-q)^2-M^2)}
\nonumber\\[2mm]
&=&\frac{L^2}{32\pisq}\,\sumn\int\dxfour\ei
\biggl\{r^\mu r^\nu \frac{{\bf n}^2}{g}\,K_2(|{\bf n}|L\sqrt{g})
+i(r^\mu n^\nu+n^\mu r^\nu)\frac{|{\bf n}|}{\sqrt{g}}\,
K_1(|{\bf n}|L\sqrt{g})
\nonumber\\[2mm]
&-&g^{\mu\nu}\frac{|{\bf n}|}{L\sqrt{g}}\,
K_1(|{\bf n}|L\sqrt{g})
-n^\mu n^\nu K_0(|{\bf n}|L\sqrt{g})\biggr\}\, ,
\nonumber\\[2mm]
&&r^\mu=(x_2+x_4)p^\mu+(x_3+x_4)q^\mu\, ,
\nonumber\\[2mm]
&&g=(x_1+x_3)m^2+(x_2+x_4)M^2+\bigl((x_2+x_4)p+(x_3+x_4)q\bigr)^2
-x_2p^2-x_3q^2-x_4(p+q)^2\, .
\en

\end{document}